\documentclass[usenatbib,iop,numberedappendix]{aeb_emulateapj_2015}
\usepackage{amsmath}
\usepackage{amssymb}
\usepackage{xspace}
\usepackage[normalem]{ulem}
\usepackage{mathrsfs}
\usepackage{longtable}
\usepackage{url}

\usepackage{natbib}

\usepackage[usenames,dvips]{xcolor}

\def\del#1{{}}

\def\yr{{\rm yr}} 

\def\pc{{\rm pc}} 
\def\kpc{{\rm k}\pc} 
\def\Mpc{{\rm M}\pc} 

\def\eV{{\rm eV}} 
\def\meV{{\rm m}\eV} 
\def\MeV{{\rm M}\eV} 
\def\GeV{{\rm G}\eV} 
\def\TeV{{\rm T}\eV} 

\def\G{{\rm G}} 

\def\rad{{\rm rad}} 

\newcommand\bmath[1] {\mbox{\boldmath$\rm #1$}}

\def\F{\mathcal{F}}
\def\Fermi{{\em Fermi}\xspace}

\def\Dpp{D_{\rm pp}}
\providecommand{\e}[1]{\ensuremath{\times 10^{#1}}}

\def\E{\mathcal{E}}

\def\grad{\bmath{\nabla}}
\def\ba{\bmath{\alpha}}
\def\bR{\bmath{R}}

\def\dOAGN{d\Omega'}
\def\dNAGN{\frac{dN_{\rm AGN}}{dt' dE' \dOAGN}}
\def\bx{\bmath{x}}
\def\bhx{\bmath{\hat{x}}}
\def\bhz{\bmath{\hat{z}}}

\def\bhp{\bmath{\hat{p}}}
\def\bhq{\bmath{\hat{q}}}
\def\bk{\bmath{k}}
\def\bp{\bmath{p}}
\def\bq{\bmath{q}}
\def\bs{\bmath{s}}
\def\bW{\bmath{W}}
\def\bv{\bmath{v}}

\def\dthxp{d^3\!x'}
\def\bhl{\bmath{\hat{\ell}}}

\def\bB{\bmath{B}}
\def\bl{\bmath{\ell}}
\def\Dpp{D_{\rm pp}}
\def\Dppn{D_{\rm pp,0}}

\def\epm{{\rm e^\pm}}
\def\e{{\rm e}}
\newcommand\circp[2]{#1^{\circ}\!\!\!.#2}

\begin{document}

\title{
  Bow Ties in the Sky I:\\
  The Angular Structure of Inverse Compton Gamma-ray Halos in the \Fermi Sky
}

\author{
Avery E.~Broderick\altaffilmark{1,2},
Paul Tiede\altaffilmark{2},
Mohamad Shalaby\altaffilmark{1,2,3},
Christoph Pfrommer\altaffilmark{4},\\
Ewald Puchwein\altaffilmark{5},
Philip Chang\altaffilmark{6},
and
Astrid Lamberts\altaffilmark{7}
}
\altaffiltext{1}{Department of Physics and Astronomy, University of Waterloo, 200 University Avenue West, Waterloo, ON, N2L 3G1, Canada}
\altaffiltext{2}{Perimeter Institute for Theoretical Physics, 31 Caroline Street North, Waterloo, ON, N2L 2Y5, Canada}
\altaffiltext{3}{Department of Physics, Faculty of Science, Cairo University, Giza 12613, Egypt}
\altaffiltext{4}{Heidelberg Institute for Theoretical Studies, Schloss-Wolfsbrunnenweg 35, D-69118 Heidelberg, Germany}
\altaffiltext{5}{Institute of Astronomy and Kavli Institute for Cosmology, University of Cambridge, Madingley Road, Cambridge, CB3 0HA, UK}
\altaffiltext{6}{Department of Physics, University of Wisconsin-Milwaukee, 1900 E. Kenwood Boulevard, Milwaukee, WI 53211, USA}
\altaffiltext{7}{TAPIR, Mailcode 350-17, California Institute of Technology, Pasadena, CA 91125, USA}

\shorttitle{Angular Structure of Gamma-ray Halos}
\shortauthors{Broderick et al.}

\begin{abstract}
Extended inverse Compton halos are generally anticipated around extragalactic sources of gamma rays with energies above 100~GeV.  These result from inverse Compton scattered cosmic microwave background photons by a population of high-energy electron/positron pairs produced by the annihilation of the high-energy gamma rays on the infrared background.  Despite the observed attenuation of the high-energy gamma rays, the halo emission has yet to be directly detected.  Here, we demonstrate that in most cases these halos are expected to be highly anisotropic, distributing the up-scattered gamma rays along axes defined either by the radio jets of the sources or oriented perpendicular to a global magnetic field. We present a pedagogical derivation of the angular structure in the inverse Compton halo and provide an analytic formalism that facilitates the generation of mock images.  We discuss exploiting this fact for the purpose of detecting gamma-ray halos in a set of companion papers. 
\end{abstract}


\maketitle

\section{Introduction} \label{sec:I}

The extragalactic gamma-ray sky above 100~GeV is dominated by the unresolved emission of a subset of active galactic nuclei (AGNs) \citep{50GevBkgnd}.  Of these, the vast majority are blazars -- objects with relativistic jets pointed in our direction \citep[see, e.g., Table 5 of ][]{2LAC}.  While the mechanisms by which these very-high energy gamma rays (VHEGRs) are produced remain unclear \citep{1993A&A...269...67M,1998MNRAS.301..451G,2007Ap&SS.309...95B}, their propagation through the cosmos has provided an invaluable means by which to probe the intervening universe \citep{Goul-Schr:67,Stec-deJa-Sala:92,deJa-Stec-Sala:94,Sala-Stec:98,Domi_etal:11,Gilm_etal:12,Vovk+12}.

Extragalactic VHEGR sources are observed to be strongly biased towards low redshifts, with the number of known sources peaking at a redshift of 0.1-0.2 \citep[see, e.g., the redshift distribution of high-syncrotron-peak sources in ][]{2LAC,3LAC}.  This is a natural consequence of the annihilation of VHEGR on the nearly-homogeneous infrared-ultraviolet extragalactic background light (EBL) that permeates the universe, generated by previous generations of stars and quasars \citep[][]{Goul-Schr:67}{, which can thus be probed by propagating VHEGRs \citep{2012Sci...338.1190A,2013APh....43..112D}}.  The center of momentum energy of the VHEGRs and EBL photon exceeds the pair-creation threshold, i.e., $E_\gamma E_{\rm IR}\gtrsim 4 m_e^2 c^4$; and thus VHEGRs can annihilate as they propagate through the EBL.  In practice, the mean free path, $\Dpp$, of VHEGRs to absorption on the EBL is both energy and redshift dependent, depending on the evolving density and spectrum of the EBL.  Higher EBL densities at larger redshifts correspond to shorter $\Dpp$, which peaks in comoving units near $z\approx1$ due to the peak in the cosmological star formation rate around that time.  As such, observations of the absorbed VHEGR spectra of nearby sources have resulted in direct measurements of the EBL \citep{2015ApJ...812...60B}.  From these it is clear that even today the universe is effectively optically thick to VHEGRs, with 
\begin{equation}
  \Dpp(E_\gamma,z) = \Dppn(z)\frac{E_0}{E_\gamma}
  =
  35 
  \left(\frac{1+z}{2}\right)^{-\zeta}
  \left(\frac{E_\gamma}{1~{\rm TeV}}\right)^{-1}
  ~{\rm Mpc}\,,
  \label{eq:Dpp}
\end{equation}
where $E_0$ is a fiducial energy and $\zeta=4.5$ for $z<1$ and $\zeta=0$ for $z\ge1$ \citep{Knei_etal:04, Nero-Semi:09}.

More recently, the evolution of the electron-positron pairs has provided a means to probe the magnetization of the intervening cosmos.  The homogeneity of the EBL coupled with the large $\Dpp$ places many of these pairs within intergalactic voids, where their propagation can be affected by the  intergalactic magnetic field (IGMF). These pairs thus ostensibly permit the only measurement of the large scale IGMF located in the mean density regions far removed from galactic activity.

The high energy of VHEGRs imply similarly high-energy pairs, which correspond to Lorentz factors of
\begin{equation}
  \gamma \approx \frac{E_\gamma}{2 m_\e c^2} \approx 10^6 \frac{E_\gamma}{1~\TeV}\,.
\end{equation}
If nothing else happens, these pairs will cool on the cosmic microwave background (CMB), producing an inverse Compton cascade (ICC) of photons with typical energies of
\begin{equation}
  E_{\rm IC} \approx 2 \gamma^2 E_{\rm CMB} =
  2 \left(\frac{E_\gamma}{1~\TeV}\right)^2 \frac{E_{\rm CMB}}{1~\meV}~\GeV\,.
  \label{eq:ICE}
\end{equation}
That is, the ICC effectively reprocesses an initial TeV gamma ray into many GeV gamma rays.  It is the non-observation of this ICC component in known VHEGR sources that have provided the strongest lower limits on the IGMF to date \citep[see, e.g.,][]{Nero-Semi:09}.

In a number of extragalactic VHEGR sources the intrinsic gamma-ray spectrum can now be constructed after making weak assumptions either about the intrinsic spectrum or the absorption on the EBL, and thus the resulting ICC emission estimated.  This is then limited directly by observations by the \Fermi gamma-ray space telescope, which has ruled out the presence of the ICC component with extraordinary confidence \citep[e.g.,][]{Nero-Semi:09}.  This is natural if an IGMF is present -- within the IGMF the electron-positron pairs deflect away from the line of sight and therefore the up-scattered gamma rays are beamed away from us.  Based upon this scenario typical estimates for the IGMF range from $10^{-17}$--$10^{-15}$~G, depending on assumptions on duty cycles \citep[see e.g.,][]{Nero-Semi:09,Nero-Vovk:10,Tave_etal:10a,Tave_etal:10b,Derm_etal:10, Tayl-Vovk-Nero:11,Taka_etal:11,Dola_etal:11,HESS:2014,Prokhorov:2016}.

This argument for non-zero IGMFs is predicated on three key assumptions:
\begin{enumerate}
\item The intrinsic TeV emission is narrowly beamed.
\item The intrinsic TeV emission spectrum can be reasonably approximated, usually by an exponentially cut-off power law.
\item No other processes control the evolution of the electron-positron pairs.
\end{enumerate}
The first is well supported by the prevalence of blazars among VHEGR-bright AGN specifically, and gamma-ray bright AGN generally \citep{2LAC,3LAC,TeVCat:2008}, which immediately implies that VHEGR emission is localized near the axis of the radio jet.  The second is reasonably well supported by the gamma-ray spectra of nearby VHEGR-bright AGN, the systematic softening of observed gamma-ray spectra with increasing redshift \citep{2012Sci...338.1190A}, and the underlying assumption that the intrinsic spectra only weakly evolve.

The third remains unclear.  Should any alternate cooling mechanism dominate inverse Compton cooling it would preempt the generation of ICCs directly.  A recently suggested example would be cooling mediated by large-scale beam plasma instabilities driven by the bulk motion of the relativistic electron-positron pairs through the ionized intergalactic medium \citep{PaperI,Schl_etal:12,Schlickeiser:2013,Chang:2014}.  While the nonlinear development of these instabilities is uncertain, there are a variety of lines of astronomical evidence that suggest a cooling mechanism with very similar properties is at work \citep{PaperII,PaperIII,PaperIV,PaperV,2015ApJ...811...19L}.  Regardless of the origin of the additional cooling, however, should the ICCs be preempted the resulting gamma-ray spectra would necessarily be consistent with the current lack of a detection of an ICC in known extragalactic VHEGR sources, independent of an IGMF.

The situation would be immediately clarified by the direct detection of the ICC component, laying squarely within the energy range probed the by large-area telescope (LAT) on \Fermi.  Not only would doing so obviate the above assumptions (importantly including the third), it would also settle questions regarding the duration of VHEGR outbursts and thereby reduce the uncertainty on the IGMF lower limits substantially \citep{Derm_etal:10}.  Even in the presence of an IGMF the ICCs are deflected away from the axis along which the original VHEGR emission is beamed.  Thus the ICC component should be visible for observers who either do not see the VHEGR emission, or see only weak VHEGR emission.

\begin{figure}
  \includegraphics[scale=.45]{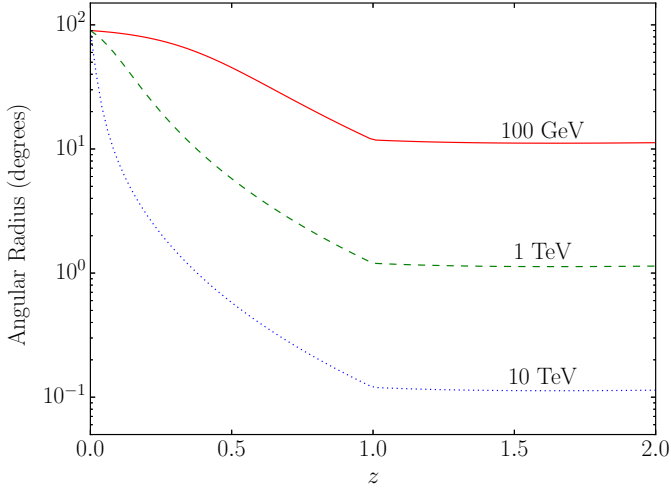}
  \caption{Angular size of VHEGR mean free path as a function of redshift and energy.}  \label{fig:size}
\end{figure}

Efforts to directly detect the ICC component are fundamentally complicated by the large mean free paths of the VHEGRs, typically resulting in halos that extend over many degrees and therefore having a low surface brightness (see, e.g., Figure \ref{fig:size}).  Therefore, all efforts to detect this emission to date have stacked multiple \Fermi gamma-ray images to increase the significance with which the halo emission can be separated from that due to the central source and background. These efforts are further complicated operationally by the uncertainties in the point spread function (PSF) of \Fermi and the spatially varying gamma-ray background \citep[e.g.,][]{Nero-Semi-Tiny-Tkac:11,FLAT-stack:2013}.  As a result, this procedure has led to now disproven detections of an excess \citep{Ando:2010}.  A more recent attempt that utilizes the most recent PSF, reported in \citet{Chen:2015}, nevertheless exhibits similar sensitivies to the uncertain instrument response.

All such efforts have ignored the possibility of structure within the gamma-ray halo.  However, such structure is a natural consequence of either the original beaming of the VHEGRs responsible for the generation of the pairs or the orientation of the IGMF where the pairs are created \citep[see, e.g.,][]{2015JCAP...09..065L}.  Thus for a wide range of parameters for an IGMF we expect highly anisotropic ICC halo.  Here we present semi-analytical computations of the halo structure, explicitly demonstrating the presence of the structure, identifying its origin, and creating the facility to generate mock images of VHEGR sources with realistic ICC halo structures.  In principle, knowledge of the halo structure can aid substantially in efforts to directly detect the ICC component.  We report on an explicit implementation of a method to do so in a companion paper \citep{BowTiesII}. In a companion letter, we apply this formalism to \Fermi-LAT data and discuss the consequences of this measurement for the IGMF \citep{BowTiesIII}.

In Section \ref{sec:qualexp} we describe qualitatively the origin of the anisotropy and how it relates to the structure of the IGMF.  General expressions describing the generation and evolution of the energetic pairs are presented in Section \ref{sec:geneqs}.  Applications to cases of highly tangled and ordered fields are described in Sections \ref{sec:GHiso} and \ref{sec:GHweak} with typical applications shown.  The construction of mock \Fermi images, including various components and instrumental effects is dicussed in Section \ref{sec:mocksgen}.  The dependence of the mock ICC halos for a typical bright \Fermi source are explored in Section \ref{sec:mocksexpl}.  Finally, concluding remarks are collected in Section \ref{sec:conc}. In order to streamline the paper, most of the demonstrations are left to the Appendix.

\section{Qualitative Origin of Structure} \label{sec:qualexp}
\begin{figure*}
  \begin{center}
    \includegraphics[width=0.49\textwidth]{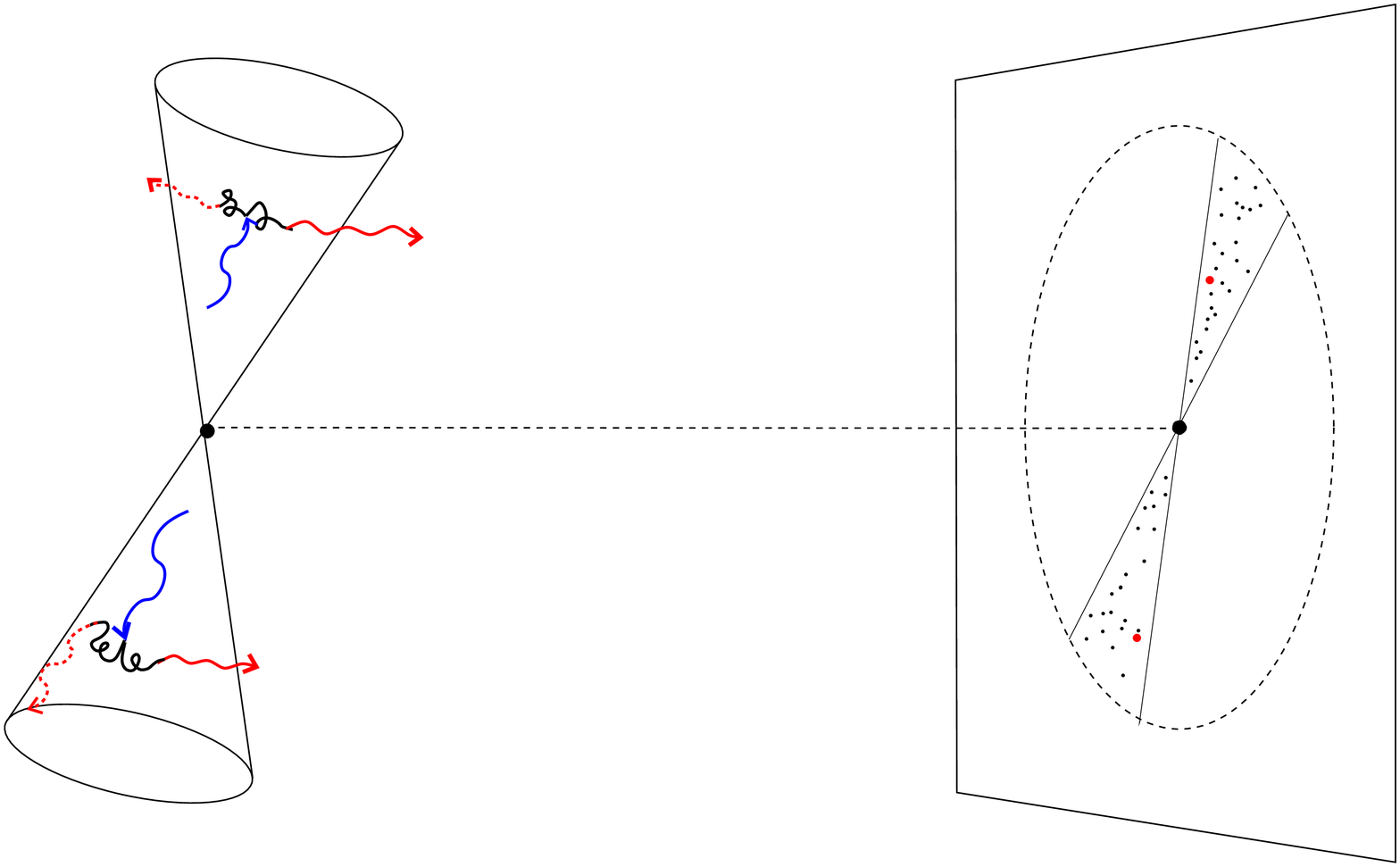}
    \includegraphics[width=0.49\textwidth]{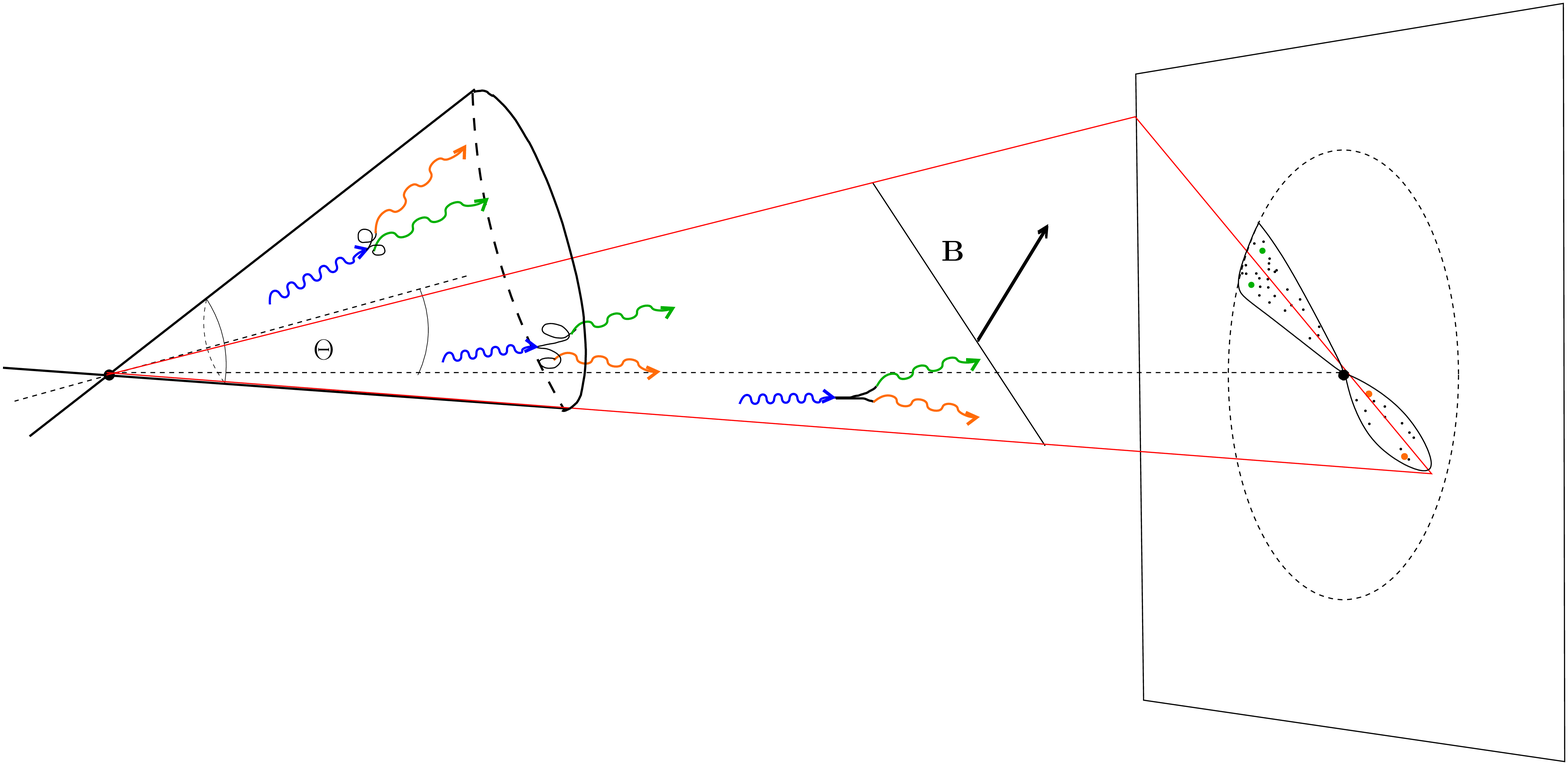}
  \end{center}
  \caption{Cartoons of the mechanisms by which anisotropy in the ICC halos is generated, distinguished by the structure of the underlying IGMF.  Left: For an IGMF tangled on small scales (correlation lengths of $\lambda_B\lesssim3~\Mpc$) the anisotropy is due to the structure of the gamma-ray jet.  Right: For an IGMF that is uniform across the gamma-ray jet (correlation length $\lambda_B\gtrsim30~\Mpc$) the anisotropy is due to the geometry of the gyrating, relativistic pairs.  In the latter, inverse Compton gamma rays from electrons and positrons are shown in red and green, respectively. All distances are not to scale and opening angles are exaggerated for visual purposes.}\label{fig:cartoons}
\end{figure*}
Before describing the creation of physically realistic ICC halos we begin with a summary of the key ideas underlying the structures we will find. This is predicated on the standard picture of the ICC halo formation described in Section \ref{sec:I}: VHEGRs are emitted from AGNs and travel cosmological distances prior to generating energetic electron-positron pairs on the EBL via photon annihilation. Those pairs then inverse Compton up-scatter the CMB to GeV energies over a comparably short distance.  However, for two independent reasons, these ICC halos are {\em not} isotropic.

First, we consider the case of a small-scale tangled IGMF that would isotropize the generated electron-positron pairs in combination with a non-aligned non-thermally dominated AGN, i.e., that the line of sight is not intersecting the jet opening angle. The VHEGRs are originally beamed along the jet axis.  This is evidenced by the overwhelming dominance of blazars in the extragalactic gamma-ray AGN sample \citep{2LAC,3LAC}.  Because the VHEGR mean free path is long in comparison to the inverse Compton cooling time of the resulting pairs this implies that the emission is essentially local, and therefore arises from a biconical region indicated by the radio jet of the source AGN.  If the inverse Compton gamma rays are isotropically emitted, arising, e.g., from a highly tangled IGMF, the spatial structure in the gamma rays generates a resultant structure in the GeV image. This is shown explicitly in the left-hand panel of Figure \ref{fig:cartoons}, along with the associated gamma-ray image of the ICC halo.

Alternatively, the process of gyration in the IGMF also can impart structure on the image if we consider a blazar where the VHEGRs are beamed towards us.  In the presence of an IGMF that is homogeneous on scales comparable to $\Dpp$ electrons and positrons will gyrate on fixed trajectories that emit towards an observer only for a subset of initial injection positions.  This is still superimposed on the jet structure, resulting in an asymmetric image structure if the line of sight does not coincide with the boresight of the jet, shown in the right-hand panel of Figure \ref{fig:cartoons}.  Gamma rays on opposite sides of the original AGN are produced predominantly by different lepton species, i.e., positrons on one side and electrons on the other.

Hence we see that different AGN populations can be used to probe the existence of small- or large-scale tangled IGMFs: while non-aligned non-thermally dominated AGN are suitable for probing small-scale fields, blazar geometries are ideal for exploring large-scale IGMFs because in both cases the associated strong anisotropy of the pair halos enables efficient stacking of the angular power spectra.  In the following sections we make these instances explicit, computing physically realistic halo flux distributions that connect the energy-dependent flux distributions to the underlying physical properties of the VHEGR emission and IGMF geometry.

\mbox{}\\
\section{Formation and Evolution of Intergalactic Pairs} \label{sec:geneqs}
The generation of the halos involves two critical steps that imprint anisotropy upon the resulting gamma-ray sky: the generation of the initial pairs by the VHEGR emission and their subsequent evolution during the inverse Compton up-scattering of the CMB.  To simplify the computation we will exploit the disparity in scales between the inverse Compton cooling length (0.01--0.1~Mpc) and the VHEGR mean free path, and thus assume that the ICC emission is generated in situ follwing pair production. Therefore, the necessary elements are: 
\begin{enumerate}
\item The spatial and energy distribution of VHEGR photons, and the
  corresponding distribution for the injected pairs. 
\item Models for the consequences of pair energy loss and deflection
  in the IGMF. 
\item The inverse Compton cascade (ICC) spectrum due to up-scattering
  the CMB. 
\end{enumerate}
We treat each generally first here, and then specialize to specific limits of interest in Sections \ref{sec:GHiso} and \ref{sec:GHweak}.

\subsection{Spatial and Energy Distribution of Injected Pairs}
\begin{figure}[h]
  \centering
  \includegraphics[scale=0.4]{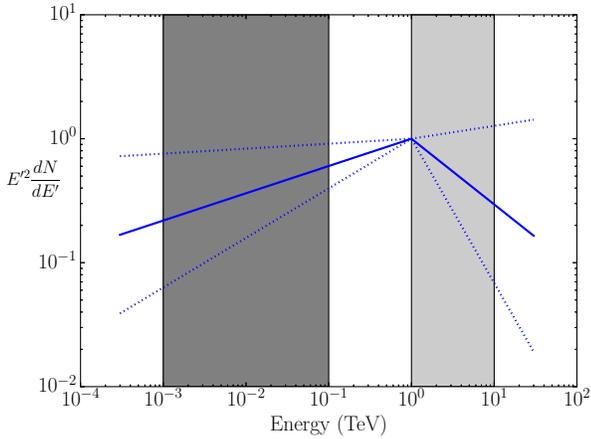}
  \caption{Typical assumed intrinsic spectrum for a bright \Fermi blazar, with a spectral break at 1~TeV.  Dotted lines show the $1\sigma$ variations in the low-energy and high-energy photon spectral indexes $\Gamma_l$ and $\Gamma_h$, respectively, among the hard gamma-ray blazars.  The dark-grey shaded region indicates the 1~GeV--100~GeV region for which we generate ICC halo realizations; the light-grey shaded region indictes the 1~TeV--10~TeV VHEGR band primarily responsible for the ICC halos.} \label{fig:spec}
\end{figure}
The origin of the ultra-relativistic pairs is the VHEGR emission from AGN.  That nearly all of the bright extragalactic TeV sources are blazars implies that the VHEGR emission from TeV-luminous AGN is strongly beamed along the jet.  Thus we begin with a description of the anisotropic VHEGR flux from the AGN itself, before annihilation on the EBL modifies it.  Specifically, we start with the number of VHEGR photons passing through a solid angle $\dOAGN$ in the direction $\bhx'$, with energies between $E'$ and $E'+dE'$, over a time interval $dt'$:
\begin{equation}
  \dNAGN(E',\bhx')\,.
\end{equation}
For our system of coordinates, primed quantities correspond to the frame of the AGN (including the sites of pair production), assumed to be at a fixed redshift, and centered upon the AGN itself.  Unprimed quantities correspond to the observer frame, centered upon the observer.
Note that $\bmath{x}'$ and $\Omega'$ correspond to the position in three-dimensional space and solid angle relative to the AGN, respectively.  The former is related to angular position on the sky, $\ba$, and distance along the line of sight, in the direction $\bhl$, by
\begin{equation}
  \bmath{x}' = D_A(z)\ba + \ell' \bhl\,.
  \label{eq:rRl}
\end{equation}
where $D_A$ is the angular diameter distance.  Hence, for example radial distance from the AGN is given by $r'^2 = D_A(z) \alpha^2 + \ell'^2$ and the angle relative to the jet axis (that is oriented along $\bhz'$), located at an angle $\Theta'$ relative to the line of sight and $\Phi'$ relative to a fiducial direction on the sky (e.g., north) defining the $x$-axis in the image plane, is given by 
\begin{equation}
  \cos\theta'
  = \bhz'\cdot\bhx'
  =
  \frac{\ell'\cos\Theta'   + D_A \sin\Theta' \left(\alpha_x\cos\Phi'+\alpha_y\sin\Phi'\right)}{r'}\,.
  \label{eq:geo}
\end{equation}
Note that along the line of sight, i.e., at $\ba=0$, $r'=\ell'$ and $\theta'=\Theta'$.

As VHEGRs propagate, they annihilate upon the EBL, and thus both the flux and spectrum of the VHEGRs deviates from that emitted by the AGN.  Due to the homogeneity of the EBL, the optical depth to annihilation is given by   
\begin{equation}
  \tau(E',z,r') 
  = 
  \frac{r'}{\Dpp(E',z)}\,,
\end{equation}
where the mean free path is given in Equation (\ref{eq:Dpp}) and depends upon both VHEGR energy and redshift ($z$)\footnote{In principle, the optical depth should be integrated over $r'$ (and thus $z'$).  However, in practice, in the cases of interest $\Dpp$ is much shorter than the Hubble length, justifying our simpler expression.}. The resulting flux distribution of VHEGRs far from the AGN is then
\begin{equation}
  \frac{dN_{\rm VHEGR}}{dt' dE' \dOAGN}(E',\bx')
  =
  \dNAGN(E',\bhx') e^{-\tau(E',z,r')}\,.
\end{equation}

For concreteness we will assume that the spectral and spatial distributions of the VHEGRs are separable.  This may not be a good approximation at large angles if the emission varies throughout the jet.  However, for the applications of primary interest, where the largest effects occur at viewing angles within the relativistic beaming pattern of the jet, i.e., $\Gamma_{\rm jet}^{-1}$, this is well motivated.

For the spectrum, we assume a broken power law, characterized by different photon spectral indexes above ($\Gamma_h$) and below ($\Gamma_l$) some pivot energy ($E_p$); an example spectrum is shown in Figure \ref{fig:spec}.  The assumed values of the photon spectral indexes are set by the typical values for the high synchrotron peak sources (HSPs) in the \Fermi sample: $\Gamma_l\approx1.8$ and $\Gamma_h\approx2.5$ \citep{2LAC,3LAC,2FHL}.  The 1$\sigma$ range about these values for HSPs is given by the dotted lines in Figure \ref{fig:spec}. For the angular flux distribution, we assume a Gaussian jet profile with opening angle $\theta_j$.  That is, we set
\begin{equation}
  \dNAGN(E',\bhx')
  =
  f_0 G(\theta')
  \begin{cases}
    (E'/E_p)^{-\Gamma_l} & E'<E_p\\
    (E'/E_p)^{-\Gamma_h} & E'\ge E_p\,,
  \end{cases}
  \label{eq:VHEGRdist}
\end{equation}
where
\begin{equation}
  G(\theta')
  =
  e^{(\cos\theta'-1)/\theta_j^2} 
  +
  e^{-(\cos\theta'+1)/\theta_j^2}\,.
\end{equation}
The two terms correspond to the two oppositely oriented jets with angular structures set by $G(\theta)$, which for $\theta_j\ll1~\rad$, as is typically the case for the AGN considered here (see Section \ref{sec:mocksexpl}), correspond to Gaussian jet profiles. Note that despite the fact that we are considering a version of the Gaussian jets, the approximate form simplifies significantly as a consequence of writing this in terms of $\cos\theta'$. 

This emission is naturally normalized by the 1~GeV--100~GeV flux observed by \Fermi.  This necessarily depends on $\Theta'$, though for \Fermi objects, in practice does so only weakly; the fact that the hard \Fermi AGN are blazars implies that the viewing angle is smaller than the relativistic beaming angle of the jet (i.e., $\Theta'\lesssim\Gamma_{\rm jet}^{-1}$).  Thus, in the absence of a dominant inverse Compton halo, assuming a roughly fixed effective area from 1~GeV to 100~GeV, the gamma-ray flux observed by \Fermi due to the intrinsic emission of the source between these energies is
\begin{equation}
  \begin{aligned}
    F_{35} 
    &=   
    \int_{1~{\rm GeV}}^{100~\GeV} dE \int_{\Omega'_{\mathrm{Fermi}}} d\Omega'
    \,\frac{dt'}{dt} \frac{dE'_{\rm VHEGR}}{dE} \dNAGN\\
    &=   
    \frac{1}{1+z}\int_{1~{\rm GeV}}^{100~\GeV} dE' \int_{\Omega'_{\mathrm{Fermi}}} d\Omega' \,\dNAGN\\
    &=
    (1+z)^{-3}
    \frac{A_{\rm eff}}{D_A^2} 
    f_0
    E_p
    N(\Gamma_l,z) G(\Theta')\,,
  \end{aligned}
  \label{eq:F35norm}
\end{equation}
where the source-frame solid angle subtended by \Fermi is $A_{\rm eff}/D_A^2(1+z)^2$ and $N(\Gamma_l,\Gamma_h,E_p,z)$ is a normalization factor depending only on the spectral shape:
\begin{equation}
  N(\Gamma_l,\Gamma_h,E_p,z)
  =
  \int_{(1+z)~\GeV/E_p}^{100(1+z)~\GeV/E_p} dx
    \left[ x^{-\Gamma_l} \Theta(1-x) + x^{-\Gamma_h} \Theta(x-1) \right]\,,
\end{equation}
in which $\Theta(x)$ is the Heaviside function.  For our purposes here we will presume that $E_p>100~\GeV$ generally, resulting in the simplified expression $N(\Gamma_l,z)=(1+z)^{1-\Gamma_l} \left[ (1~\GeV/E_p)^{1-\Gamma_l} - (100~\GeV/E_p)^{1-\Gamma_l} \right]/\left(\Gamma_l-1\right)$.  Therefore, the normalization is related to the observed flux via
\begin{equation}
  f_0 = \frac{F_{35}}{E_p} \frac{D_A^2}{A_{\rm eff}}
  \frac{(1+z)^{3}}{N(\Gamma_l,z) G(\Theta')}\,.
\end{equation}

The rate at which pairs are created by the VHEGR is set by the rate at which VHEGRs are annihilated. Each generated pair has an energy $\E'=E'/2$.\footnote{To avoid confusion we will refer to the energies of the emitted gamma rays, including the VHEGRs, as $E'$, lepton energies as $\E'$, and observed gamma-ray energies as $E$.}  Thus, the production rate of electrons/positrons with energies between $\E'$ and $\E'+d\E'$ at position $\bhx'$ is
\begin{equation}
  \begin{aligned}
    &\frac{dN_{\epm}}{dt'd\E'\dthxp}(\E',\bx)\\
    &\quad=
    \frac{\dOAGN dr'}{\dthxp} \frac{dE'}{d\E'} \frac{dN_{\rm VHEGR}}{dt' dE' \dOAGN}(2\E',\bx')
    \frac{d\tau}{dr'}(2\E',z)\\
    &\quad=
    \frac{2}{r'^2}  \frac{e^{-\tau(2\E',z)}}{\Dpp(2\E',z)} \dNAGN(2\E',\bhx')\,, 
  \end{aligned}
  \label{eq:pair_injection_spectrum}
\end{equation}
where the additional factor of $d\tau/dr'$ is the rate at which pairs are being produced locally.  It is straightforward to show that power and number flux are conserved in the above relation, and is shown explicitly in Appendix \ref{app:pairs_energy}.

At this point we have only the injection spectrum of pairs as a function of position.  To progress we need a model for how the pairs evolve and their subsequent inverse Compton emission.

\subsection{Evolution of the Pair Distribution Functions}
We have greatly simplified the computation of the gamma ray flux at the expense of pushing the difficulty onto the determination of the evolved electron distribution.  This evolution is occurring under the action of three processes: injection by the VHEGRs, possible gyration in an IGMF, and inverse Compton scattering.  Different assumptions regarding the scale and strength of the IGMF will impact the evolution and therefore the image.  Throughout we will make use of the ultra-relativistic approximation everywhere permissible, and thus $pc$ and $E$ are used interchangeably. 

Generally the evolution of the distribution function of the pairs, $f_{e^{\pm}}$, is described by the Boltzmann equation,
\begin{equation}
  \dot{f}_{\epm} + \bv\cdot\grad f_{\epm} + \dot{\bp}\cdot\grad_p f_{\epm} = \dot{f}_{\rm scat}+\dot{f}_{\rm inj}\,,
\end{equation}
where $\dot{f}_{\rm scat}$ describes the impact of inverse Compton scattering and $\dot{f}_{\rm inj}$ is the injected distribution of pairs. {$\mathbf{p}$ and $\mathbf{v}$ are the momentum and velocity of the pairs, respectively.}

In general the scattering term must be written in terms of an integral over initial and final particle states, describing the multitude of ways in which particles may scatter into and out of the state of interest.  However, in the soft seed photon limit the scattering term can be substantially simplified as a consequence of the small momentum changes during each scattering (see Appendix \ref{app:Boltzmann}).  That is, in this limit
\begin{equation}
  \dot{f}_{\rm scat}
  \approx
  -\grad_p\cdot\left(\bW f_{\epm}\right)\,,
\end{equation}
where 
\begin{equation}
  \bW=- \frac{4}{3}\frac{\sigma_T u_s}{m_\e^2c^2} p \bp \equiv -\frac{p\bp}{t_{\rm IC} m_\e c}\,,
\end{equation}
in which $t_{\rm IC} = 3 m_\e c / 4 \sigma_T u_s \approx 2.4\times10^{12}(1+z)^{-4}~\yr$ is the asymptotic inverse Compton cooling time for a nonrelativistic lepton.  As a result, the Boltzmann equation simplifies to a strictly partial differential equation.

Strong variability has been observed in many extragalactic VHEGR sources, indicating that variable emission is characteristic.  However, when this occurs over timescales much shorter than the typical cooling time, the impact on the pair distribution is small.  That is, even in the presence of a rapidly varying $\dot{f}_{\rm inj}$, the pair distribution can be well approximated by a stationary solution.  Thus we set $\dot{f}_{\epm}=0$.

The large disparity between the inverse Compton cooling length of the pairs ($700~\kpc$ at 0.5~TeV) and the pair production mean free path of the VHEGR photon (currently $800~\Mpc$ at 1~TeV) implies that this is a fundamentally local and nearly homogeneous process.  The latter immediately implies that $\bv\cdot\grad f_{\epm}$ is small, and may be neglected henceforth.  The former implies that the energy distribution of $\dot{f}_{\rm inj}$ is given by Equation (\ref{eq:pair_injection_spectrum}); we defer a discussion of the angular description to Sections \ref{sec:GHiso} and \ref{sec:GHweak}.

Therefore, in practice we seek to solve
\begin{equation}
  \dot{\bp}\cdot\grad_p f_{\epm} + \grad_p\cdot\left( \bW f_{\epm} \right)= \dot{f}_{\rm inj}\,.
\end{equation}
For non-pathological $\dot{\bp}(\bp)$ and $\bW(\bp)$, this is immediately solvable via the method of characteristics (see Appendix \ref{app:chars}).  For deflections in a locally fixed magnetic field described by a local spherical coordinate system ($q', \vartheta', \varphi'$), it is convenient to express this in terms of a set of polar momentum coordinates aligned with the field ($p',\vartheta'_p,\varphi'_p$), yielding
\begin{multline}
  f_{\epm}(\bx',\bp')
  =
  \frac{m_\e c t_{\rm IC}}{{p'}^4} \int_{p'}^\infty dq'  {q'}^2\\
  \times
  \dot{f}_{\rm inj}\left[\bx',q',\vartheta'_p,\varphi'_p\pm\frac{\omega_B t_{\rm IC} \sin\vartheta'_p}{2}\left(\frac{m_\e^2 c^2}{{q'}^2}-\frac{m_\e^2 c^2}{{p'}^2}\right)\right]\,,
  \label{eq:BES}
\end{multline}
where $\omega_B=eB/m_\e c$ is the cyclotron frequency.

\subsection{Inverse Compton Emission} \label{sec:ICEG}
While we will consider a variety of potential models for the evolution of the pairs after production, their inverse Compton emission may be described by a single framework in which this evolution enters solely as an unknown electron and positron distribution function, $f_{\epm}(\bx',\bp')$.  We construct this framework here, paying particular attention to identifying a number of simplifying assumptions.
  
The first of these is stationarity, which in turn corresponds to a presumption regarding the duty cycle of VHEGR sources.  Short time variability will necessarily impart similar variability on the energy dependence and spatial distribution of the pair population.  Both of these will be smoothed, however, if the source is active for a sufficient period.  In the case of the energy dependence this is set by the inverse Compton cooling timescale, roughly $10^6~\yr$. For the spatial distribution this is determined by the propagation-dependent time delay, which for $2^\circ$ halos ranges from $10^2~\yr$ to $4\times10^3~\yr$, depending on redshift and gamma-ray energy.  Thus, if VHEGR activity persists for longer than $10^6~\yr$ we may assume the underlying pair population has reached steady state in both terms.  Note that this is comfortably short in comparison to typical radio duty cycles of a few times $10^7~\yr$ to a few times $10^8~\yr$ \citep{1987MNRAS.225....1A,2005ApJ...628..629N,2005Natur.433...45M,2008MNRAS.388..625S}, suggesting that this is well justified.  As such, we will not consider non-stationary transients \citep[though see][]{Menz-Schl:15}.

The inverse Compton process is simplified by three additional assumptions, relaxing any of which will make at most small changes to the results.  The first is that a single generation of pairs is created, i.e., that inverse Compton scattered gamma rays do not themselves annihilate on the EBL and generate subsequent generations of additional pairs.  This corresponds to a joint constraint on the energy of the VHEGRs considered and the seed photon population that is inverse Compton scattered, effectively limiting it to the CMB.  The former constitutes a conservative assumption, limiting our attention to VHEGRs with energies less than 10~TeV.  The latter is suppressed by the ratio of the number densities of CMB and EBL photons -- the EBL provides the only substantial alternative population -- of at least $10^4$, and thus is rare in practice \citep{2013APh....43..112D}.  Including additional generations of pairs will modify the GeV signal by a comparable amount, justifying its neglect.

The inverse Compton cooling of the pairs is primarily due to the CMB, with typical energies $\epsilon'\approx 10^{-3}(1+z)$~eV.  For VHEGR energies $E'\ll m_\e^2c^4/\epsilon' \sim 3\times10^2/(1+z)$~TeV this is sufficiently low that the energy of the upscattered photon is well approximated by Equation (\ref{eq:ICE}).  Typically we will assume that only VHEGRs with energies less than 10~TeV are relevant, hence this is well justified.  We will further assume that the CMB is characterized by a monoenergetic photon distribution.  That is the seed photon distribution function is 
\begin{equation}
  f_s(\bx',\bq'_s) = \frac{u_s}{4\pi {q_s'}^3 c} \delta\left(q'_s-\frac{E_{\rm CMB}'}{c}\right)\,,
  \label{eq:fseed}
\end{equation}
where $E_{\rm CMB}'=0.7(1+z)~\meV$ and $u_s$ are the typical energy and energy density of CMB photons, respectively.  Relaxing this makes little difference to the resulting emission spectrum at the expense of complicating the analysis substantially. 

The high energies of the pairs in comparison to the soft seed photons responsible for their cooling implies that the resulting inverse Compton emission is highly beamed in the direction of the lepton propagation.  We will assume that this is exclusively the case, i.e., the inverse Compton photons propagate along the direction of the scattering electrons and positrons.\footnote{This is an excellent approximation for a single scattering due to the high energies of the leptons.  That this remains true after many scatterings is shown in Appendix \ref{app:fsl}.}   In the isotropic approximation this implies that the differential scattering cross section for up-scattering a seed photon with momentum $\bmath{q}'_s$ to a gamma-ray with momentum $\bmath{q'}$ by an electron with initial and final momentum $\bp'$ and $\bp_f'$ is 
\begin{equation}
  \frac{d\sigma}{d^3\!q' d^3\!q'_s d^3\!p' d^3\!p'_f}
  =
  \sigma_T 
  \delta^3\left(\bmath{q}'- \frac{2 q_s p'}{m_\e^2 c^2} \bmath{p}'\right)
  \delta^3(\bmath{p}'_f-\bmath{p}'+\bmath{q}')\,,
  \label{eq:dsig_example}
\end{equation}
where the first $\delta$-function encodes both the forward-propagation approximation and the soft-seed limit, while the second $\delta$-function enforces momentum conservation.  Note that since the pair distribution functions will evolve this need not be in the direction of the original VHEGR. 

Given these assumptions, the rate at which ICC gamma rays with momenta $\bq'$ are produced is 
\begin{equation}
  \frac{dN_{\rm IC,\epm}}{dt' d^3\!x' d^3\!q'}
  =
  \frac{\sigma_T c u_s}{2 q_s' c}
  \left(\frac{m_\e^2 c^2}{2 q_s' q'}\right)^{3/2}
  f_{\epm}\left.\left(\bx',\frac{m_\e c \bq'}{\sqrt{2q_s'q'}}\right)
  \right|_{q_s'=E_{\rm CMB}'/c}\,,
  \label{eq:dNICC}
\end{equation}
the derivation of which can be found in Appendix \ref{sec:DICF}.  From this we may construct the flux as seen at Earth from a source along a line of sight in the direction $\bhl'$, and thus corresponding to momenta $\bq'=E' \bhl'/c$,
\begin{equation}
  \frac{dN_\gamma}{dt' d^2\!x' d^3\!q'}
  =
  \sum_{\rm e^+,e^-}
  \int d\ell' \frac{dN_{{\rm IC},\epm}}{dt' d^3\!x' d^3\!q'} \left(\bx',\frac{E'\bhl'}{c}\right)\,,
  \label{eq:dNg}
\end{equation}
obtained by directly integrating the Vlasov equation \citep[see, e.g.,][]{2006MNRAS.366L..10B}.  In the above we have neglected any subsequent scattering or absorption of the ICC gamma rays, which is well justified by their typical energy, given by Equation (\ref{eq:ICE}) and generally $\lesssim100~\GeV$. 

The above may immediately be converted into a surface brightness as measured by \Fermi: 
\begin{equation}
  \begin{aligned}
    \frac{dN_\gamma}{dt dE d^2\!\alpha}
    &=
    \frac{dt'}{dt} \frac{dE'}{dE}
    \int_{\Omega'_{\rm Fermi}} \frac{q'^2 d\Omega'_q}{c} \frac{d^2\!x'}{d^2\!\alpha} \frac{dN_\gamma}{dt' d^2\!x' d^3\!q'}\\
    &=
    \sum_{\rm e^+,e^-}
    \frac{A_{\rm eff}}{(1+z)^4}
    \frac{3 m_\e^4 c^5}{16 t_{\rm IC} E_{\rm CMB}^2}
    \sqrt{\frac{E}{2E_{\rm CMB}}}\\
    &\qquad\times
    \int d\ell' 
    f_{\epm}\left(\bx',m_\e c \sqrt{\frac{E}{2 E_{\rm CMB}}}\bhl'\right)\,,
  \end{aligned}
  \label{eq:image_master}
\end{equation}
where $E_{\rm CMB}'=(1+z) E_{\rm CMB}$ and $E'/E_{\rm CMB}'=E/E_{\rm CMB}$ were used.  At this point it is necessary to explicitly define the evolution model for the electrons and positrons and the corresponding $f_{\epm}(\bx,\bp)$.  Thus, we now turn to explicitly considering the injection model described by Equation (\ref{eq:pair_injection_spectrum}) in the context of two limiting evolutionary models.  We defer until Sections \ref{sec:mocksgen} and \ref{sec:mocksexpl} the construction of mock images, which include additional contributions (central source and gamma-ray background), instrumental effects (i.e., the PSF), and draw explicit subsamples with the appropriate statistics. 

The physical picture of ICCs reprocessing the VHEGR emission from the central source to lower-energy gamma rays permits a variety of integral relationships between the two.  These are used to check both the results of this section as well as the following two where the energy-dependent halo structures, differing in the assumed structure of the IGMF, in Appendix \ref{app:checks}.  At the same time a clear discussion of the anticipated redshift dependence may be found in Appendix \ref{app:redshift}.

\section{Gamma-ray Halos from Tangled Fields} \label{sec:GHiso}
The first limit we consider assumes rapid isotropization of the pairs, i.e., the pair momenta become isotropic on a timescale short in comparison to the inverse Compton cooling time.  This naturally occurs if the IGMF is sufficiently strong and tangled.  This occurs in two steps, first isotropizing in the azimuthal direction about the magnetic field due to gyration, and second isotropizing in the poloidal direction due to variations in the local magnetic field orientation.  In practice, this may occur only statistically should the gyration radius be small in comparison to the correlation length, i.e., after coarse-graining over the cooling scale, or via diffusion if the gyration radius is large in comparison to the correlation length, $\lambda_B$.

Quantitatively, the IGMF strength and value of $\lambda_B$ at which the pairs effectively isotropize depends on the gamma-ray energy and source redshift.  Pairs with high energies will gyrate the slowest, placeing the strongest limit on IGMF strength; at 10~TeV a lepton will make a full gyration within the inverse Compton cooling time when  $B\ge10^{-15}~\G$.  If the current lower-limits on the strength of the IGMF are taken at face value, they imply this is likely to be the case \citep[however, cf.][]{Derm_etal:10}.  The statistical isotropization requires tangling of the IGMF on scales comparable to the jet opening angle at distances of $\Dpp$.  In turn, this requires $\lambda_B\ll\theta_j \Dpp$, which is smallest at 10~TeV and $z=0$: $\lambda_B\ll3~\Mpc$.  Therefore, nominally the applicability of the rapid isotropization limit depends solely on the structure of the IGMF.

\begin{figure*}
  \begin{center}
    \includegraphics[width=0.24\textwidth]{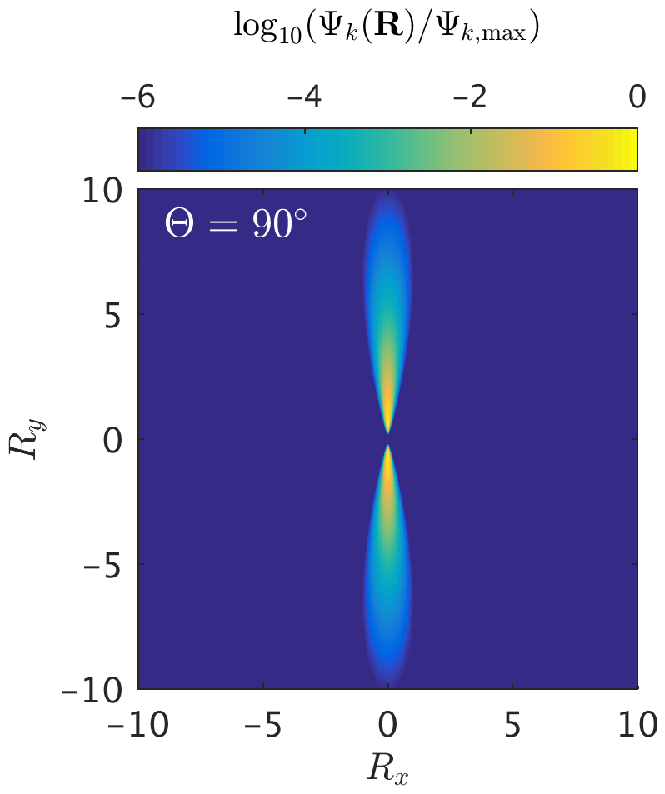}
    \includegraphics[width=0.24\textwidth]{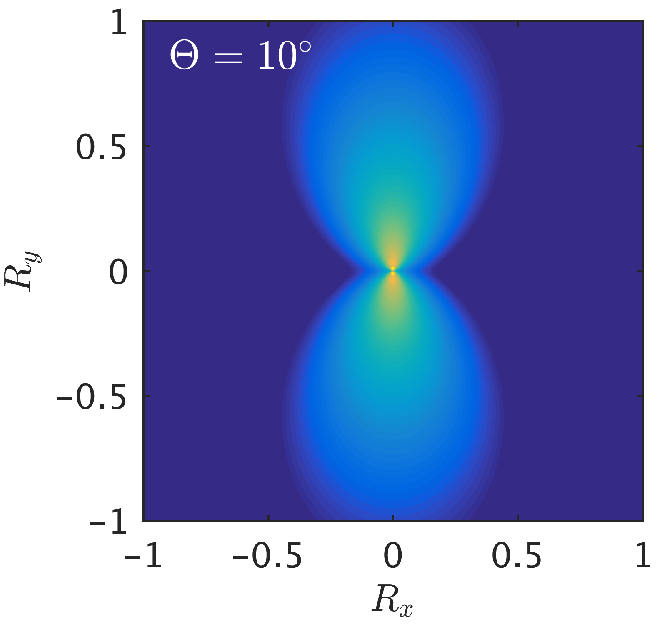}
    \includegraphics[width=0.24\textwidth]{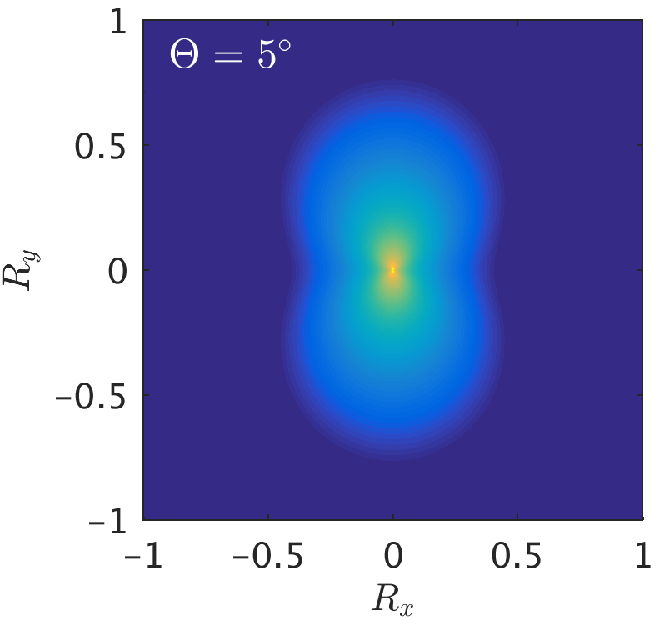}
    \includegraphics[width=0.24\textwidth]{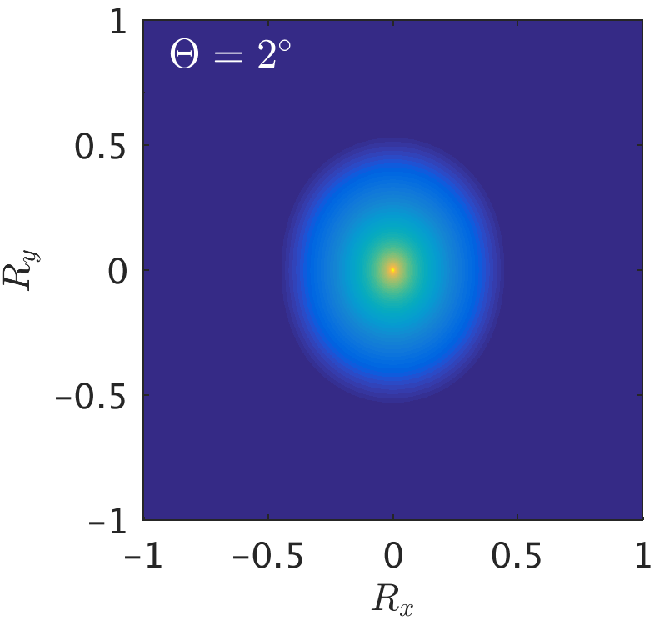}
  \end{center}
  \caption{$\Psi_k(\bR)$ for various viewing angles, $\Theta$, and $k=-0.6$ (i.e. $\Gamma_h=2.6$).  At oblique viewing angles ($\Theta\gtrsim10^\circ$) the anisotropic structure of follows closely that of the jet.  At acute viewing angles ($\Theta\lesssim10^\circ$) geometric foreshortening curtails this structure, and by $\Theta\approx2^\circ$ it is nearly isotropic.}\label{fig:Psi}
\end{figure*}

\begin{figure*}
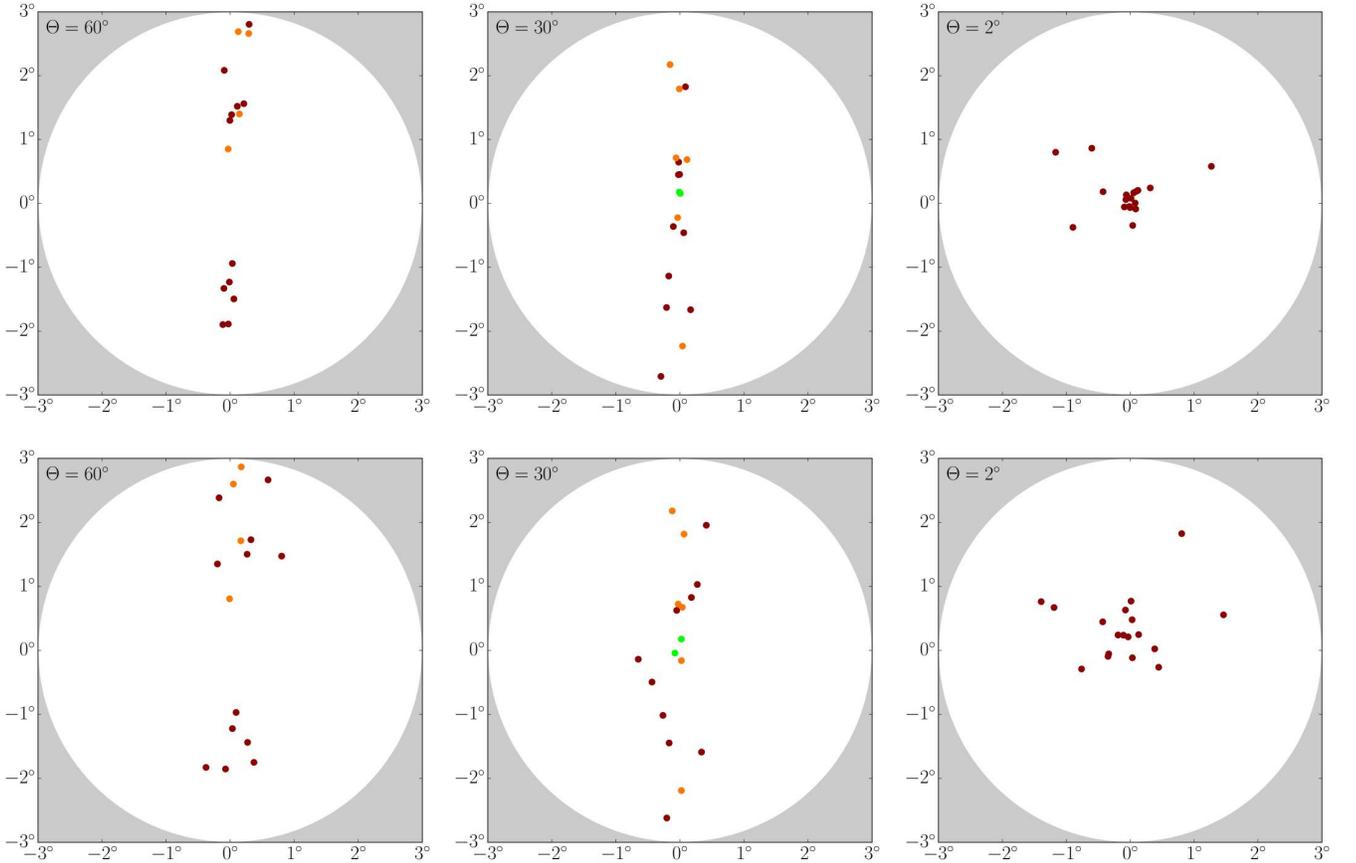

  \begin{center}
    \includegraphics[width=0.32\textwidth]{fig5a.eps2}
    \includegraphics[width=0.32\textwidth]{fig5b.eps2}
    \includegraphics[width=0.32\textwidth]{fig5c.eps2}\\
    \includegraphics[width=0.32\textwidth]{fig5d.eps2}
    \includegraphics[width=0.32\textwidth]{fig5e.eps2}
    \includegraphics[width=0.32\textwidth]{fig5f.eps2}
  \end{center}
  \caption{Example realization of ICC halo in the presence of a small-scale tangled magnetic field viewed at a jet inclination of $60^\circ$ (left), $30^\circ$ (middle), and $2^\circ$ (right) before (top) and after (bottom) convolving with the \Fermi Pass 8R2\_V6-front PSF.  In all cases, the on-axis number of gamma rays was assumed to be 5000 and the source placed at $z=0.3$, $\theta_j=3^\circ$, and $\Gamma_h=2.6$.  The energy of the gamma rays are indicated by color: with 1~GeV-3~GeV in dark red, 3~GeV-10~GeV in orange, 10~GeV-30~GeV in green.}\label{fig:isoex} 
\end{figure*}

\subsection{Pair Distribution Function}
Within this context we approximate the effect of rapid isotropization as an instantaneous redefinition of the pair injection model: 
\begin{equation}
  \dot{f}_{\rm inj, \epm}(\bx',\bp') = \frac{c}{4\pi {p'}^2}\frac{dN_{\epm}}{dt'd\E'\dthxp}\left(p'c,\bx'\right)\,.
\end{equation}
The isotropy renders the azimuthal evolution of the distribution function during inverse Compton cooling moot, and thus 
\begin{equation}
  f_{\epm}(\bx',\bp') = \frac{t_{\rm IC} m_\e c^2}{4\pi {p'}^4} \int_{p'}^\infty d\tilde{p}' \frac{dN_{\epm}}{dt'd\E'\dthxp}\left(\tilde{p}'c,\bx'\right)\,.
\end{equation}
For the expression for the pair injection spectrum in Equation (\ref{eq:pair_injection_spectrum}) this may be integrated explicitly at each $r'$, producing 
\begin{equation}
  \begin{aligned}
    f_{\epm}(\bx',\bp')
    &=
    \frac{t_{\rm IC} m_\e c^5}{4 \pi E_p^4}
    \frac{dN_{\epm}}{dt'd\E'\dthxp}\left(\frac{E_p}{2},\bx'\right)
    e^{\tau(E_p,z)}\\
    &\quad~~
    \times \left(\frac{\E'}{E_p}\right)^{-4}\int_{\E'}^\infty d\E'\, \left(\frac{2\E'}{E_p}\right)^{1-\Gamma_h} e^{-\tau(2\E',z)}\\
    &=
    \frac{t_{\rm IC} m_\e c^5}{8 \pi E_p^3}
    \frac{dN_{\epm}}{dt'd\E'\dthxp}\left(\frac{E_p}{2},\bx'\right)
    e^{\tau(E_p,z)}\\
    &~~~
    \times
    \left(\frac{\E'}{E_p}\right)^{-(\Gamma_h+2)}
    \left(\frac{\E'}{E_D}\right)^{\Gamma_h-2}
    \Gamma\left(2-\Gamma_h,\frac{2\E'}{E_D}\right)\,.
  \end{aligned}
\end{equation}
where $E_D\equiv E_p/\tau(E_p,z)$ and $\Gamma(k,x)$ is the incomplete Gamma function of order $k$ \citep{A&S}.

\subsection{Gamma-ray Halo Structure}
With the distribution function in hand we may now compute the gamma-ray halo flux distribution from Equation (\ref{eq:image_master}) explicitly.  This is simplified in this case by the fact that $f_{\epm}(\bx',\bp')$ is separable in $\bx'$ and $\bp'$.  Up to an energy-dependent coefficient, the flux distribution can be expressed as an energy-dependent rescaling of a single spatial function, i.e., the images at different energies form a homologous class of images described by 
\begin{equation}
  \Psi_k(\bR') \equiv
  \int_{-\infty}^\infty d\ell' r'^{-k-2} \Gamma(k,r')
  G(\theta')\,,
  \label{eq:Psidef}
\end{equation}
where $\bR'=D_A(z)\ba$, $r'\equiv\sqrt{R'^2+\ell'^2}$, and $\theta'$ is given by Equation (\ref{eq:geo}).  Hence, in practice, for a given source structure and geometry (i.e., $\theta_j$, $\Theta$, $\Phi$, and $\Gamma_h$) it is sufficient to numerically compute $\Psi_k(\bR)$.  This function is shown in Figure \ref{fig:Psi} for various viewing angles.

In terms of $\Psi_k$, the full energy-dependent, gamma-ray surface brightness distribution is given by 
\begin{multline}
  \frac{dN_{\gamma}}{dt dE d^2\!\alpha}
  =
  F_{35}
  \frac{m_\e c^2}{\delta_A^2 E_{\rm CMB}^2}
  \frac{3 (1+z)^{-1}}{4\pi N(\Gamma_l,z) G(\Theta')}
  \left( \frac{m_\e c^2}{E_p}\right)^{1-\Gamma_h}\\
  \times
  \left(\frac{2E}{E_{\rm CMB}}\right)^{-\Gamma_h/2}
  \Psi_{2-\Gamma_h}\left(
  \sqrt{\frac{2E}{E_{\rm CMB}}}
  \frac{\ba}{\delta_A}
  \right)\,,
  \label{eq:Fiso}
\end{multline}
where we have summed over the two species and $\delta_A\equiv (E_p/m_\e c^2) \Dpp(E_p,z)/D_A(z)$ is a measure of the angular size of mean free path of VHEGRs, modulo the large energy ratio appearing in the prefactor.  Note that the normalization is fully determined by the observed $F_{35}$ and the assumed intrinsic spectral shape.  That this reproduces the integrated flux in the parent VHEGRs is verified in Appendix \ref{app:check_iso}.  

The spatial structure of the ICC halo follows that of the underlying gamma-ray jet.  That is, since the pairs are isotropized, and therefore emit via inverse Compton isotropically, their gamma-ray emission maps out the pair injection sites.  Since the VHEGR emission is confined to jets, so are the pairs, and consequently so is their emission.  As a result, the orientation of the ICC halos in this limit follows that of the gamma-ray jet and presumably the much smaller-scale radio jets. 

Generally the ICC halo spectrum will vary across the image due to the different mean free paths of the underlying VHEGRs.  However, it is possible to characterize the large-scale ICC halo spectrum.  $\Psi_k(\bR)$ peaks near a fixed $\bR$, falling rapidly thereafter, corresponding to the exponential suppression of the injection of pairs by the VHEGR after a mean free path.  As a result, for sufficiently extended images the internal structure of $\Psi_k(\bR)$ is unimportant and the ICC halo spectrum is proportional to $E^{-1-\Gamma_h/2}$.  The resolved spectra are generally harder due to the smaller $\Dpp$. 

For the nearly flat spectrum sources under consideration here typical values for the low- and high-energy photon spectral indexes are $\Gamma_l\approx1.8$ and $\Gamma_h\approx2.5$.  Given the latter the typical ICC halo photon spectral index is $2.25$, intermediate to the two, and importantly, intermediate between that of the source and the soft background (which has a typical photon spectral index of 2.4).  This has two relevant consequences.  First, it is difficult to spectrally separate the ICC halo emission from the source or the background.  Second, since the ICC halos are marginally harder than the background the marginal value of low-energy gamma rays ($\lesssim1~\GeV$) is small, typically adding more noise than ICC halo signal.  For techniques that leverage the anisotropic structure of the halos this is compounded by the growth of the PSF at low energies.  Hence, our decision to restrict our attention to gamma rays with energies above 1~GeV.

\subsection{Numerical Implementation}
The creation of mock \Fermi images will ultimately require the rapid generation of samples drawn from the energy-dependent flux distribution in Equation (\ref{eq:Fiso}).  Despite its simplicity in comparison to Monte Carlo propagation schemes, this remains nontrivial, requiring a handful of practical optimizations.

We begin by constructing $\Psi_{2-\Gamma_h}(\bR)$ numerically and tabulating the result logarithmically in $R$, ranging from $10^{-8}$ to $10^4$.  This is then integrated over $\bR$ and $E$ between 1~GeV and 100~GeV to obtain the ratio between the number of gamma rays in the halo to that contained within the direct source component, i.e., the halo-core flux ratio.  We then construct the following additional marginalized probability distributions for a halo photon on the image:
\begin{equation}
  \begin{aligned}
    P(E) &\propto \int d^2\!\alpha \frac{dN_{\gamma}}{dt dE d^2\!\alpha}\\
    P(\alpha|E) &\propto \int d\theta_\alpha \frac{dN_{\gamma}}{dt dE d^2\!\alpha}\\
    P(\theta_\alpha|E,\alpha) &\propto \frac{dN_{\gamma}}{dt dE d^2\!\alpha}\,,
  \end{aligned}
\end{equation}
where $\alpha=|\bmath{\alpha}|$, $\theta_\alpha$ is the polar angle on the sky relative to the projected jet axis, and the normalizations are set appropriately.  The first is already known analytically, $P(E)\propto E^{-1-\Gamma_h/2}$.  The latter two are constructed numerically in terms of integrals over $\Psi_{2-\Gamma_h}(\bR)$ and tabulated.  To generate a gamma-ray event the procedure is then
\begin{enumerate}
\item Draw a random number to determine if the gamma ray is in the halo or the source based upon the halo-core flux ratio.  If it is in the halo:
\item Draw an energy from $P(E)$.
\item Draw a value for $\alpha$ from $P(\alpha|E)$.
\item Draw a value for $\theta_\alpha$ from $P(\theta_\alpha|\alpha,E)$.
\end{enumerate}

\subsection{Example Halo Realizations}
Example realizations of ICC halos associated with a small-scale, tangled IGMF are shown in Figure \ref{fig:isoex} for various viewing angles.  As anticipated implicitly by the energy dependence of $\Dpp$ and explicitly by Equation (\ref{eq:Fiso}), the energy of the gamma rays are strongly correlated with the distance from the central source.  High-energy halo photons are found at the smallest angular offsets, while low-energy halo photons compose the bulk of the distant halo.

At large viewing angles, the expected angular structure of the halo is easily visible, both before and after convolution with the \Fermi Pass P8R2\_V6 PSF for front-converted events (see Section \ref{sec:PSF}).  Small viewing angles, comensurate with those implied by observations of gamma-ray blazars, result in an extreme foreshortening that eliminates much of the anisotropy.

In all cases the total number of intrinsic source gamma rays seen by an observer looking down the gamma-ray jet axis, i.e., the on-axis source flux, was held fixed, corresponding to a fluence of 5000~ph, chosen to match a typical, bright, hard \Fermi source.  The comparatively small halo fluence is a result of the isotropy of the ICC component, which follows from the rapid isotropization of the pairs, and dilutes the energy flux from the initially highly-beamed VHEGR jet by a factor of $2\theta_j^{-2}$.  For the parameters employed here, this corresponds to a reduction by a factor of nearly $7\times10^2$.

\section{Gamma-ray Halos from Ordered Fields} \label{sec:GHweak}
The second limit we consider assumes that the distribution function in minimally isotropized, evolving over the entire cooling time under the action of a uniform IGMF.  We further assume that this occurs coherently across the jet, requiring large IGMF coherence lengths.  Both the magnitude and nature of the limit on $\lambda_B$ depends on redshift and energy.  For sufficiently nearby objects $\lambda_B$ is limited by the proper distance, $D_P$ -- for Mkn 421 this is $130~\Mpc$.  For more distant objects the relevant limit is set by $\Dpp$.  For a typical ICC gamma ray (10~GeV) from a typical moderate-redshift source ($z=0.2$) this is also roughly $130~\Mpc$.  Thus, typically, the IGMF may be treated as uniform if $\lambda_B\gg10^2~\Mpc$.

This presents a natural counter-point to the isotropized case.  It also permits a simplified geometric picture since only a small subset of the gyrating pairs will gyrate into our line of sight and therefore generate gamma rays in our direction.

The characteristic radial extent of halos in this case is determined by both the deflection angle of typical pairs and the opening angle of the jet.  For observed gamma-ray energy $E$, the former typically limits the angular extent of the halos to
\begin{equation}
  \Delta\alpha_{\rm def}
  \approx
  \frac{\omega_B t_{\rm IC}}{\gamma^2}
  \approx
  \circp{1}{5}
  \left(\frac{B}{10^{-16}~\G}\right) \left(\frac{E}{10~\GeV}\right)^{-1} \left(1+z\right)^{-4}\,,
  \label{eq:adef}
\end{equation}
which for $z\lesssim 1$ is generally substantial for fields consistent with the arguments of \citet{Nero-Vovk:10}.  This is not surprising given that those require similar size deflections to explain the absence of an ICC component in the spectra of VHEGR sources.  The latter sets a typical angular scale for blazar jets independent of IGMF strength of
\begin{equation}
  \Delta\alpha_{\rm jet}
  \approx
  \frac{\theta_j \Dpp}{D_A-\Dpp}
  \approx
  2^\circ \frac{\Dpp}{D_A-\Dpp}\,,
\end{equation}
which is comparable to the limit from the magnetic deflections for nearby sources ($z\lesssim0.15$) and dominates the ICC halo sizes for distance sources.  As a result, once again the structure of the IGMF dominates the gross features of the ICC halo for a large-scale, homogeneous IGMF that is consistent with the current lower limits on its strength.

\begin{figure*}
  \begin{center}
    \includegraphics[width=0.32\textwidth]{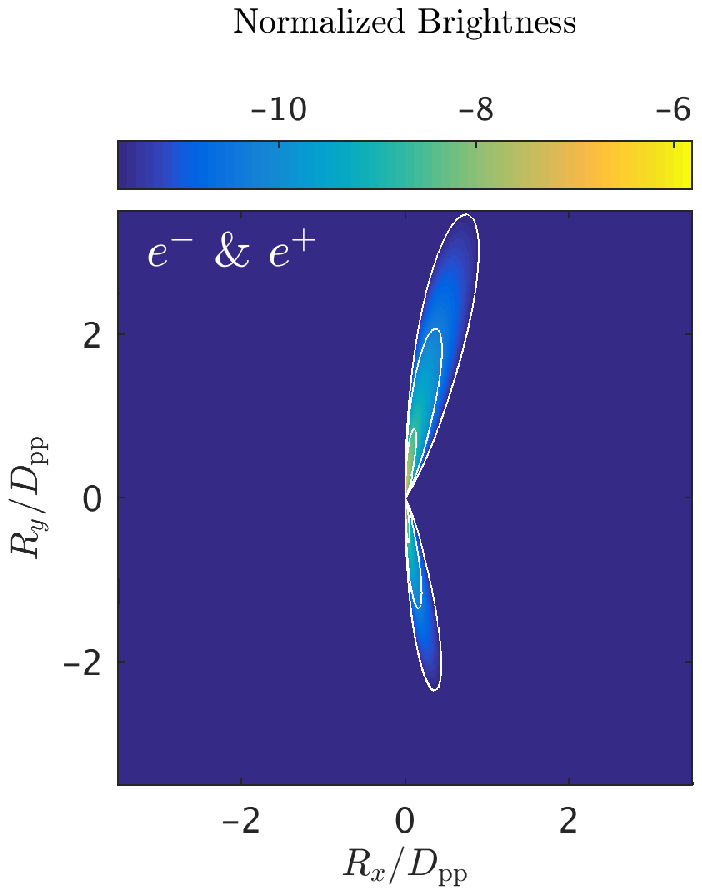}
    \includegraphics[width=0.32\textwidth]{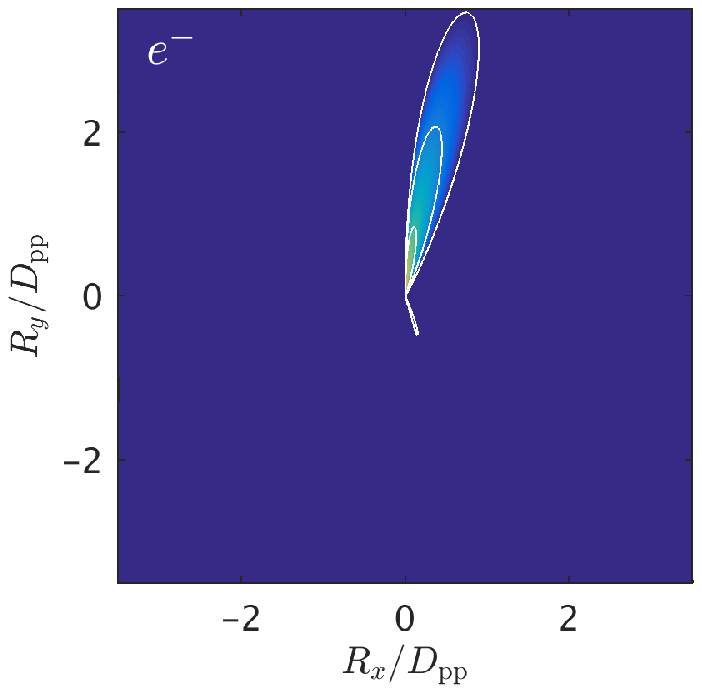}
    \includegraphics[width=0.32\textwidth]{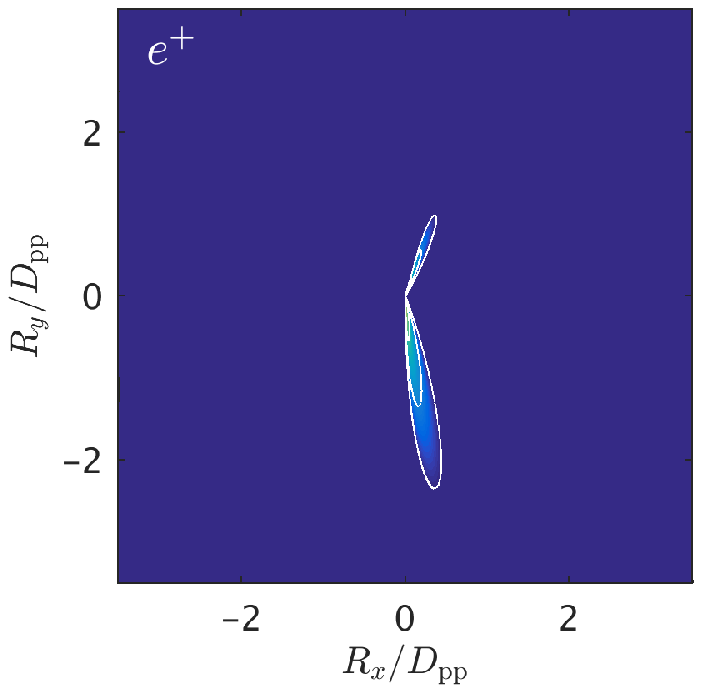}
  \end{center}
  \caption{1~GeV gamma-ray brightness distribution associated with a uniform IGMF before any instrumental response for the ICC halo generated by both electrons and positrons (left), only electrons (center), and only positrons (right).  The color indicates the logarithmic brightness, normalized relative to the highest value (concentrated at the source); in white are the iso-brightness contours enclosing 90\%, 99\%, and 99.9\% of the total flux.  Extreme values of the source parameters were chosen to accentuate key qualitative features ($B=2.5\times10^{-14}~\G$ located at $45^\circ$ to the line of sight, $\theta_j=10^\circ$, $\Theta=\circp{6}{7}$).}\label{fig:whf}
\end{figure*}

\begin{figure*}
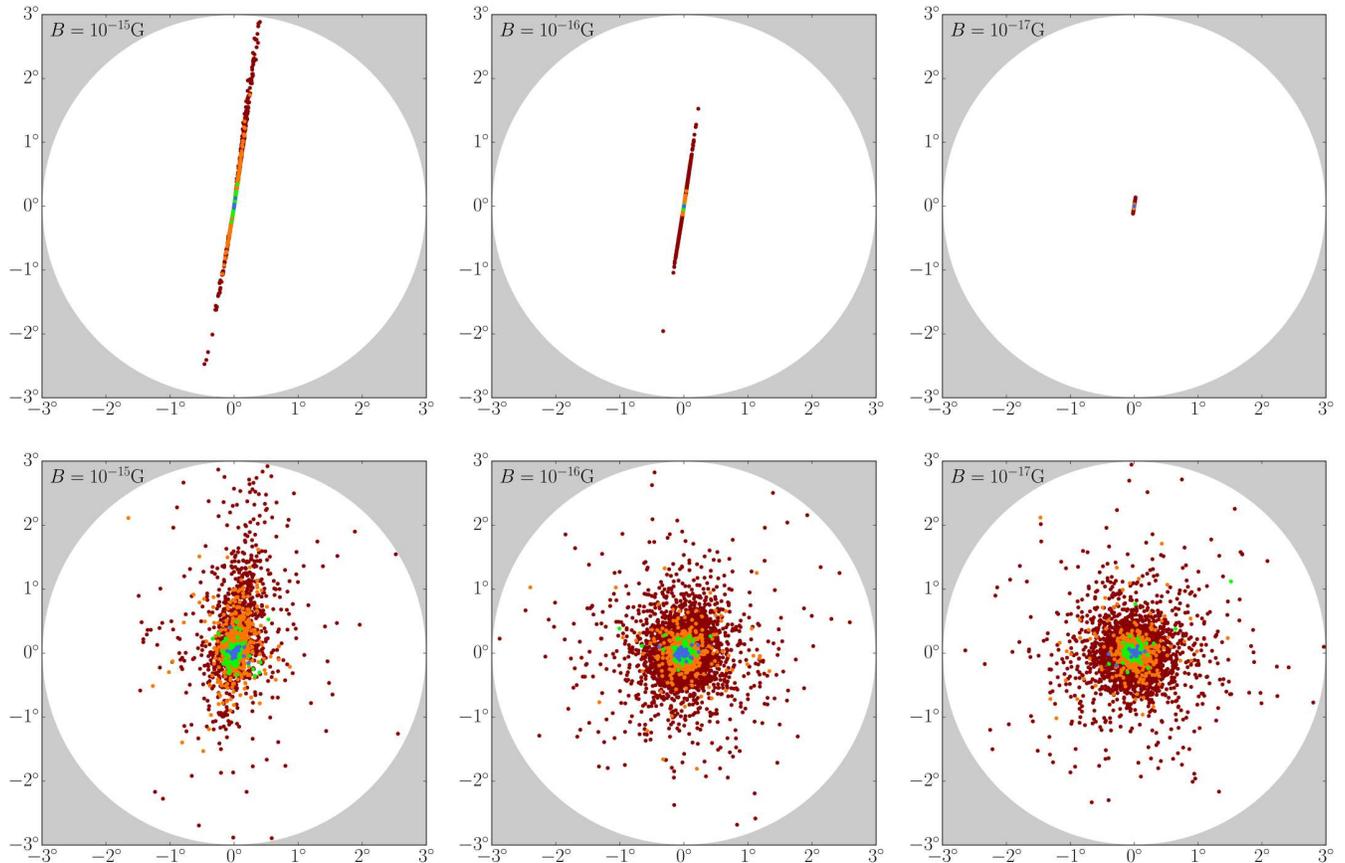

  \begin{center}
    \includegraphics[width=0.32\textwidth]{fig7a.eps2}
    \includegraphics[width=0.32\textwidth]{fig7b.eps2}
    \includegraphics[width=0.32\textwidth]{fig7c.eps2}\\
    \includegraphics[width=0.32\textwidth]{fig7d.eps2}
    \includegraphics[width=0.32\textwidth]{fig7e.eps2}
    \includegraphics[width=0.32\textwidth]{fig7f.eps2}
  \end{center}
  \caption{Example realization of ICC halo in the presence of a large-scale ordered magnetic field with strengths $10^{-15}$~G, $10^{-16}$~G, $10^{-17}$~G, before (top) and after (bottom) convolving with the \Fermi Pass 8R2\_V6 ULTRACLEANVETO-front PSF.  In all cases, the on-axis number of 1~GeV--100~GeV source gamma rays was assumed to be 5000 and the source placed at $z=0.3$, $\theta_j=3^\circ$, $\Theta=2^\circ$, and $\Gamma_h=2.6$.  The energy of the gamma rays are indicated by color: with 1~GeV-3~GeV in dark red, 3~GeV-10~GeV in orange, 10~GeV-30~GeV in green, and 30~GeV-100~GeV in blue.}\label{fig:homex} 
\end{figure*}

\subsection{Pair Distribution Function}
In this case the pairs retain the memory of their injection momentum, and thus,
\begin{multline}
  \dot{f}_{\rm inj, \epm}(\bx',\bp') = \frac{c}{{p'}^2\sin\vartheta'_p} \frac{dN_{\epm}}{dt'd\E'\dthxp}\left(p'c,\bx'\right)\\
  \times \delta(\vartheta'_p-\vartheta') \delta\left[\varphi'_p-\varphi'\right]\,.
  \label{eq:homogeneous_pdf}
\end{multline}
Perhaps counter-intuitively the $\varphi_p(E')$ depend on $E'$ as a result of the definition of $\bp_0$ adopted in Appendix \ref{app:chars}, which effectively describes the orientation of the injected leptons not at injection but by that of a fictitious lepton injected at infinite energy and inverse Compton cooled until $E'$ (which necessarily occurs over a finite time).  The corresponding stationary distribution function during inverse Compton cooling is
\begin{multline}
  f_{\epm}(\bx',\bp') = \frac{t_{\rm IC} m_\e c^2}{{p'}^4\sin\vartheta'_p} \delta(\vartheta'_p-\vartheta') \\
  \times \int_{p'}^\infty dq' \frac{dN_{\epm}}{dt'd\E'\dthxp}\left(q'c,\bx'\right) \delta\left[\varphi'_p(q')-\varphi'\right]\,,
\end{multline}
where we have subsumed all of the azimuthal argument in Equation \ref{eq:BES} into $\varphi'_p(q')$, i.e.,
\begin{equation}
  \varphi'_p(q')
  =
  \varphi'_p+\frac{\omega_B t_{\rm IC} \sin\vartheta'_p}{2}\left(\frac{m_\e^2 c^2}{{q'}^2}-\frac{m_\e^2 c^2}{{p'}^2}\right)\,.
\end{equation}
Again the integration may be performed explicitly for the injection spectrum in Equation (\ref{eq:pair_injection_spectrum}), now with the aid of the $\delta$-function.  Some care must be taken regarding the infinite number of zeros in the argument of the $\delta$-function, corresponding to the infinite number of times the momentum electron/positron gyrates into azimuthal (but not necessarily poloidal) alignment with a given direction, yielding 
\begin{multline}
  f_{\epm}(\bx',\bp')
  =
  \frac{c^3}{8\omega_B m_\e c^2 E_p}
  \frac{\delta(\vartheta_p-\vartheta)}{\sin^2\vartheta_p}
  \left(\frac{\E'}{E_p}\right)^{-4}\\
  \times
  \frac{dN_{\epm}}{dt'd\E'\dthxp}\left(\frac{E_p}{2},\bx'\right) e^{\tau(E_p,z)}\\
  \times
  \sum_{k=-\infty}^\infty
  \left(\frac{2\E^{\epm}_k}{E_p}\right)^{4-\Gamma_h}
  \!e^{-2\E^{\epm}_k/E_D}
  \,\Theta(\E^{\epm}_k-\E')\,,
  \label{eq:homogeneous_evolved_pdf}
\end{multline}
where $\Theta(x)$ is the Heaviside function and
\begin{equation}
  \E^{\epm}_k
  \equiv
  m_\e c^2 \sqrt{\frac{\omega_B t_{\rm IC} \sin\vartheta'/2}{2\pi k - (\varphi'_p-\varphi') \pm \omega_B t_{\rm IC} \sin\vartheta' m_\e^2 c^4/2 {\E'}^2} }\,.
  \label{eq:EKS}
\end{equation}
For a lepton injected at infinite energy with azimuthal angle $\varphi'$ the $\E^{\epm}_k$ are the energies at which it will have gyrated $k$ times through an azimuthal angle $\varphi'_p$.  Thus, these represent the energies at which the multiplicity of leptons oriented in a given azimuthal direction increases by one.  Note that the $\E^{\epm}_k$ depend on species through the sign of $\omega_B$, corresponding to the direction of the gyration.

\subsection{Gamma-ray Halo Structure}
To construct the spatial distribution of ICC gamma rays we again employ Equation (\ref{eq:image_master}), now with the distribution function in Equation (\ref{eq:homogeneous_evolved_pdf}).  Unlike the rapidly isotropized limit, a large-scale homogeneous IGMF evolves the lepton distribution functions on a cone of fixed opening angle about the IGMF, $\bB$.  As a result, the integration along the line of sight receives contributions only where the line of sight, $\bl'$, lies on this cone.

This admits a simple geometric identification of the gross geometry of the contributing regions, that may most easily be imagined by time-reversing the inverse Compton process.  The observed ICC gamma rays are produced by leptons directed along $\bl'$, and thus leptons that are gyrating about $\bB$ on a cone defined by $\bl'$.  Since the initial VHEGRs are assumed to propagate radially from the source, this implies that $\bmath{r}'$ must also lie on this cone.  Thus, the ICC halo will receive contributions only from leptons located on a cone with opening angle $\cos^{-1}(\bB\cdot\bl'/B\ell')$ and centered at the VHEGR source.  In practice, this is enforced via the remaining $\delta$-function in $f_{\epm}(\bx',\bp')$ in Equation (\ref{eq:homogeneous_evolved_pdf}).  

Performing the integration along $\bl'$ is then reduced to the determination of the relevant position on the contributing cone at a given angular position and performing the integration over the $\delta$-function.  Once again it is convenient to define a coordinate system aligned with the line of sight as in Equation (\ref{eq:rRl}) and following Equation (\ref{eq:Psidef}).  In terms of this basis, we also define the three projections $B_\ell\equiv\bB\cdot\bhl$, $B_R\equiv\bB\cdot\bR'/R'$, and $B_X\equiv\bB\cdot(\bR'\times\bhl')/R'$.  Then, the contributing cone has
\begin{equation}
  \begin{gathered}
    r' = R' \left| \frac{B_\ell^2 + B_R^2}{2 B_\ell B_R} \right|\,,\quad
    \sin\vartheta'_\ell = \frac{\sqrt{B^2-B_\ell^2}}{B}\,,\\
    \tan(\varphi'_\ell-\varphi') = \frac{2 B |B_\ell B_R| B_X}{B^2(B_\ell^2-B_R^2)-B_\ell^2(B_\ell^2+B_R^2)}\,.
  \end{gathered}
  \label{eq:cone}
\end{equation}
where $(\vartheta'_\ell,\varphi'_\ell)$ is the orientation of $\bl'$ as seen from the source.  The angle $\theta'$ relative to the jet axis is then determined via Equation (\ref{eq:geo}).  Note that while $\bl'$, and thus $B_\ell$, is fixed, $\bR'$, and thus both $B_R$ and $B_X$ depend on $\ba$, i.e., they vary across the image.  The integration over the $\delta$-function produces a Jacobian in addition to restricting the contributing region to the above cone:
\begin{equation}
  \begin{aligned}
    \int d\ell' \delta(\vartheta'-\vartheta'_\ell) f(\vartheta')
    &\equiv
    \frac{\sin\vartheta'_\ell}{\mathcal{J}} f(\vartheta'_\ell)\\
    &=
    \sin\vartheta'_\ell \left[ \frac{1}{R'} \left| \frac{4B_\ell^2 B_R^3}{B\left( B_\ell^2+B_R^2\right)^2 } \right| \right]^{-1} f(\vartheta'_\ell)\,,
    \end{aligned}
\end{equation}
where $f(\vartheta')$ is any function of $\vartheta'$.  Implementing the above in Equation (\ref{eq:image_master}) yields
\begin{multline}
  \frac{dN_{\gamma}}{dt dE d^2\!\alpha}
  =
  F_{35}
  \frac{m_\e c^2}{\delta_A^2 E_{\rm CMB}^2}
  \frac{3 (1+z)^{-1}}{64 N(\Gamma_l,z) G(\Theta')}\\
  \times
  \frac{1}{ \omega_B t_{\rm IC} }
  \left( \frac{E_p}{m_\e c^2}\right)^{4}
  \left( \frac{E}{2 E_{\rm CMB}} \right)^{-3/2}
  \frac{\Dpp(E_p,z)}{r'^2 \sin\vartheta'_\ell \mathcal{J}} \, G(\theta') \\
  \times
  \sum_{\rm e^+,e^-} \sum_{k=-\infty}^\infty
  \left(\frac{2\tilde{\E}^{\epm}_k}{E_p}\right)^{4-\Gamma_h}
  \!\! e^{-2\tilde{\E}^{\epm}_k/E_D}
  \,\Theta\left(\frac{\tilde{\E}^{\epm}_k}{m_\e c^2}- \sqrt{\frac{E}{2E_{\rm CMB}}}\right)\,,
  \label{eq:homogeneous_image}
\end{multline}
where all spatially dependent terms are evaluated at the positions in Equation (\ref{eq:cone}) given $\bR$, and $\tilde{\E}^{\epm}_k\equiv \left.\E^{\epm}_k\right|_{\E'=m_\e c^2\sqrt{E/2E_{\rm CMB}}}$.  That this reproduces the integrated flux in the parent VHEGRs is verified in Appendix \ref{app:check_weak}.

This is shown for an extreme case (chosen to exhibit key qualitative features) in Figure \ref{fig:whf}.  Here the bipolar symmetry is only approximate, broken by the asymmetry in the VHEGR flux and the geometry of the gyrating leptons.  When the contributions of the electrons and positrons are separated it is clear that apart from a small contribution due to leptons that complete a full orbit (present only for very strong IGMFs) the origin of the two lobes is the opposite sense of gyration of the two lepton species.

\subsection{Numerical Implementation}
Despite the explicitly performed integration over the line-of-sight, generating samples of the energy-dependent flux distribution in Equation (\ref{eq:homogeneous_image}) presents distinct challenges as a result of the inseparability of the energy and spatial probability distributions and the typically narrow features in the image that must be resolved.  As before, we numerically construct a halo-core flux ratio and tabulated tables of
\begin{equation}
  \begin{aligned}
    P(E) &\propto \int d^2\!\alpha \frac{dN_{\gamma}}{dt dE d^2\!\alpha}\\
    P(\alpha|E) &\propto \int d\theta_\alpha \frac{dN_{\gamma}}{dt dE d^2\!\alpha}\\
    P(\theta_\alpha|E,\alpha) &\propto \frac{dN_{\gamma}}{dt dE d^2\!\alpha}\,.
  \end{aligned}
\end{equation}
Unlike the highly-tangled IGMF case these can neither be determined analytically nor tabulated in terms of a single function.  As a result, the procedure is
\begin{enumerate}
\item Draw a random number to determine if the gamma ray is in the halo or the source.  If it is in the halo:
\item Draw an energy from $P(E)$.
\item Tabulate $P(\alpha|E)$ and draw a value for $\alpha$.
\item Tabulate $P(\theta_\alpha|\alpha,E)$ and draw a value for $\theta_\alpha$.
\end{enumerate}
In the tabulation of $P(\alpha|E)$ we take care to step logarithmically in $\alpha$ between limits many orders of magnitude smaller and larger than the projected $\Dpp$, to which we have verified that the resulting image is insensitive.  Because the halo features are typically extremely narrow, prior to tabulating $P(\theta_\alpha|\alpha,E)$ we first obtain estimated limits on $\theta_\alpha$ that define the regions of non-vanishing probability density (in practice, there are two, one for the halo in each direction).  We do this iteratively, sampling in $\theta_\alpha$ with increasing resolution until many points appear across the non-vanishing portion of $P(\theta_\alpha|\alpha,E)$ at a fixed value of $\alpha$, set to the projected angular scale of $0.01\Dpp$ at 1~GeV; low energies produce the broadest halos.  These are then used to generate a non-contiguous table for $P(\theta_\alpha|\alpha,E)$ that resolves the halo features accurately.  In practice, for most cases we could have adopted a linear approximation for the halo features; here we nevertheless generate the full angular structure.  Again we have verified that expanding the regions in $\theta_\alpha$ sampled makes no difference to the resulting images.

\subsection{Example Halo Realizations}
Example realizations of ICC halos associated with a large-scale, uniform IGMF are shown in Figure \ref{fig:homex}.  Again the on-axis intrinsic fluence was held fixed at 5000~ph.  In comparison to the tangled-IGMF case there are far more halo photons.  This is a direct result of the effective isotropy, or in this case lack thereof, in the halo energy flux.  In the absence of a mechanism to isotropize the lepton distribution functions, the comparatively weak deflections in the uniform IGMF result in a lepton distribution function that continues to be primarily beamed along the axis of the VHEGR jet, along which the ICC halo photons are then also beamed.  As a result, the halo fluence and morphology is a strong function of viewing angle (see Section \ref{sec:VA}).

Typically, prior to convolution with an instrument response the halos appear as striking linear features.  The narrow width of these features is indicative of the projected range of positions which satisfy the geometric constraints imposed by the forward-scattering limit and particle gyration, e.g., Equation (\ref{eq:cone}).  In principle, this can be broadened by modifications to the permitted magnetic field geometry and/or evolution of the particle momenta beyond gyration in the magnetic field.  By assumption we will neglect the former; we show that the latter is not relevant in Appendix \ref{app:fsl}.

As with the tangled-IGMF case, the high-energy photons are concentrated at small angular displacements from the central source and low-energy photons more extended.  Thus, there remains a clear energy-dependence of the halo structure, as anticipated by the form of $\Dpp$.  This continues to be true after convolution with the \Fermi Pass 8R2\_V6 PSF for front-converted events (see Section \ref{sec:PSF}).  Because high-energy leptons gyrate less, their halo emission is less diffused by the phase-space spreading driven by the IGMF.  Combined with their intrinsic concentration near the source, this results in a halo that is far more centrally concentrated than for the tangled-IGMF.

As anticipated by Equation (\ref{eq:adef}), the angular extent of the halos is a strong function of IGMF strength: large IGMFs produce large halos while small IGMFs produce small halos.  This is strongly impacts the ability to observe ICC halos in the presence of a weak, large-scale IGMF .

\mbox{}\\
\section{Mock Image Modeling} \label{sec:mocksgen}
Quantitatively assessing the ICC halo structure requires the ability to generate realizations of gamma-ray images of prospective sources.  The energies of the gamma rays that comprise the putative ICC halos are well matched to those at which the \Fermi LAT is sensitive, and we restrict our efforts to energies between 1~GeV and 100~GeV.  This has the added benefit that the \Fermi LAT response is nearly energy independent throughout this energy range, simplifying the subsequent analysis substantially.

At the most granular level \Fermi images consist of collections of individual photons, numbered in the thousands for a single bright source, each with a reported sky location and energy.  Thus, in principle this procedure consists of first identifying the joint probability distribution of photons from various emission components with a given energy and location, $dF/dEd^2x$, and second efficiently drawing a random realization from this, $\{E_j,\bmath{x}_j\}$.  In practice, this is further modified by the \Fermi LAT response, which primarily impacts the images via the PSF.  We begin by discussing this process for each emission component separately, following a discussion of the PSF.

We consider a three component model comprised of a uniform background, an intrinsic point source, and a putative ICC halo.  The former two are relatively straightforward and have parameters reasonably fixed by \Fermi directly.  That is, the background is well defined by a well-measured spectral shape and normalization, and for many gamma-ray bright AGN the spectra, total flux, and redshift of the point source are known.

Less clear are the ICC halos.  Their brightness and morphology depend on the unobserved VHEGRs, and thus require some spectral and collimation model that extends the properties observed by \Fermi to TeV energies.  This introduces a variety of additional poorly known parameters, e.g., VHEGR jet opening angle, VHEGR jet orientation, and intrinsic VHEGR spectrum.  Here we construct images assuming these are known, adopting typical values for these in practice.  

\begin{figure}[t]
  \begin{center}
  \includegraphics[width=\columnwidth]{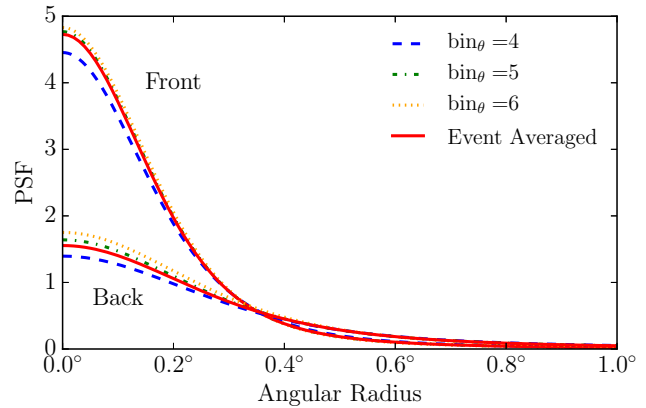}
  \end{center}
  \caption{Average Front and Back detector PSF for all events in our sample (red) in comparison to the bore-angle-specific PSFs reported by the Pass 8R2\_V6 IRFs for a handful of bore-angle ranges.  The event-averaged PSF is similar in both cases to the PSF at $\circp{36}{9}-\circp{45}{6}$ (i.e., bin$_\theta$=5).  For comparison this is shown for PSFs associated with bore-angle ranges $\circp{45}{6}-\circp{53}{1}$ (bin$_\theta$=4) and $\circp{25}{8}-\circp{36}{9}$ (bin$_\theta$=6).} \label{fig:PSFs}
\end{figure}

\subsection{\Fermi Point Spread Function} \label{sec:PSF}
Because we are interested in previously unresolved features we restrict our attention to the Pass 8R2\_V6 ULTRACLEANVETO sample, corresponding to those events that are most confidently associated with astronomical sources and not necessarily associated with nearby bright sources.

The PSF of the \Fermi LAT has evolved substantially since Pass 6v3, after which the first empirical calibrations were preformed using AGN \citep{P6V11,LAT_perf}.  This was motivated in part by the claims of the detection of bright, extended ICC halos based on earlier PSF estimates that implied substantial narrowing at high energies \citep{Ando:2010}.  These were belied by the discovery of similarly broad halos about Galactic gamma-ray sources \citep{Nero-Semi-Tiny-Tkac:11} and subsequently understood by follow-up studies of the ground-based calibration process of the LAT \citep{LAT_perf,FLAT-stack:2013}.  As a result we necessarily pay close attention to the structure of the PSF and its impact.

The form of the Pass 8R2\_V6 PSF is different for the Front and Back detectors.  For both it is described by a King function
\begin{equation}
K(x,\sigma,\gamma) = \frac{1}{2\pi\sigma^2}\left(1-\frac{1}{\gamma} \right)
    \,\left[1 + \frac{1}{2\gamma}\,\frac{x^2}{\sigma^2}\right]^{-\gamma},
\end{equation}
where $x$ is a scaled angular deviation given by
\begin{equation}
  \begin{gathered}
      x = \frac{\delta p}{S_P(E)} \\
      \delta p = 2\sin^{-1}\left(\frac{|\hat{p}' - \hat{p}|}{2}\right)\\
      S_p(E) = \sqrt{\left[c_0\,\left(\frac{E}{100\ \MeV}\right)^{-\beta}\right]^2 + c_1^2},
  \end{gathered}
\end{equation}
and $\hat{p}, \hat{p}'$ are the true and reconstructed direction of the photon. Note that for all event types $\beta=0.8$. The Pass 8R2\_V6 also includes a tail component which is constructed from a second King function
\begin{equation}
  \begin{gathered}
    P(x,\vec{\alpha}_P) = f_{\rm core}K(x,\sigma_{\rm core}, \gamma_{\rm core}) 
    + (1-f_{\rm core})K(x,\sigma_{\rm tail}, \gamma_{\rm tail})\\
    f_{\rm core} = \frac{1}{1 + N_{\rm tail}\sigma^2_{\rm tail}/\sigma^2_{\rm core}}
  \end{gathered}
\end{equation}
 Parameters $\gamma_{\rm core},\, \gamma_{\rm tail},\,  \sigma_{\rm core},\, \sigma_{\rm tail},\,  c_0,\,  c_1,\,  N_{\rm tail}$ are dependent on energy, bore angle, and differ for the front and back PSF.\footnote{More detailed information is provided by \textit{Cicerone} \url{http://fermi.gsfc.nasa.gov/ssc/data/analysis/documentation/Cicerone}. Note that there is a discrepancy between the \Fermi documentation where $\beta$ is replaced by $-\beta$ in the actual paramter files.}  This is simplified substantially by the relatively modest PSF dependence on energy above 1~GeV and the fact that the collection of events within the Pass 8R2\_V6 ULTRACLEANVETO sample is distributed among a large number of potential bore angles.  As a result the collective PSF for the Front and Back detectors are well approximated by that at a single bore angle, which we show for each detector in Figure \ref{fig:PSFs}, and corresponding to $\circp{36}{9}$--$\circp{45}{6}$ in both cases.  This is the set of PSFs we adopt here.

In principle the square geometry of the LAT imposes a strong dependence on the azimuthal angle of the photon \citep{LAT_perf}.  However, in practice the long duration of the \Fermi observations (8 years) combined with the varying roll angle of the telescope as \Fermi tracks the Sun and the eight-fold symmetry of the LAT result in a nearly cylindrical symmetry.  This may be broken for short duration or bursty events, and thus if the gamma-ray AGN of interest underwent periods of substantial variability (i.e., on timescales short in comparison to 45 days) a small residual angular structure may appear.

Imposing the PSF is done by imparting a random shift to the photon locations in the intrinsic image.  That is, given the photon energy and LAT detector to be modeled, we choose a random orientation and distance from the appropriate PSF for each photon individually.  We have verified that this gives qualitatively similar images for bright Galactic sources to those obtained by \Fermi.

\subsection{Central Source}
The 3LAC \citep{3LAC} provides 1~GeV--100~GeV fluxes ($F_S$) and photon spectral indexes ($\Gamma_S$) for all current point sources in the collected $6.9$ years of \Fermi data.  There is no evidence to date of significant spatial substructure within the source, and as we are interested in features at large distances in comparison to the width of the \Fermi PSF, we model the intrinsic emission as a point source.  Thus we assume that all gamma rays from the central source are located at the origin.

The number of photons drawn for a particular object is set by a Poisson deviate with mean $N_S=F_S T_F A_F$ where $T_F$ is the \Fermi live time on source and $A_F$ is the effective area, which we assume to be fixed over the energy range of interest.  The requisite number of photons are then assigned energies between 1~GeV and 100~GeV from a random deviate drawn from a power law distribution with index $-\Gamma_S$. 

These are then stochastically shifted by an amount set by the PSF as described in Section \ref{sec:PSF} to produce the set of source gamma rays: $\{E_j^S,\bmath{x}_j^S\}$.

\begin{figure*}[th]
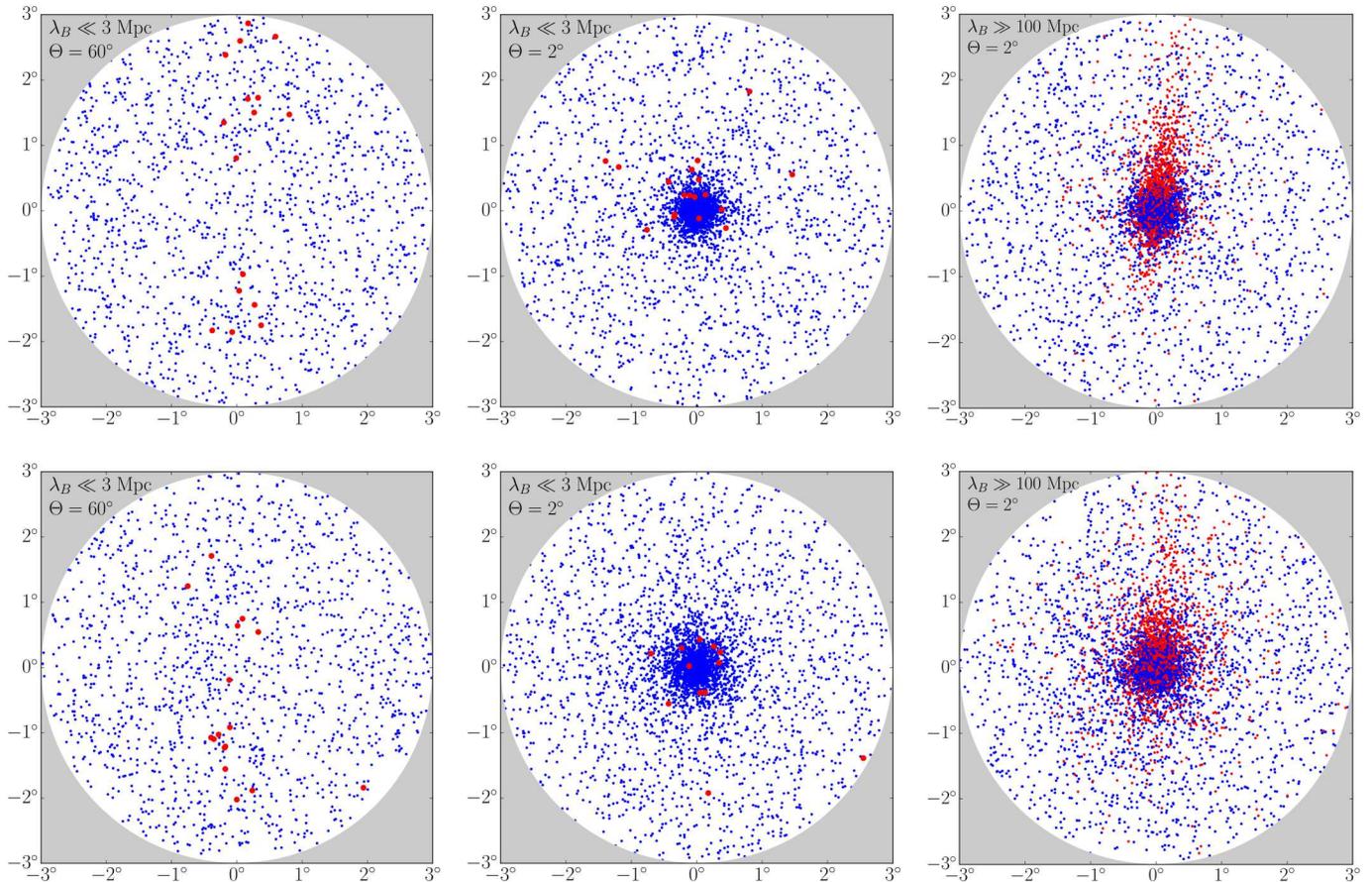

  \begin{center}
    \includegraphics[width=0.33\textwidth]{fig9a.eps2}
    \includegraphics[width=0.33\textwidth]{fig9b.eps2}
    \includegraphics[width=0.33\textwidth]{fig9c.eps2}\\
    \includegraphics[width=0.33\textwidth]{fig9d.eps2}
    \includegraphics[width=0.33\textwidth]{fig9e.eps2}
    \includegraphics[width=0.33\textwidth]{fig9f.eps2}
    \caption{Comparison of ICC halos for a typical bright, hard gamma-ray blazar as seen by events converted in the Front (top) and Back (bottom) portion of the LAT.  ICC halos are shown for an IGMF characterized by small-scale tangled fields (left and center) and for a large-scale uniform $10^{-15}~\G$ IGMF (right).  The viewing angles are $60^\circ$ (left) and $2^\circ$ (center and right) corresponding to an edge-on jet (and unlikely to be seen by \Fermi) and a typical \Fermi source, respectively. ICC halo photons are indicated in red, while source and background photons are shown in blue.} \label{fig:fbmock}
  \end{center}
\end{figure*}

\begin{figure*}[th]
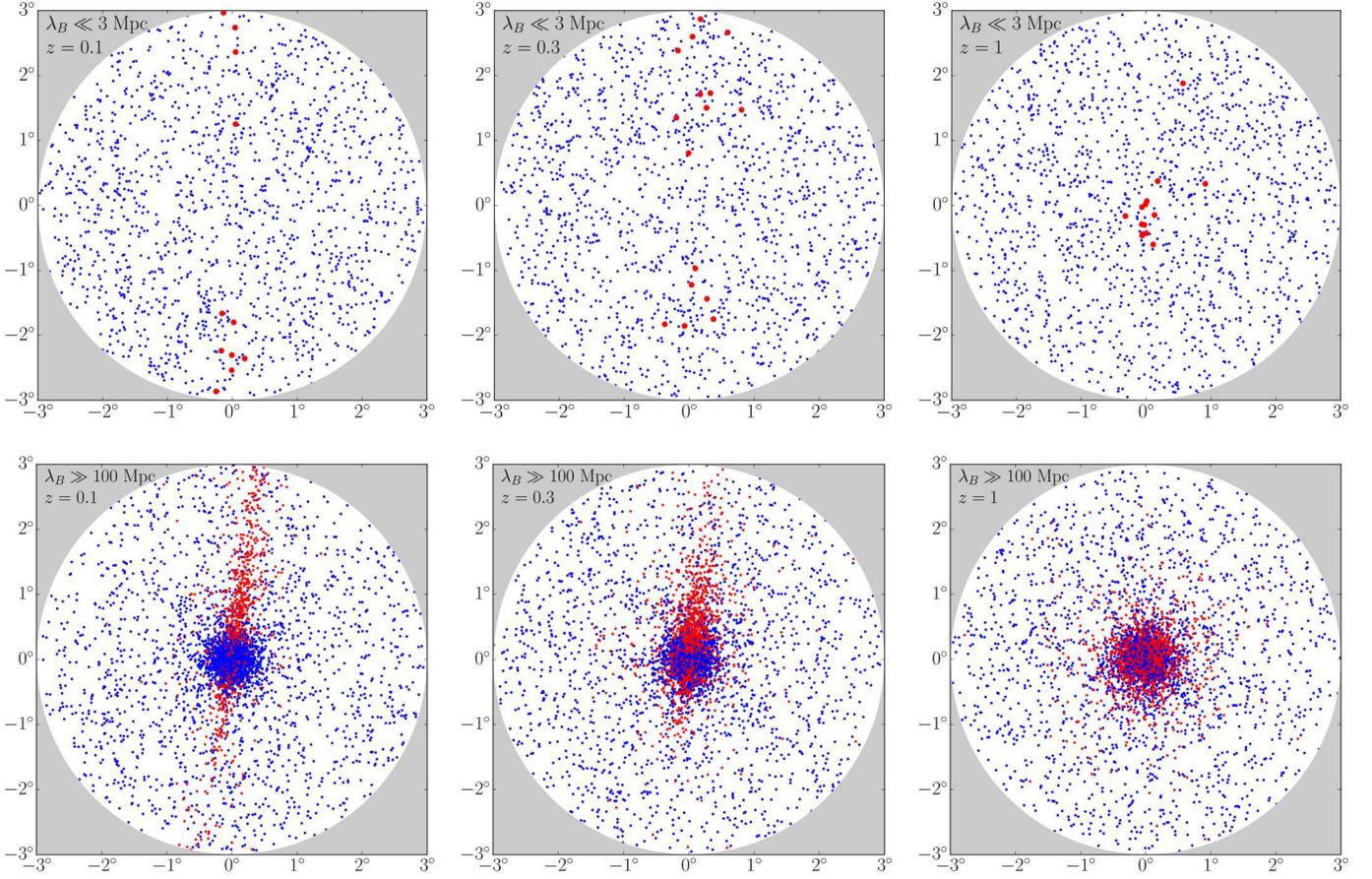

  \begin{center}
    \includegraphics[width=0.33\textwidth]{fig10a.eps2}
    \includegraphics[width=0.33\textwidth]{fig10b.eps2}
    \includegraphics[width=0.33\textwidth]{fig10c.eps2}\\
    \includegraphics[width=0.33\textwidth]{fig10d.eps2}
    \includegraphics[width=0.33\textwidth]{fig10e.eps2}
    \includegraphics[width=0.33\textwidth]{fig10f.eps2}
    \caption{Comparison of ICC halos for a typical bright, hard gamma-ray blazar located at a redshift of $0.1$ (left), $0.3$ (center), and $1$ (right).  Top panels show an ICC halo for a small-scale, tangled IGMF viewed at $60^\circ$; bottom panels show an ICC halo for a large-scale, uniform $10^{-15}~\G$  IGMF viewed at $2^\circ$.  In all cases $\theta_j=3^\circ$ and the on-axis fluence is set to 5000~ph.  For both cases we show only Front-converted events. ICC halo photons are indicated in red, while source and background photons are shown in blue.} \label{fig:zmock}
  \end{center}
\end{figure*}

\begin{figure*}[th]
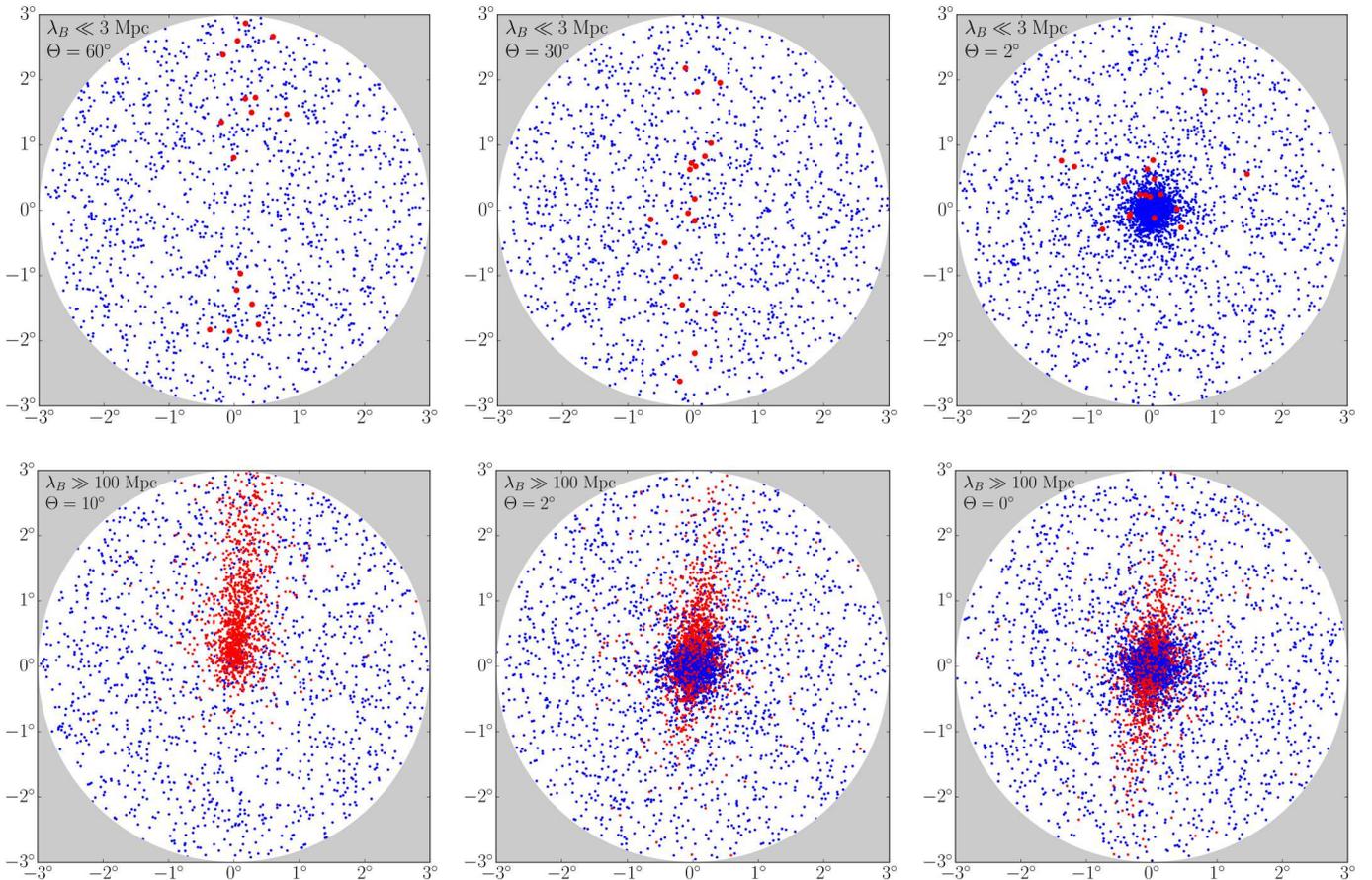

  \begin{center}
    \includegraphics[width=0.33\textwidth]{fig11a.eps2}
    \includegraphics[width=0.33\textwidth]{fig11b.eps2}
    \includegraphics[width=0.33\textwidth]{fig11c.eps2}\\
    \includegraphics[width=0.33\textwidth]{fig11d.eps2}
    \includegraphics[width=0.33\textwidth]{fig11e.eps2}
    \includegraphics[width=0.33\textwidth]{fig11f.eps2}
    \caption{Comparison of ICC halos for a typical bright, hard gamma-ray blazar viewed from a number of different angles.  Top panels show an ICC halo for a small-scale, tangled IGMF viewed at $60^\circ$ (left), $30^\circ$ (center), and $2^\circ$ (right).  Bottom panels show an ICC halo for a large-scale, uniform $10^{-15}~\G$  IGMF viewed at $10^\circ$, $2^\circ$, and $0^\circ$.  In all cases $z=0.1$, $\theta_j=3^\circ$, and the on-axis fluence is set to 5000~ph.  For both cases we show only Front converted events. ICC halo photons are indicated in red, while source and background photons are shown in blue.} \label{fig:Tmock}
  \end{center}
\end{figure*}

\begin{figure*}[th]
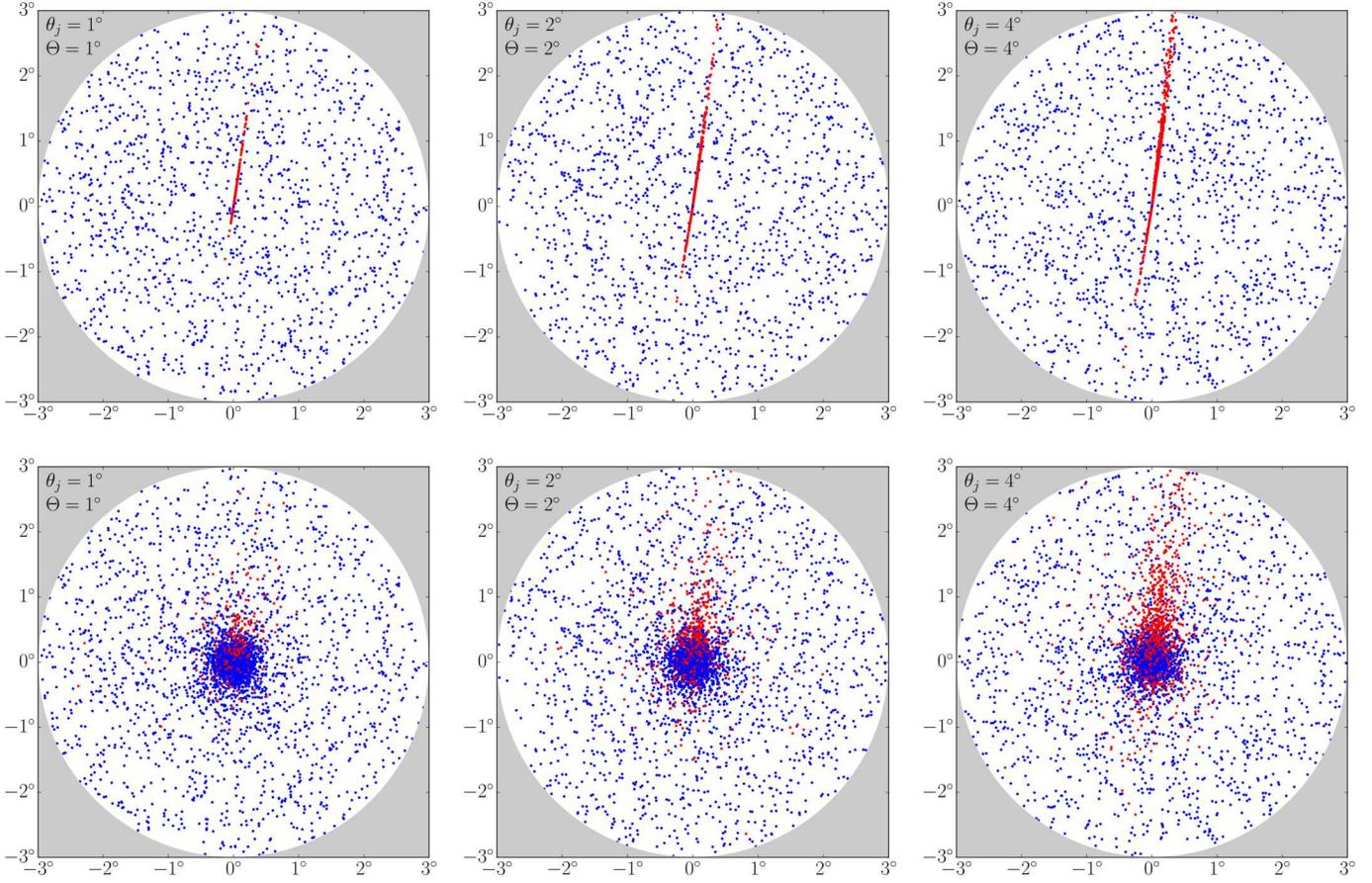

  \begin{center}
    \includegraphics[width=0.33\textwidth]{fig12a.eps2}
    \includegraphics[width=0.33\textwidth]{fig12b.eps2}
    \includegraphics[width=0.33\textwidth]{fig12c.eps2}\\
    \includegraphics[width=0.33\textwidth]{fig12d.eps2}
    \includegraphics[width=0.33\textwidth]{fig12e.eps2}
    \includegraphics[width=0.33\textwidth]{fig12f.eps2}
    \caption{Comparison of ICC halos in the presence of a large-scale, uniform IGMF for a typical bright, hard gamma-ray blazar with a number of different jet opening angles, with viewing angle to jet opening angle ratio fixed to $1$. Jet opening angle set to $\theta_j=1^\circ$ (left), $\theta_j=2^\circ$ (middle), $\theta_j=4^\circ$ (right).  In all cases $z=0.3$, $B=10^{-15}\G$, $\Theta=\theta_j$ and the  on-axis fluence is set to 5000~ph.  We show realizations of ICC halos before (top) and after (bottom) convolving with the \Fermi Pass 8R2\_V6-front PSF. ICC halo photons are indicated in red, while source and background photons are shown in blue.} \label{fig:tjmock}
  \end{center}
\end{figure*}

\begin{figure*}[th]
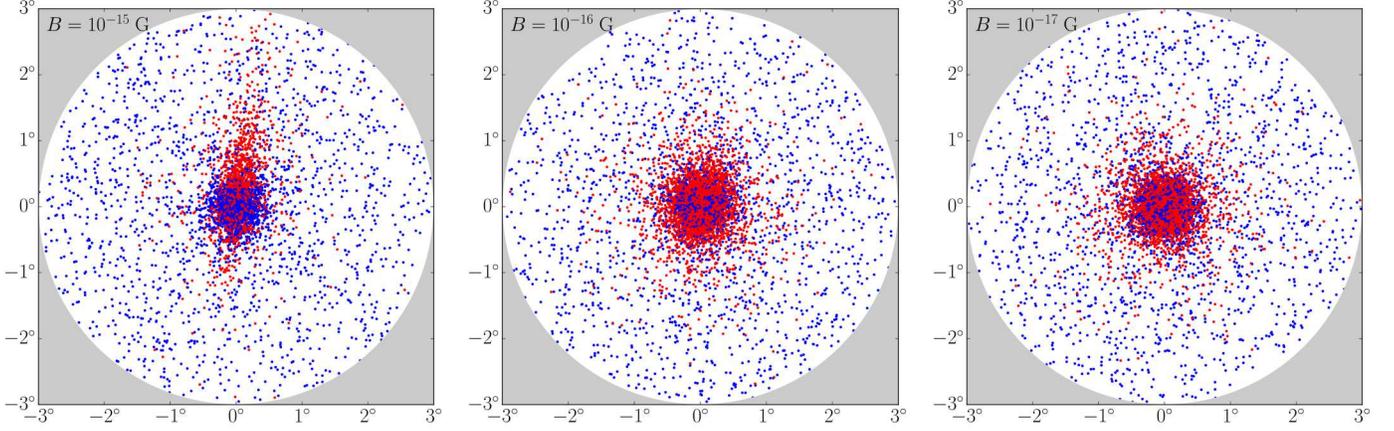

  \begin{center}
    \includegraphics[width=0.33\textwidth]{fig13a.eps2}
    \includegraphics[width=0.33\textwidth]{fig13b.eps2}
    \includegraphics[width=0.33\textwidth]{fig13c.eps2}
    \caption{Comparison of ICC halos in the presence of a large-scale, uniform IGMF for a typical bright, hard gamma-ray blazar for $10^{-15}~\G$ (left), $10^{-16}~\G$ (center), and $10^{-17}~\G$ (right) large-scale, uniform IGMFs.  In all cases $z=0.3$, $\theta_j=3^\circ$, $\Theta=2^\circ$, and the on-axis fluence is set to 5000~ph.  We show only Front converted events. ICC halo photons are indicated in red, while source and background photons are shown in blue.} \label{fig:Bmock}
  \end{center}
\end{figure*}
\begin{figure*}[th]
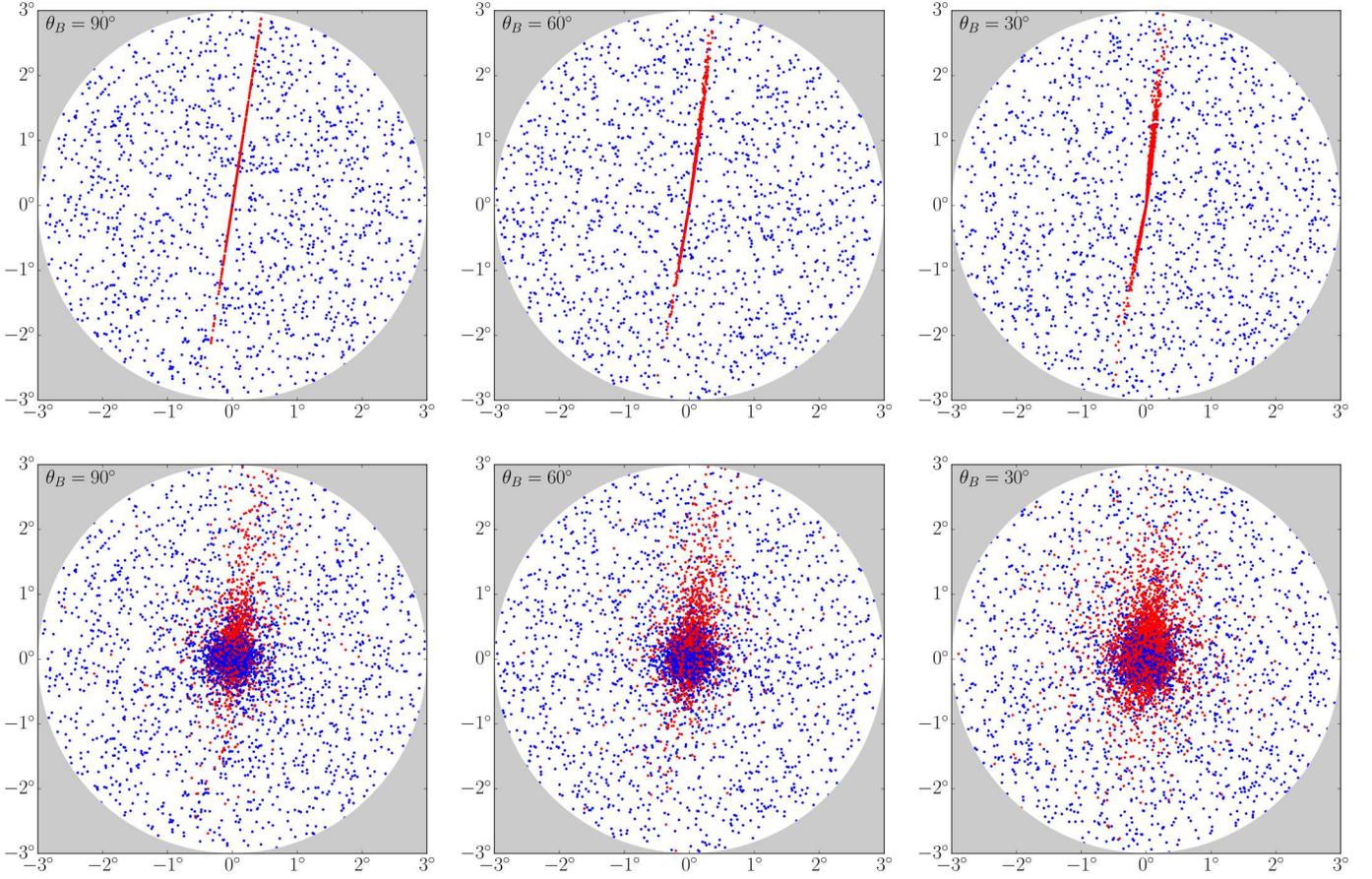

  \begin{center}
    \includegraphics[width=0.33\textwidth]{fig14a.eps2}
    \includegraphics[width=0.33\textwidth]{fig14b.eps2}
    \includegraphics[width=0.33\textwidth]{fig14c.eps2}\\
    \includegraphics[width=0.33\textwidth]{fig14d.eps2}
    \includegraphics[width=0.33\textwidth]{fig14e.eps2}
    \includegraphics[width=0.33\textwidth]{fig14f.eps2}
    \caption{Comparison of ICC halos in the presence of a large-scale, uniform IGMF for a typical bright, hard gamma-ray blazar for an IGMF oriented at $90^\circ$ (left), $60^\circ$ (center), and $30^\circ$ (right) to the line of sight.  In all cases $z=0.3$, $\theta_j=3^\circ$, $\Theta=2^\circ$, and the on-axis fluence is set to 5000~ph. We show realizations of ICC halos before (top) and after (bottom) convolving with the \Fermi Pass 8R2\_V6-front PSF. ICC halo photons are indicated in red, while source and background photons are shown in blue.} \label{fig:tBmock}
  \end{center}
\end{figure*}

\begin{figure*}[th]
  \begin{center}
    \includegraphics[width=0.33\textwidth]{fig15a.eps2}
    \includegraphics[width=0.33\textwidth]{fig15b.eps2}
    \includegraphics[width=0.33\textwidth]{fig15c.eps2}
    \caption{Comparison of ICC halos in the presence of a large-scale, uniform IGMF for a typical bright, hard gamma-ray blazar for $\Gamma_l=1.7$ (left), $\Gamma_l=1.9$ (middle), $\Gamma_l=2.1$ (right) with large-scale, uniform IGMFs.  In all cases $z=0.3$, $\theta_j=3^\circ$, $\Theta=2^\circ$, and the on-axis fluence is set to 5000~ph.  We show only Front converted events. ICC halo photons are indicated in red, while source and background photons are shown in blue.} \label{fig:GlMock}
  \end{center}
\end{figure*}
\begin{figure*}[th]
  \begin{center}
    \includegraphics[width=0.33\textwidth]{fig16a.eps2}
    \includegraphics[width=0.33\textwidth]{fig16b.eps2}
    \includegraphics[width=0.33\textwidth]{fig16c.eps2}
    \caption{Comparison of ICC halos in the presence of a large-scale, uniform IGMF for a typical bright, hard gamma-ray blazar for $\Gamma_h=1.9$ (left), $\Gamma_h=2.5$ (middle), $\Gamma_h=3.1$ (right) with large-scale, uniform IGMFs.  In all cases $z=0.3$, $\theta_j=3^\circ$, $\Theta=2^\circ$, and the on-axis fluence is set to 5000~ph.  We show only Front converted events. ICC halo photons are indicated in red, while source and background photons are shown in blue.} \label{fig:GhMock}
  \end{center}
\end{figure*}

\subsection{Background Modeling}
Immediately evident in \Fermi images is the presence of a currently unresolved background.  This background is sufficiently bright that within a few degrees nearly as many background gamma rays are present as source gamma rays.  At high Galactic latitudes the background gamma-ray spectrum is well approximated by a power law with photon spectral index $2.4$ \citep{Fermi_EGRB2013}.  This remains true for the local background about individual sources.

The unresolved background is not globally uniform, exhibiting large gradients at low Galactic latitudes.  Nevertheless, at high Galactic latitudes, the gamma-ray background is homogeneous.  Thus we restrict our attention to the cases where a uniform background remains a good approximation.  In principle, while the entirety of the background may ultimately be resolved into individual point sources in the future \citep[see, e.g.,][]{50GevBkgnd} we model it here as a locally uniformly distributed component with a power-law spectrum.

Note that anisotropy within the background has been previously discussed as a means to measure the unresolved gamma-ray bright AGN population \citep{Fermi_aniso,Cuoco}.  While considerable uncertainty regarding the meaning of those measurements persists \citep{PaperVa}, it is true that if a substantial fraction of the background is collected in marginally unresolved sources (e.g., sources from which 2-5 photons were detected) the statistical properties of the background could be quite different.

Therefore, we generate a uniform background with the locally desired fluence.  Again, the photons are then stochastically shifted by an amount set by the PSF as described in Section \ref{sec:PSF} to produce the set of background gamma rays: $\{E_j^B,\bmath{x}_j^B\}$.  While formally superfluous, this ensures that we treat the entirety of the intrinsic image uniformly.

\section{Mock Images of Hard \Fermi Sources} \label{sec:mocksexpl}
Here we present mock \Fermi images of bright, hard sources.  These necessarily depend on the source properties and IGMF structure.  Thus we show a number of comparisons for a typical fluence for a bright hard source in Pass 8R2\_V6, corresponding to the $5000$ photons, and explore variations in each parameter direction.

Because we will be interested in applying this to the particular source classes appearing in future publications, i.e., a bright, hard subset of the \Fermi AGN sample, we limit the range of possibilities to those that are of observational interest.  Unless otherwise specified we adopt $\theta_j=3^\circ$, $\Theta=2^\circ$, $\Gamma_l=1.7$, $\Gamma_h=2.6$, $E_p=1~\TeV$, and $z=0.3$ \citep[see, e.g., the discussions surrounding the relevant \Fermi population in][]{BowTiesII}.  

\subsection{Front vs. Back Conversion}
The reconstruction of events in the Front and Back parts of the LAT have different PSFs and thus different sensitivities to anisotropic structure in the ICC halos.  Figure \ref{fig:fbmock} shows this explicitly for the three cases of most direct interest: an oblique ICC halo in the presence of a tangled IGMF, presumably associated with a gamma-ray dim object, an ICC halo from a blazar for a tangled IGMF, and an ICC halo from a blazar for a uniform IGMF.  In all cases the broader PSF of the Back-converted events is evident, reducing the prominence of the ICC halo structure. Nevertheless, where significant anisotropy in the halo is present it remains so in both at levels that are visually identifiable.  In practice, the additional information afforded by the increase in event statistics upon combining Front- and Back-converted events substantially outways the reduction in intrinsic halo structure due to the PSF.

\subsection{Redshift}
Halos from more distant objects are typically smaller as a result of the decreasing $\Dpp$ and growing distance.\footnote{While the angular diameter distance does indeed fall at high-$z$, this causes the angular scale to grow only very weakly over the redshifts of interest, e.g., those shown in Figure \ref{fig:size}.}  As a result the prominence of the halos' anisotropic structure after convolution with the \Fermi PSF is a function of redshift.  This is shown for both kinds of IGMF geometries we consider in Figure \ref{fig:zmock}, for which the extent of the halos decrease systematically with increasing $z$.  By $z=1$ the extended halo emission is effectively within the \Fermi Pass 8R2\_V6 PSF, implying that in general the halo component is spatially distinguished only for closer objects.

\subsection{Viewing Angle} \label{sec:VA}
Both the direct emission and halo components are strongly dependent on viewing angle primarily through the angular structure of the underlying gamma-ray jet.  The direct emission falls precipitously as $\Theta$ grows beyond $\theta_j$ due to the assumed Gaussian jet profile.  Thus, at large oblique angles (i.e., $\Theta\gg3^\circ$) there is no discernable source emission.  However, depending on the structure of the IGMF there may still be halo emission, detectable around AGN otherwise not detected by \Fermi or Cerenkov telescopes. 

For a small-scale, tangled IGMF the dependence of the halo on $\Theta$ arises from the foreshortening of the conical halo region.  For $\Theta=2^\circ$ (i.e., our default value) this all but erases the halo anisotropy.  In stark contrast, for $\Theta\gtrsim30^\circ$ the halo photons are clearly anisotropic, as seen in the top panels of Figure \ref{fig:Tmock}.  The anisotropy does not grow substantially from $\Theta=30^\circ$ to $\Theta=60^\circ$ due to the narrowness of the gamma-ray jet ($\theta_j=3^\circ$).  This suggests that gamma-ray studies of oblique radio-jet sources will provide a sensitive diagnostic for small-scale IGMFs.

For large-scale, uniform IGMFs the halo emission vanishes for viewing angles that substantially exceed the gyration angle at low energies, i.e., $\Theta>\Delta\alpha_{\rm def}$.  In principle, this places a mild constraint on the $\Theta$ for which the extended halo emission is visible.  In practice, as shown in the bottom panels of Figure \ref{fig:Tmock}, clear halo structure persists even to moderate values of $\Theta$.  Rather the primary impact of viewing angle in this case is found in the bipolar symmetry of the halo structure: larger $\Theta$ result in less symmetric halos.

The reason for this is again found in the structure of the gamma-ray jet.  For large $\Theta/\theta_j$ the intrinsic VHEGR emission exhibits a strong gradient across the field of view, i.e., pairs are not generated uniformly about the source.  Therefore, there will by many more pairs generated on the side of the projected, approaching jet.  The deficit of leptons on the other side results in a correspondingly dimmer halo component, and hence an asymmetry between the two sides of the otherwise bipolar halo.  When $\Theta/\theta_j\gg1$ this produces a one-sided halo, as seen for $\Theta=10^\circ$ in Figure \ref{fig:Tmock}.  For $\Theta/\theta_j\ll1$ the two components have similar strengths, with $\Theta/\theta_j\approx1$ only moderately asymmetric.

\subsection{VHEGR Jet Opening angle}
The impact of varying the jet width is shown in Figure \ref{fig:tjmock} for the large-scale, uniform IGMF models.  Because we already considered the dependence of halo structure on the viewing angle in the previous subsection, we keep $\Theta/\theta_j$ fixed at unity.  As a result, the asymmetry of the halo is similar for all panels.  What differs is the extent of the halos.  Because the halos are limited by the extent of the jet, smaller $\theta_j$ result in smaller halos.  This remains true after the LAT response is considered.  For typical values of $\theta_j$ we anticipate halos extending for many degrees.

\mbox{}\\
\subsection{IGMF}
Unsurprisingly, the impact of the IGMF strength depends on the particular IGMF geometry under consideration.  For ICC halos associated with small-scale, tangled IGMFs the strength must be sufficient to isotropize the pairs, as discussed in Section \ref{sec:GHiso}.  Thus, beyond requiring an IGMF strength above $10^{-15}~\G$, the ICC halos are independent of any additional parameters of the IGMF.

For large-scale, uniform IGMFs, however, the strength and orientation of the IGMF has a significant impact on the halo structure.  Stronger IGMFs produce larger deflections and therefore larger halos.  This is clearly evident in Figure \ref{fig:Bmock}, which shows halo realizations for a variety of IGMF strengths with fixed orientation.  For $B=10^{-16}~\G$ the anisotropic halo structure is marginally visible, becoming more so as $B$ increases.  In contrast, at $10^{-17}~\G$ and below the halo extent has reduced well within the PSF of front- and back-converted events in Pass 8R2\_V6, making it indistinguishable from an additional direct-source contribution.

\subsection{Spectral Shape}
The shape of the instrinsic spectrum affects the ICC halo through the relative normalizations of the halo and direct gamma-ray components.  This is clearly evident in Figures \ref{fig:GlMock} and \ref{fig:GhMock} which show the impact of varying the low- and high-energy photon spectral indexes.  Of these, the low-energy spectral shape appears to be significantly more important.  The reason for this, however, is closely related to the class of spectra we considered.

The luminosity in VHEGRs, and therefore in halo photons, is typicall sensitive primarily to the intrinsic VHEGR flux at 1~TeV.  While we do show one inverted VHEGR spectrum ($\Gamma_h=1.9$), it is only mildly so, again making it a strong function of the flux at a TeV, where most of the VHGER photons reside.  In comparison, it is only modestly dependent on $\Gamma_h$, roughly as $(\Gamma_h-2)^{-1}$, typically making a factor of few difference.

In contrast, the low-energy spectral shape has a large impact.  This is a direct result of the normalizing the halo by the observed 1-100~GeV intrinsic emission.  Despite the flat/inverted spectral energy distribution of the hard sources that are of most interest, the number of gamma rays is still overwhelmingly dominated by those at low energies, effectively pinning the spectrum near 1~GeV.  Thus, small variations in $\Gamma_l$ result in large variations in the number of gamma rays at the presumed spectral break, i.e., $\left[E_p/(1~\GeV)\right]^{-\Gamma_l}=10^{-3\Gamma_l}$, where in the latter we have assumed $E_p=1~\TeV$.  Unlike the dependence on $\Gamma_h$, this is exponential on $\Gamma_l$.  Generally, softer spectra produce less VHEGR power and thus weaker ICC halos.

\mbox{}\\
\section{Conclusions} \label{sec:conc}
The putative gamma-ray halos that surround bright VHEGR sources are generally highly structured.  The reason for and degree of structure within the ICC halo depends most strongly on the strength and geometry of the IGMF and jet orientation.

For small-scale, tangled IGMFs ($\lambda_B\ll3~\Mpc$) the structure arises solely from the jetted nature of the VHEGR emission.  Oblique viewing angles ($\Theta\gg\theta_j$) produce clear bimodal gamma-ray features.  In contrast, acute viewing angles ($\Theta\lesssim\theta_j$) result in foreshortened ICC halos, suppressing the anisotropy.

For large-scale, uniform IGMFs ($\lambda_B\gg3~\Mpc$) the structure arises from the ordered gyration of the pairs, i.e., their limited phase-space evolution.  This results in a pair of thin halo features whose size and strength are indicative of the magnetic field strength and viewing angle, respectively.  Strong fields result in large deflection angles and therefore extended halos.  Acute viewing angles produce nearly-symmetric bimodal halos; oblique viewing angles generate increasing asymmetry.

For nearby VHEGR sources, i.e., $z<1$, the asymmetric halo structures typically remain visible after convolution with the LAT instrument response.  This is a consequence of their typically large angular extent.  However, for large-scale, uniform IGMFs with strength smaller than $10^{-17}~\G$ this is not true, generating halos that are fully contained within the effective PSF of the intrinsic source photons.  As a result, for acute viewing angles the ICC halos arising in a sufficiently strong, large-scale, uniform IGMF are readily identifiable.  Similarly,  for oblique viewing angles a sufficiently strong, small-scale, tangled IGMF is clearly evident.

These conclusions are quantitatively but not qualitatively dependent on the remaining source parameters: redshift, jet opening angle, where the event converted in the LAT (i.e., front- vs back-conversion).  The structure is independent of the total source flux, background, and angular structure of the LAT.

In \citet{BowTiesII} we present a statistical scheme to identify the presence of gamma-ray halos around \Fermi sources that exploits their nearly symmetric, bimodal structure.  In \citet{BowTiesIII} we apply this to the existing sample of suitable \Fermi blazars, placing constraints on the geometry and strength of the IGMF.

\acknowledgments A.E.B.~receives financial support from the Perimeter Institute for Theoretical Physics and the Natural Sciences and Engineering Research Council of Canada through a Discovery Grant.  Research at Perimeter Institute is supported by the Government of Canada through Industry Canada and by the Province of Ontario through the Ministry of Research and Innovation.  C.P.~acknowledges support by the European Research Council under ERC-CoG grant CRAGSMAN-646955 and by the Klaus Tschira Foundation. P.C. gratefully acknowledges support from the NASA ATP program through NASA grant NNX13AH43G, and NSF grant AST-1255469. A.L. receives financial support from an Alfred P. Sloan Research Fellowship, NASA ATP Grant NNX14AH35G, and NSF Collaborative Research Grant 411920 and CAREER grant 1455342. E.P. acknowledges support by the Kavli Foundation.

\begin{appendix}

\section{Conservation Cross-Checks} \label{app:checks}
In a number of places we relate spectra of various kinds.  Here we demonstrate explicitly that the anticipated energy and/or number is conserved in these relations.  These take two forms: first is the explicit numerical comparison in the transition from one spectrum to the next and second based upon the physical expectation that the ICCs serve as a mechanism to reprocess the VHEGR luminosity to lower energies.  This latter condition also permits an assessment of the various redshift dependencies.  

\subsection{General Formulas}
We begin with the validation of key general formula.  In Appendix \ref{app:ICCcons} we consider situations specific to the small- and large-scale IGMF halo models.

\subsubsection{From VHEGR to pairs} \label{app:pairs_energy}
The first general relation we check the total power injected into the pairs.  That is, we compare the power in generated pairs to that in the incident gamma rays. The total kinetic power injected into the pairs is
\begin{equation}
  \begin{aligned}
    &\iint d\E' d^3\!x' \E' \left[
      \frac{dN_{e^-}}{dt'd\E'\dthxp}(\E',\bx')
      +
      \frac{dN_{e^+}}{dt'd\E'\dthxp}(\E',\bx')
      \right] =
    2 \iint d\E'  d^3\!x' \E' \frac{2}{r'^2} \dNAGN(2\E',\bhx') e^{-\tau(2\E',z,r')}
    \frac{d\tau}{dr'}(2\E',z,r')\\
    &\qquad\qquad=
    \iint dE'  \dOAGN E' \dNAGN(E',\bhx') 
    \int dr'
    e^{-\tau(E',z,r')}
    \frac{d\tau}{dr'}(E',z,r')
    =
    \iint dE'  \dOAGN E' \dNAGN(E',\bhx')\,,
  \end{aligned}
\end{equation}
which is the total power in VHEGRs from the AGN, as expected.

We also expect numbers to be conserved, in a sense, in this case as every VHEGR produces two pairs.  Again this is easy to show: the rate at which pairs are produced is
\begin{equation}
  \begin{aligned}
    &\iint d\E' d^3\!x' \left[
      \frac{dN_{e^-}}{dt' d\E' \dthxp}(\E',\bx')
      +
      \frac{dN_{e^+}}{dt' d\E' \dthxp}(\E',\bx')
      \right]
    =
    2 \iint d\E'  d^3\!x' \frac{2}{r'^2} \dNAGN(2\E',\bhx') e^{-\tau(2\E',z,r')}
    \frac{d\tau}{dr'}(2\E',z,r')\\
    &\qquad =
    \iint dE'  \dOAGN 2 \dNAGN(E',\bhx') 
    \int dr'
    e^{-\tau(E',z,r')}
    \frac{d\tau}{dr'}(E',z,r')
    =
    2 \iint d\E'  \dOAGN \dNAGN(\E',\bhx')\,,
  \end{aligned}
\end{equation}
which is twice the rate at which VHEGR photons are produced by the AGN.

\subsubsection{From pairs to halo photons} \label{app:ICCs_energy}
Inverse Compton scattering reprocesses energy from the pairs to the up-scattered gamma rays.  As such, the total power injected into the pairs should equal the total power leaving in gamma rays.  Note that this does not imply that the band-specific powers need be identical.  Even for matched bands (i.e., the injection energy band in pairs corresponds to the range of energies that produce inverse Compton gamma rays in the band of interest) this balance will be violated as a result of pairs cooling out of or into the band, modifying the balance by factors of order unity.

Combining Equations (\ref{eq:dNICC}) and (\ref{eq:BES}), and using the definition of $t_{\rm IC}$, the total power in the ICC emission from electrons is
\begin{equation}
  \begin{aligned}
    &\int d^3\!q' \, \E' \frac{dN_{\rm IC,\epm}}{dt' d^3\!x' d^3\!q'}
    =
    \int d^3\!q' q'c \frac{3 m_\e c}{8 t_{\rm IC} q_s'}
    \left(\frac{m_\e^2 c^2}{2 q_s'q'}\right)^{3/2}
    f_\epm \left(\bx',\frac{m_\e c \bq'}{\sqrt{2q_s'q'}}\right)\\
    &\qquad\qquad=
    \int
    d^3\!\left(\frac{m_\e c q'}{\sqrt{2 q_s' q'}}\right)
    \frac{3 m_\e c^2 q'}{4 t_{\rm IC} q_s'}
    f_\epm \left(\bx',\frac{m_\e c \bq'}{\sqrt{2q_s'q'}}\right)  
    =
    \int
    d^3\!p'
    \frac{3 p'^2}{2 t_{\rm IC} m_\e}
    f_\epm \left(\bx',\bp'\right)\\
    &\qquad\qquad=
    \int
    d^3\!p'
    \frac{3 p'^2}{2 t_{\rm IC} m_\e}
    \frac{m_\e c t_{\rm IC}}{p'^4}
    \int_{p'}^\infty dq' q'^2
    \dot{f}_{\rm inj,\epm}[\bx',\bq'(\bp')]
    =
    \frac{3c}{2}
    \int d\Omega'_p
    \int_0^\infty dp'
    \int_{p'}^\infty dq' q'^2
    \dot{f}_{\rm inj,\epm}[\bx',\bq'(\bp')]\\
    &\qquad\qquad=
    \frac{3}{2} c
    \int_0^\infty dq' q'^2
    \int d\Omega_q'
    \int_0^{q'} dp'
    \dot{f}_{\rm inj,\epm}(\bx',\bq')
    =
    \frac{3}{2}
    \int d^3\!q'
    q'c
    \dot{f}_{\rm inj,\epm}(\bx',\bq')\,,
  \end{aligned}
\end{equation}
where we used the fact that $d\Omega_p'=d\Omega_q'$.  Up to the factor of $3/2$ that arises from the isotropic-scattering approximation\footnote{The isotropic-scattering approximation cannot simultaneously reproduce the correct number of scattering events and the power of scattered gamma rays.  Because we are interested in the distribution of gamma-ray events associated with the ICC halos, we have chosen to maintain the former at the expense of the latter.} the final expression is the power injected in each species, as anticipated.

\subsection{From VHEGR to halo photons} \label{app:ICCcons}
Because inverse Compton reprocesses the luminosity in VHEGRs, the total luminosity in the ICC halos must match that of the driving VHEGR emission.  Hence, this provides a natural check of the halo gamma-ray maps.

For simplicity we will limit our attention to $z=0$, suppress the primes, assume the VHEGR spectrum has a single power-law (i.e., $\Gamma_l=\Gamma_h$), and that the observation band has only a lower energy limit, $E_m$.  In this case we generally anticipate 
\begin{equation}
  L_{\rm ICC}
  =
  \frac{3}{2} \frac{L_{\rm VHEGR}}{\Gamma-1}
  =
  \frac{3}{2} \frac{1}{\Gamma-1}
  \int d\Omega \int_{2 m_\e c^2\sqrt{E_m/2 E_{\rm CMB}}}^\infty dE E
  \frac{dN_{\rm AGN}}{dt dE d\Omega}
  =
  \frac{3}{2} \frac{1}{\Gamma-1}
  \left( \int d\Omega G \right) \frac{f_0 E_p^2}{\Gamma-2}
  \left(\frac{2m_\e c^2}{E_p} \sqrt{\frac{E_m}{2E_{\rm CMB}}}\right)^{2-\Gamma}\,.
  \label{eq:ICCVHEGRbalance}
\end{equation}
The additional factor of $\Gamma-1$ arises due to an excess of pairs generated at the low-energy limit that subsequently cool out of the relevant energy band, and the $3/2$ is a result of the isotropic-scattering approximation.  The remainder of the expression is the VHEGR luminosity emitted at energies sufficient to generate pairs who will initially inverse Compton up-scatter the CMB to $E_m$.

\subsubsection{Tangled Fields} \label{app:check_iso}
While normally we are interested in relating the ICC halo to the observed source flux, in this case we must fix the intrinsic flux.  We do this in principle by choosing a value for $F_{35}$ along the jet axis; in practice we simply use Equation (\ref{eq:F35norm}) to remove $F_{35}$ altogether in favor of an orientation independent normalization, $f_0$.  Thus, the total ICC luminosity, i.e., the total energy flux integrated over all directions is 
\begin{equation}
  \begin{aligned}
    L_{\rm ICC}
    &=
    \int d\Omega_p
    \int_{E_m}^\infty dE
    \int \frac{d^2\!R}{D_A^2}
    \frac{D_L^2}{A} E
    \frac{A}{D_A^2} f_0 E_p
    \frac{m_\e c^2}{\delta_A^2 E_{\rm CMB}^2} \frac{3}{4\pi}
    \left(\frac{m_\e c^2}{E_p}\right)^{1-\Gamma}
    \left(\frac{2E}{E_{\rm CMB}}\right)^{-\Gamma/2}
    \Psi_{2-\Gamma}\left(\sqrt{\frac{2E}{E_{\rm CMB}}}\frac{\ba}{\delta_A}\right)\\
    &=
    4\pi
    \frac{3}{4\pi}
    \frac{f_0 E_p m_\e c^2}{E_{\rm CMB}^2 D_p^2} \left(\frac{m_\e c^2}{E_p}\right)^{2+1-\Gamma}
    \frac{E_{\rm CMB}^2}{4}
    \int_{2E_m/E_{\rm CMB}}^\infty d\left(\frac{2E}{E_{\rm CMB}}\right)
    \left(\frac{2E}{E_{\rm CMB}}\right)^{1-\Gamma/2}
    \int d^2\!R \Psi_{2-\Gamma}\left(\frac{m_\e c^2}{E_p} \sqrt{\frac{2E}{E_{\rm CMB}}}\frac{\bR}{D_p}\right)\\
    &=
    \frac{3}{4} \frac{f_0 E_p m_\e c^2}{D_p^2} \left(\frac{m_\e c^2}{E_p}\right)^{3-\Gamma}
    \int_{2E_m/E_{\rm CMB}}^\infty dx x^{(2-\Gamma)/2}
    \left(\frac{E_p}{m_\e c^2}\right)^2 x^{-1} D_p^2
    \int d^2\!y \Psi_{2-\Gamma}(\bmath{y})\,.
  \end{aligned}
\end{equation}
At this point a useful identity is
\begin{equation}
  \int d^2\!R \Psi_k(\bR) = \int d^3\!x r^{-k-2} \Gamma(k,r) G
  = \left(\int d\Omega G \right) \int dr r^{-k} \Gamma(k,r)
  = \left(\int d\Omega G \right) \frac{1}{1-k}\,,
\end{equation}
where we have suppressed the angular arguments of $G(\theta')$.  Therefore,
\begin{equation}
  \begin{aligned}
    L_{\rm ICC}
    &=
    \frac{3}{4} f_0 E_p^2 \left(\frac{m_\e c^2}{E_p}\right)^{2-\Gamma}
    \int_{2E_m/E_{\rm CMB}}^\infty dx x^{-\Gamma/2}
    \left(\int d\Omega G\right) \frac{1}{\Gamma-1}\\
    &=
    \frac{3}{4} f_0 E_p^2
    \frac{2}{\Gamma-2} \left(\frac{m_\e c^2}{E_p}\right)^{2-\Gamma}\left(\frac{2E_m}{E_{\rm CMB}}\right)^{(2-\Gamma)/2}
    \left(\int d\Omega G\right) \frac{1}{\Gamma-1}\\
    &=
    \frac{3}{2} \frac{1}{\Gamma-1} \left(\int d\Omega G\right) \frac{f_0 E_p^2}{\Gamma-2}
    \left(\frac{m_\e c^2}{E_p} \sqrt{\frac{2E_m}{E_{\rm CMB}}}\right)^{2-\Gamma}
    =
    \frac{3}{2} \frac{1}{\Gamma-1} L_{\rm VHEGR}\,.
  \end{aligned}
\end{equation}
Thus, we find the expected result.

We have also verified the expected equality numerically when employing the numerically generated realizations.  This was done not only in the special case analytically assessed above, but for arbitrary $\Gamma_h$, $\Gamma_l$, and $E_p$.

\subsubsection{Uniform Fields} \label{app:check_weak}
We now turn to the case of uniform IGMFs, for which the inverse Compton luminosity is
\begin{equation}
  \begin{aligned}
    L_{\rm ICC}
    &=
    \int d\Omega_p
    \int_{E_m}^\infty dE
    \int \frac{d^2\!R}{D_A^2}
    \frac{D_L^2}{A} E
  \frac{A}{D_A^2} f_0 E_p
  \frac{mc^2}{\epsilon^2} \left(\frac{m_\e c^2}{E_p}\right)^2 \frac{D_A^2}{D_p^2} \frac{3}{64}
  \frac{1}{\omega_B t_{\rm IC}}
  \left(\frac{E_p}{m_\e c^2}\right)^4
  \left(\frac{E}{2 E_{\rm CMB}}\right)^{-3/2}\\
  &\qquad\qquad\qquad\qquad\qquad\qquad\times
  \frac{D_p G_c}{r_c^2\sin\vartheta_\ell \mathcal{J}_c}
  \sum_{\rm e^+,e^-} \sum_k 
  \left(\frac{2\tilde{\E}^{\epm}_k}{E_p}\right)^{4-\Gamma}
  e^{-2\tilde{\E}^{\epm}_k/E_D}
  \Theta\left(\tilde{\E}^{\epm}_k - m_\e c^2\sqrt{\frac{E}{2 E_{\rm CMB}}}\right)\\
  &=
  \frac{3}{32} f_0 E_p^2 \left(\frac{E_p}{m_\e c^2}\right)
  \frac{1}{\beta}
  \int d\Omega_p \int d^2R \int_{E_m/2 E_{\rm CMB}}^\infty
  d\left(\frac{E}{2 E_{\rm CMB}}\right)
  \left(\frac{E}{2 E_{\rm CMB}}\right)^{-1/2}\\
  &\qquad\qquad\qquad\qquad\qquad\qquad\times
  \frac{G_c}{D_p r_c^2 \mathcal{J}_c}
  \sum_{\rm e^+,e^-} \sum_k
  \left(\frac{2\tilde{\E}^\epm_k}{E_p}\right)^{4-\Gamma}
  e^{-2\tilde{E}^\epm_k/E_D}
  \Theta\left(\tilde{\E}^\epm_k - m_\e c^2\sqrt{\frac{E}{2 E_{\rm CMB}}}\right)\,,
  \end{aligned}
\end{equation}
where for compactness we have defined $\beta=\omega_B t_{\rm IC} \sin\vartheta_\ell/2$ and we have explicitly noted quantities that are evaluated at the roots of $\delta(\vartheta-\vartheta_\ell)$ with a subscript $c$.  Without loss of generality we may redefine the azimuthal integration to subsume the $\varphi$ in Equation (\ref{eq:EKS}), and thus
\begin{equation}
  \tilde{\E}^\epm_k = m_\e c^2 \sqrt{\frac{\beta}{2\pi k-\varphi_p \pm 2E_{\rm CMB} \beta/E}}\,.
\end{equation}
Note that this differs from Equation (\ref{eq:EKS}) because $E$ is the energy of the up-scattered gamma ray instead of that of the lepton.

At this point we will re-order the integrations and make a change of variables from $E$ to $\tilde{\E}^\epm_k$.  To do this, note that 
\begin{equation}
  \frac{E}{2E_{\rm CMB}} = \frac{\beta}{\varphi-2\pi k \pm \beta m_\e^2c^4/\tilde{\E}^{\epm\,2}_k}
  \quad\rightarrow\quad
  d\left(\frac{E}{2E_{\rm CMB}}\right)
  =
  2 \left(\frac{E}{2E_{\rm CMB}}\right)^2
  \frac{m_\e^2 c^4}{{\tilde{\E}^{\epm\,3}_k}} d\tilde{\E}^\epm_k\,.
\end{equation}
Therefore, effecting the limits on the energy integration through Heaviside functions,
\begin{equation}
  \begin{aligned}
    L_{\rm ICC}
    &=
    \frac{3}{16} f_0 E_p^2 \left(\frac{E_p}{m_\e c^2}\right)
    \frac{1}{\beta}
    \int d\Omega_p  \int d^2\!R
    \frac{G_c}{D_p r_c^2 \mathcal{J}_c}
    \sum_{\rm e^+,e^-} \sum_k
    \int
    d\tilde{\E}^{\epm}_k
    \left(\frac{E}{2E_{\rm CMB}}\right)^{3/2}
    \frac{m_\e^2c^4}{\tilde{\E}^{\epm\,3}_k}
    \left(\frac{2\tilde{\E}^{\epm}_k}{E_p}\right)^{4-\Gamma}
    e^{-2\tilde{\E}^{\epm}_k/E_D}\\
    &\qquad\qquad\qquad\qquad\qquad\times
    \Theta\left(\tilde{\E}^\epm_k - m_\e c^2\sqrt{\frac{E}{2E_{\rm CMB}}}\right)
    \Theta\left(E-E_m\right) \Theta\left(\infty-E\right) \\    
    &=
    \frac{3}{16} f_0 E_p^2 \left(\frac{E_p}{m_\e c^2}\right)
    \beta^{1/2}
    \int d\Omega_p  \int d^2\!R
    \frac{G_c}{D_p r_c^2 \mathcal{J}_c}
    \sum_{\rm e^+,e^-} \sum_k
    \int
    d\left(\frac{2\tilde{\E}^{\epm}_k}{E_p}\right)
    \left(\varphi-2\pi k + \frac{\beta m_\e^2c^4}{\tilde{\E}^{\epm\,2}_k}\right)^{-3/2}\\
    &\qquad\qquad\qquad\qquad\qquad\times
    \frac{4 m_\e^2c^4}{E_p^2}
    \left(\frac{2\tilde{\E}^{\epm}_k}{E_p}\right)^{1-\Gamma}
    e^{-2\tilde{\E}^{\epm}_k/E_D}
    \Theta\left(\tilde{\E}^{\epm}_k - m_\e c^2\sqrt{\frac{E}{2E_{\rm CMB}}}\right)
    \Theta\left(E-E_m\right) \Theta\left(\infty-E\right) \\
    &=
    \frac{3}{4} f_0 E_p^2 \left(\frac{E_p}{m_\e c^2}\right)^{-1}
    \beta^{1/2}
    \sum_{\rm e^+,e^-} \sum_k
    \int_0^\pi \sin\vartheta_\ell d\vartheta_\ell \int d^2\!R
    \frac{G_c}{D_p r_c^2 \mathcal{J}_c}
    \int 
    dx
    x^{1-\Gamma}
    e^{-(r_c/D_p)x}\\
    &\qquad\qquad\qquad\qquad\qquad\times
    \int_0^{2\pi} d\varphi
    \left(\varphi-2\pi k + \frac{4\beta m_\e^2c^4}{E_p^2 x^2}\right)^{-3/2}
    \Theta\left(\tilde{\E}^\epm_k - m_\e c^2\sqrt{\frac{E}{2E_{\rm CMB}}}\right)
    \Theta\left(E-E_m\right) \Theta\left(\infty-E\right) \,.    
  \end{aligned}
\end{equation}
The limits on $\varphi$ integral are set in part by the Heaviside functions, and depend on $k$.  We will assume that $2\epsilon\beta/E_m \ll 2\pi$, corresponding to the assumption that the leptons that generate the IC photons of interest make only a small deflection prior to doing so.  Furthermore, we will consider only the species for which $\beta>0$ -- the final contribution to the luminosity for both are equal.  Then,
\begin{equation}
  \begin{aligned}
    &\tilde{\E}^{\epm}_k \ge m_\e c^2\sqrt{\frac{E}{2E_{\rm CMB}}}
    = m_\e c^2\sqrt{\frac{\beta}{\varphi-2\pi k \pm \beta m_\e^2c^4/\tilde{\E}^{\epm\,2}_k}}
    &&\quad\rightarrow\quad
    \varphi \ge 2\pi k\\
    &\infty \ge E = \frac{2E_{\rm CMB}\beta}{\varphi-2\pi k + \beta m_\e^2c^4/\tilde{\E}^{\epm\,2}_k}
    &&\quad\rightarrow\quad
    \varphi \ge 2\pi  - \frac{\beta m_\e^2c^4}{\tilde{\E}^{\epm\,2}_k}\\
    &E=\frac{2E_{\rm CMB}\beta}{\varphi-2\pi k + \beta m_\e^2c^4/\tilde{\E}^{\epm\,2}_k} \ge E_m
    &&\quad\rightarrow\quad
    \varphi \le 2\pi k + \frac{2E_{\rm CMB} \beta}{E_m} - \frac{\beta m_\e^2c^4}{\tilde{\E}^{\epm\,2}_k}.
  \end{aligned}
\end{equation} 
These are supplemented with the integration range $0\le\varphi\le2\pi$.  Given the assumptions above, there are no values of $k$ for which the allowed interval in $\varphi$ is finite when $E_k< mc^2 \sqrt{E_m/2E_{\rm CMB}}$. This is expected because such leptons would not emit into the \Fermi-LAT band.  For $E_k>mc^2\sqrt{E_m/2E_{\rm CMB}}$ we have only $k=0$ contributions, for which 
\begin{equation}
  0
  \le \varphi \le
  \frac{2E_{\rm CMB}\beta}{E_m} - \frac{\beta m_\e^2c^4}{E_k^2}
  =
  \frac{2E_{\rm CMB}\beta}{E_m} - \frac{4 \beta m_\e^2c^4}{E_p^2 x^2}\,.
\end{equation}
Therefore, upon performing the $\varphi$ integral, 
\begin{equation}
  \begin{aligned}
    L_{\rm ICC}
    &=
    \frac{3}{4} f_0 E_p^2 \left(\frac{E_p}{m_\e c^2}\right)^{-1}
    \beta^{1/2}
    \sum_S
    \int_0^\pi \sin\vartheta_\ell d\vartheta_\ell \int d^2\!R
    \frac{G_c}{D_p r_c^2 \mathcal{J}_c}
    \int_{(2m_\e c^2/E_p)\sqrt{E_m/2E_{\rm CMB}}}^\infty
    dx
    x^{1-\Gamma}
    e^{-(r_c/D_p)x}\\
    &\qquad\qquad\qquad\qquad\qquad\qquad\qquad\qquad\qquad\qquad\qquad\qquad\times
    2
    \left[
    \left(\frac{4\beta m_\e^2c^4}{E_p^2 x^2}\right)^{-1/2}
    -
    \left(\frac{2E_{\rm CMB}\beta}{E_m}\right)^{-1/2}
    \right]\\
    &=
    \frac{3}{2} f_0 E_p^2 \left(\frac{E_p}{m_\e c^2}\right)^{-1}
    \sum_S
    \int_{(2m_\e c^2/E_p)\sqrt{E_m/2E_{\rm CMB}}}^\infty
    dx
    x^{1-\Gamma}
    \left[
    \left(\frac{4m_\e^2c^4}{E_p^2 x^2}\right)^{-1/2}
    -
    \left(\frac{2E_{\rm CMB}}{E_m}\right)^{-1/2}
    \right]\\
    &\qquad\qquad\qquad\qquad\qquad\qquad\qquad\qquad\qquad\qquad\qquad\qquad\times
    \int_0^\pi \sin\vartheta_\ell d\vartheta_\ell \int d^2\!R
    \frac{G_c}{D_p r_c^2 \mathcal{J}_c}
    e^{-(r_c/D_p)x}\\
    &=
    \frac{3}{2} f_0 E_p^2 \left(\frac{E_p}{m_\e c^2}\right)^{-1}
    \sum_S
    \int_{(2m_\e c^2/E_p)\sqrt{E_m/2E_{\rm CMB}}}^\infty
    dx
    x^{1-\Gamma}
    \left[
    \left(\frac{4m_\e^2c^4}{E_p^2 x^2}\right)^{-1/2}
    -
    \left(\frac{2E_{\rm CMB}}{E_m}\right)^{-1/2}
    \right]\\
    &\qquad\qquad\qquad\qquad\qquad\qquad\qquad\qquad\qquad\qquad\qquad\qquad\times
    \int_0^\pi \sin\vartheta_\ell d\vartheta_\ell \int d^2\!R d\ell
    \frac{G}{D_p r^2 \sin\vartheta_\ell}\delta(\vartheta-\vartheta_\ell)
    e^{-(r/D_p)x}\,,
  \end{aligned}
\end{equation}
where we have re-inserted the explicit integration over the line of sight in the final expression.  Rearranging the volume integral we find, after some simplification
\begin{equation}
  \begin{aligned}
    L_{\rm ICC}
    &=
    \frac{3}{2} f_0 E_p^2 \left(\frac{E_p}{m_\e c^2}\right)^{-1}
    \sum_S
    \int_{(2m_\e c^2/E_p)\sqrt{E_m/2E_{\rm CMB}}}^\infty
    dx
    x^{1-\Gamma}
    \left[
    \left(\frac{4m_\e^2c^4}{E_p^2 x^2}\right)^{-1/2}
    -
    \left(\frac{2E_{\rm CMB}}{E_m}\right)^{-1/2}
    \right]
    \int d\Omega G
    \int dr \frac{e^{-(r/D_p)x}}{D_p}\\
    &=
    \frac{3}{2} f_0 E_p^2 \left(\frac{E_p}{m_\e c^2}\right)^{-1}
    \left( \int d\Omega G \right)
    \sum_S
    \int_{(2m_\e c^2/E_p)\sqrt{E_m/2E_{\rm CMB}}}^\infty
    dx
    x^{1-\Gamma}
    \left[
    \left(\frac{4m_\e^2c^4}{E_p^2 x^2}\right)^{-1/2}
    -
    \left(\frac{2E_{\rm CMB}}{E_m}\right)^{-1/2}
    \right] \frac{1}{x}\\
    &=
    \frac{3}{4} f_0 E_p^2 
    \left( \int d\Omega G \right)
    \sum_S
    \int_{(2m_\e c^2/E_p)\sqrt{E_m/2E_{\rm CMB}}}^\infty
    dx
    x^{-\Gamma}
    \left(
      x
      -
      \frac{2m_\e c^2}{E_p} \sqrt{\frac{E_m}{2E_{\rm CMB}}}
    \right)\\
    &=
    \frac{3}{4} f_0 E_p^2 
    \left( \int d\Omega G \right)
    \sum_S
    \left[
      \frac{1}{\Gamma-2}
      \left( \frac{2m_\e c^2}{E_p}\sqrt{\frac{E_m}{2E_{\rm CMB}}} \right)^{2-\Gamma}
      -
      \frac{1}{\Gamma-1} 
      \left( \frac{2m_\e c^2}{E_p}\sqrt{\frac{E_m}{2E_{\rm CMB}}} \right)^{2-\Gamma}
    \right]\\
    &=
    \frac{3}{2} f_0 E_p^2 
    \left( \int d\Omega G \right)
    \left( \frac{2m_\e c^2}{E_p}\sqrt{\frac{E_m}{2E_{\rm CMB}}} \right)^{2-\Gamma}
    \frac{1}{(\Gamma-2)(\Gamma-1)}\\
    &=
    \frac{3}{2} \frac{1}{\Gamma-1}
    L_{\rm VHEGR}\,,
  \end{aligned}
\end{equation}
which again satisfies the anticipated relation.  As before, we have also explicitly verified the expected equality numerically when employing the numerically generated realizations for a variety of spectral and jet parameters.

\subsection{Redshift Dependence of ICC Halos} \label{app:redshift}
Finally we consider the redshift dependence of the ICC halos generally and compare with that found in Sections \ref{sec:GHiso} and \ref{sec:GHweak}.  Again we will exploit the fact that ICC halos are the reprocessed VHEGR emission, and thus with the exception of the negligible energy cost of creating the pairs (i.e., $2m_\e c^2$) we expect the luminosity of the halo and VHEGR components to be identical.

We will denote quantities in the source frame (at redshift $z$) with primes and quantities in the observer frame (at $z=0$) without.  We will further assume that
\begin{equation}
L'_{E'} = L_0 \left(\frac{E'}{E_p}\right)^{1-\Gamma}\,,
\end{equation}
i.e., that $\Gamma_l=\Gamma_h$ and a spectrum in accordance with Equation (\ref{eq:VHEGRdist}).

We begin with the well-known relation for the {\em energy flux}
\begin{equation}
  \F = \frac{L'}{4\pi D_L^2}\,,
\end{equation}
which implies that the specific {\em photon} flux (i.e., flux) is related to the intrinsic specific luminosity via
\begin{equation}
  \F_E = \frac{d\F}{dE} = \frac{1}{4\pi D_L^2} \frac{dL'}{dE}
  = \frac{1}{4\pi D_L^2} \frac{dE'}{dE} \frac{dL'}{dE'}
  = (1+z) \frac{L'_{E'}}{4\pi D_L^2}\,.
\end{equation}
This then gives for the flux
\begin{equation}
  F_E \equiv \frac{\F_E}{E} = (1+z) \frac{\F_E}{E'} = (1+z)^2 \frac{(L'_{E'}/E')}{4\pi D_L^2}\,.
\end{equation}
We can now integrate this up over various bands.

The band-specific flux from $E_m$ to $E_M$ is
\begin{equation}
  \F_{mM}
  = \int_{E_m}^{E_M} dE \F_E
  = \int_{E_m}^{E_M} dE (1+z) \frac{L'_{E'}}{4\pi D_L^2}
  = \int_{(1+z)E_m}^{(1+z)E_M} dE' \frac{dE}{dE'} (1+z) \frac{L'_{E'}}{4\pi D_L^2}
  = \frac{1}{4\pi D_L^2} \int_{(1+z)E_m}^{(1+z)E_M} dE' L'_{E'}\,.
\end{equation}
Inserting the assumed form of the specific luminosity gives
\begin{equation}
  \F_{mM}
  =
  \frac{1}{4\pi D_L^2} \int_{(1+z)E_m}^{(1+z)E_M} dE' L_0 \left(\frac{E'}{E_p}\right)^{1-\Gamma}
  =
  \frac{(1+z)^{2-\Gamma}}{4\pi D_L^2} \frac{L_0 E_p}{2-\Gamma}
  \left[ \left(\frac{E_M}{E_p}\right)^{2-\Gamma} - \left(\frac{E_m}{E_p}\right)^{2-\Gamma} \right]\,.
\end{equation}
Here the additional factors of $(1+z)$ arise from the band-correction.

Similarly, the band-specific flux from $E_m$ to $E_M$ is
\begin{equation}
  F_{mM}
  = \int_{E_m}^{E_M} dE F_E
  = \int_{E_m}^{E_M} dE (1+z)^2 \frac{(L'_{E'}/E')}{4\pi D_L^2}
  = \frac{(1+z)}{4\pi D_L^2} \int_{(1+z)E_m}^{(1+z)E_M} dE' \frac{L'_{E'}}{E'}\,.
\end{equation}
Inserting the assumed form of the specific luminosity gives
\begin{equation}
  F_{mM}
  =
  \frac{(1+z)}{4\pi D_L^2} \int_{(1+z)E_m}^{(1+z)E_M} dE' \frac{1}{E'} L_0 \left(\frac{E'}{E_p}\right)^{1-\Gamma}
  =
  \frac{(1+z)^{2-\Gamma}}{4\pi D_L^2} \frac{L_0}{1-\Gamma}
  \left[ \left(\frac{E_M}{E_p}\right)^{1-\Gamma} - \left(\frac{E_m}{E_p}\right)^{1-\Gamma} \right]\,.
\end{equation}
Here again the additional factors of $(1+z)$ arise from the band-correction.

Similarly, for the ICC halo,
\begin{equation}
  \F_{\rm ICC} = \frac{L'_{IC}}{4\pi D_L^2}
  \quad\rightarrow\quad
  \F_{{\rm ICC},E} = (1+z) \frac{L'_{{\rm ICC},E'}}{4\pi D_L^2}
  \quad\text{and}\quad
  F_{{\rm ICC},E} = (1+z)^2 \frac{(L'_{{\rm ICC},E'}/E')}{4\pi D_L^2}\,.
\end{equation}
The band-specific IC flux is
\begin{equation}
  \F_{{\rm ICC},mM}
  =
  \int_{E_m}^{E_M} dE \F_{{\rm ICC},E}
  =
  \frac{1}{4\pi D_L^2} \int_{(1+z)E_m}^{(1+z)E_M} dE' L'_{{\rm ICC},E'}\,.
\end{equation}
However, the ICC luminosity is equal to the VHEGR luminosity at the corresponding energy range up to a factor of $3/2(\Gamma-1)$ (due to particle cooling, see Equation (\ref{eq:ICCVHEGRbalance})).  That is,
\begin{equation}
  \int_{E'_m}^{E'_M} dE' L'_{{\rm ICC},E'}
  =
  \frac{3}{2(\Gamma-1)}
  \int_{2m_\e c^2\sqrt{E'_m/2E_{\rm CMB}'}}^{2m_\e c^2\sqrt{E'_M/2E_{\rm CMB}'}} dE' L'_{E'}\,.
\end{equation}
Note that since both the CMB and gamma rays of interest redshift at equal rates $E'_M/E_{\rm CMB}' = E_M/E_{\rm CMB}$, simplifying the limits in principle.  Inserting the power-law intrinsic spectrum gives
\begin{equation}
  \begin{aligned}
    \F_{{\rm ICC},mM}
    &=
    \frac{1}{4\pi D_L^2}
    \frac{3}{2(\Gamma-1)}
    \frac{L_0}{2-\Gamma}
    \left[
      \left(\frac{2m_\e c^2}{E_p}\sqrt{\frac{E_M}{2E_{\rm CMB}}}\right)^{2-\Gamma}
      -
      \left(\frac{2m_\e c^2}{E_p}\sqrt{\frac{E_m}{2E_{\rm CMB}}}\right)^{2-\Gamma}
    \right]\\
    &=
    \frac{1}{4\pi D_L^2}
    \frac{3}{2(\Gamma-1)}
    \frac{L_0}{2-\Gamma}
    \left(\frac{2m_\e c^2}{E_P}\right)^{2-\Gamma}
    \left[
      \left(\frac{E_M}{2E_{\rm CMB}}\right)^{1-\Gamma/2}
      -
      \left(\frac{E_m}{2E_{\rm CMB}}\right)^{1-\Gamma/2}
    \right]\,.
  \end{aligned}
\end{equation}
Note that there is no band correction here because the band limits are fixed in physical units independent of redshift as a result of the redshifting of the CMB.

The band-specific IC flux is
\begin{equation}
  F_{{\rm ICC},mM}
  =
  \int_{E_m}^{E_M} dE F_{{\rm ICC},E}
  =
  \frac{(1+z)}{4\pi D_L^2} \int_{(1+z)E_m}^{(1+z)E_M} dE' \frac{L'_{{\rm ICC},E'}}{E'}\,.
\end{equation}
Again we can use the equality of the IC and VHEGR luminosities to determine the integrand:
\begin{equation}
  \begin{aligned}
    \frac{L'_{{\rm ICC},E'}}{E'}
    &=
    \frac{3}{2(\Gamma-1)}
    \frac{1}{E'} \frac{d}{dE'_M} \int_{2m_\e c^2\sqrt{E'_m/2E_{\rm CMB}'}}^{2m_\e c^2\sqrt{E'_M/2E_{\rm CMB}'}} dE' L'_{E'}
    =
    \frac{3}{2(\Gamma-1)}
    \frac{1}{E'} \frac{d}{dE'_M}
    \int_{E'_m}^{E'_M} dE' \frac{m_\e c^2}{\sqrt{2E'E_{\rm CMB}'}} L'_{2m_\e c^2\sqrt{E'/2E_{\rm CMB}'}}\\
    &=
    \frac{3}{2(\Gamma-1)}
    \frac{1}{E'} \frac{m_\e c^2}{\sqrt{2E'E_{\rm CMB}'}} L'_{2m_\e c^2\sqrt{E'/2E_{\rm CMB}'}}\,.
  \end{aligned}
\end{equation}
Therefore, the flux of the ICC halo is given by
\begin{equation}
  \begin{aligned}
    F_{{\rm ICC},mM}
    &=
    \frac{(1+z)}{4\pi D_L^2}
    \frac{3}{2(\Gamma-1)}
    \int_{(1+z)E_m}^{(1+z)E_M} dE' 
    \frac{1}{E'} \frac{m_\e c^2}{\sqrt{2E'E_{\rm CMB}'}} L'_{2m_\e c^2\sqrt{E'/2E_{\rm CMB}'}}\\
    &=
    \frac{(1+z)}{4\pi D_L^2}
    \frac{3}{2(\Gamma-1)}
    \int_{2m_\e c^2\sqrt{E'_m/2E_{\rm CMB}'}}^{2m_\e c^2\sqrt{E'_M/2E_{\rm CMB}'}} dE' \left(\frac{2m_\e c^2}{E'}\right)^2 \frac{1}{2E_{\rm CMB}'} L'_{E'}\\
    &=
    \frac{1}{4\pi D_L^2}
    \frac{3}{2(\Gamma-1)}
    \left(\frac{2 m_\e c^2}{E_p}\right)^2 \frac{1}{2E_{\rm CMB}}
    \int_{2m_\e c^2\sqrt{E'_m/2E_{\rm CMB}'}}^{2m_\e c^2\sqrt{E'_M/2E_{\rm CMB}'}} dE' L_0 \left(\frac{E'}{E_p}\right)^{-(1+\Gamma)}\\
    &=
    \frac{1}{4\pi D_L^2}
    \frac{3}{2(\Gamma-1)}
    \left(\frac{2 m_\e c^2}{E_p}\right)^2 \frac{1}{2E_{\rm CMB}}
    \frac{L_0}{(-\Gamma)}
    \left[
      \left( \frac{2m_\e c^2}{E_p}\sqrt{\frac{E_M}{2E_{\rm CMB}}}\right)^{-\Gamma}
      -
      \left( \frac{2m_\e c^2}{E_p}\sqrt{\frac{E_m}{2E_{\rm CMB}}}\right)^{-\Gamma}
    \right]\\
    &=
    \frac{1}{4\pi D_L^2}
    \frac{3}{2(\Gamma-1)}
    \frac{1}{2E_{\rm CMB}}
    \frac{L_0}{(-\Gamma)}
    \left(\frac{2m_\e c^2}{E_p}\right)^{2-\Gamma}
    \left[
      \left( \frac{E_M}{2E_{\rm CMB}}\right)^{-\Gamma/2}
      -
      \left( \frac{E_m}{2E_{\rm CMB}}\right)^{-\Gamma/2}
    \right]\,.
  \end{aligned}
\end{equation}
Again the absence of redshift factors is due to the fact that the band limits are independent of $z$ themselves.

Therefore, we find generally that we will expect
\begin{equation}
  \frac{\F_{{\rm ICC},mM}}{\F_{mM}} \propto (1+z)^{\Gamma-2}
  \qquad\text{and}\qquad
  \frac{F_{{\rm ICC},mM}}{F_{mM}} \propto (1+z)^{\Gamma-2}\,,
\end{equation}
consistent with what was found in Equations (\ref{eq:Fiso}) and (\ref{eq:homogeneous_image}) with $\Gamma=\Gamma_l$ and after including the $(1+z)^{1-\Gamma_l}$ in $N(\Gamma_l,z)$.  The reason for the restriction to $\Gamma_l$ is that the spectral index enters only into the band-correction of the low-energy flux, i.e., the definition of $F_{35}$; for the ICC component the lack of redshifting the lepton energy, and thus source VHEGR, responsible for the observed gamma ray eliminates any dependence on $\Gamma_h$.

\section{Boltzmann Equation in the Presence of Dissipation}\label{app:Boltzmann}
Here we derive general expressions for the Boltzmann equation in the presence of dissipation in the soft scattering limit that is appropriate for the inverse Compton cooling of an ultrarelativistic population of leptons via the CMB.  Intuitively, the result may be presaged as follows: (i) in the soft scattering limit there is a negligble change in the direction of the lepton's momentum, (ii) the change in the magnitude is given by the standard inverse Compton cooling formula \citep[see, e.g.,][]{1986rpa..book.....R} and may treated as as drag, (iii) the drag force can be uniquely included in a manifestly conservative form of the Boltzmann equation which corresponds to the inclusion of a scattering term.  However, given the apparent arbitrariness of this procedure, we instead present a rigorous derivation here, highlighting which assumptions are made where.

The standard development of the Boltzmann equation is dependent upon Liouville's theorem, 
\begin{equation}
  \frac{\partial f}{\partial t} 
  + \bv\cdot\frac{\partial f}{\partial\bx} 
  + \dot{\bp}\cdot\frac{\partial f}{\partial\bp} = 0
  \quad\Rightarrow\quad
  \frac{d}{dt} \int d^3\!x \, d^3\!p \, f
  =
  0\,,
\end{equation}
namely the conservation of phase space volume during the evolution of the distribution function.  This is, however, not true generically, itself failing to hold when dissipation is present.

We begin with the standard proof of the Liouville's theorem via Hamilton's equations.  Let $\mathcal{H}(\bx,\bp)$ be a Hamiltonian describing the dynamics of the particles we wish to characterize with $f(\bx,\bp)$.  Then, from Hamilton's equations we have,
\begin{equation}
  \bv = \frac{\partial\mathcal{H}}{\partial \bp}
  \quad\text{and}\quad
  \dot{\bp} = -\frac{\partial\mathcal{H}}{\partial \bx}\,,
\end{equation}
and thus,
\begin{equation}
  \begin{aligned}
    \frac{\partial f}{\partial t} 
    + \bv\cdot\frac{\partial f}{\partial\bx} 
    + \dot{\bp}\cdot\frac{\partial f}{\partial\bp} 
    &=
    \frac{\partial f}{\partial t} 
    + \frac{\partial\mathcal{H}}{\partial\bp}\cdot\frac{\partial f}{\partial\bx} 
    - \frac{\partial\mathcal{H}}{\partial\bx}\cdot\frac{\partial f}{\partial\bp}
    =
    \frac{\partial f}{\partial t} 
    + \frac{\partial}{\partial\bx}\cdot\frac{\partial\mathcal{H}}{\partial\bp} f 
    - \frac{\partial}{\partial\bp}\cdot\frac{\partial\mathcal{H}}{\partial\bx} f
    =
    0\,,
  \end{aligned}
\end{equation}
which is explicitly in flux conservative form.  Supplemented with the requirement that $f$ vanishes at infinity (i.e., infinite spatial extent and at infinite energy), this implies conservation of the integrated distribution function.

The above proof explicitly fails if the forces cannot be obtained from a Hamiltonian formalism.  More generally, Liouville's theorem does not hold if we cannot subsume the force term into the momentum derivative (assuming $\bv=\bp/m$), at which point the Boltzmann equation cannot be placed in a flux conservative form.  Dissipative systems are frequently non-Hamiltonian, e.g., friction in which $\dot{\bp}\propto-\bp$.  However, such systems are generally only non-Hamiltonian as a consequence of an incomplete description: when the entire system is considered microscopically energy should remain conserved, and the individual constituents well described by some microphysical Hamiltonian.

Of particular importance here is the case of inverse Compton cooling, where there is a clear microphysical mechanism responsible for the dissipative terms in the evolution of the electrons/positrons.  Correspondingly, there is an obvious extension of the lepton Boltzmann equations, including the scattering terms associated with the lepton-photon interactions.  Here we derive a general formula for these sorts of scattering-induced dissipative contributions and demonstrate explicitly lepton conservation.

\subsection{General Considerations in the Soft Scattering Limit}
We begin by noting that the dissipation process here is not a uniform force in momentum, and therefore not associated with a $\dot{\bp}$ obtained from some Hamiltonian, but rather due to a scattering process.  In the most general case the scattering terms can be written as:
\begin{equation}
  \frac{\partial f}{\partial t}
  + \bv\cdot\frac{\partial f}{\partial\bx} 
  + \dot{\bp}\cdot\frac{\partial f}{\partial\bp} 
  =
  \int d^3\!q \left[ W(\bp,\bq) f(\bq) - W(\bq,\bp) f(\bp) \right]\,,
\end{equation}
where $W(\bp,\bq)$ is the scattering kernel that takes particles from momentum $\bq$ to momentum $\bp$.\footnote{We are interested in the extreme low-density limit, and thus have ignored terms which arise when the state occupancy is high, e.g., Fermi blocking or stimulated emission.}  Note that there are two terms, the first associated with scattering into $\bp$ and the second associated with scattering out of $\bp$.  It is straightforward to show that upon integrating over $d^3\!p$ the term on the right-hand side vanishes identically.  Thus if only conservative forces are considered on the left-hand side, this will still satisfy Liouville's theorem.

We will ultimately be interested in nearly continuous scattering processes, i.e., ones that extract small amounts of energy per interaction.  Formally, we may define such a process by the condition that $W(\bp,\bq)$ is significant only when the momentum change, $\bs\equiv\bp-\bq$, satisfies $|\bs|\ll|\bp|$, i.e., scattering in the soft limit.  For example, we might imagine a quantized scattering process that takes $\bp$ to $\bp-\bs$, with:
\begin{equation}
  W(\bp,\bq) 
  =
  W(\bp)\left[ \delta^3(\bq+\bs-\bp) - \delta^3(\bp) \right]\,.
\end{equation}
In the limit in which $\bs\rightarrow0$, this may be expressed as
\begin{equation}
  W(\bp,\bq) 
  = 
  \bW(\bp) \cdot \frac{\partial}{\partial\bs}\delta^3(\bs)
\end{equation}
for some vector function $\bW(\bp)$.

This is necessarily only the first term in a gradient expansion of the scattering kernel that may be made formal in Fourier space.  That is, set
\begin{equation}
  \tilde{W}(\bp,\bk) = \int d^3\!s e^{-2\pi i \bk\cdot\bs} W(\bp,\bp+\bs)\,,
\end{equation}
and expand around $\bk=0$ to obtain
\begin{equation}
  \tilde{W}(\bp,\bk) = \tilde{W}(\bp,0)
  + \bk\cdot\frac{\partial \tilde{W}}{\partial\bk} (\bp,0)
  + \frac{1}{2} k^i k^j \frac{\partial^2 \tilde{W}}{\partial k^i \partial k^j} (\bp,0)
  + \dots\,.
\end{equation}
Inverting the Fourier transform yields
\begin{equation}
  W(\bp,\bp+\bs)
  =
  U_0(\bp) \delta(\bs)
  + U_1^i(\bp) \cdot \frac{\partial}{\partial s^i} \delta^3(\bs)
  + U_2^{ij}(\bp) \frac{\partial}{\partial s^i \partial s^j} \delta^3(\bs)
  + \dots
\end{equation}
where
\begin{equation}
  U_0(\bp) = 2\pi \tilde{W}(\bp,0)\,,\quad
  U_1^i(\bp) = - 2\pi i \frac{\partial \tilde{W}}{\partial k^i}(\bp,0)\,,\quad
  U_2^{ij}(\bp) = - 2\pi \frac{\partial \tilde{W}}{\partial k^i \partial k^j}(\bp,0)\,,\quad
  \dots
\end{equation}
Number conservation requires that $U_0=0$ identically, leaving $U_1$ as the lowest order non-zero term.  We are interested in the soft limit, i.e., long-wavelength limit, and hence only when the first term dominates, for which we identify
\begin{equation}
  \bW(\bp) = -2\pi i \frac{\partial \tilde{W}}{\partial\bk}(\bp,0)\,.
\end{equation}
In this limit the scattering term looks like
\begin{equation}
  \begin{aligned}
    &\int d^3\!q \left[ W(\bp,\bq) f(\bq) - W(\bq,\bp) f(\bp) \right]
    =
    -\int d^3\!s\,
    \left[
      f(\bp-\bs)
      \bW(\bp) \cdot \frac{\partial}{\partial\bs} \delta^3(\bs) 
      +
      f(\bp)
      \bW(\bp-\bs) \cdot \frac{\partial}{\partial\bs} \delta^3(\bs) 
      \right]\\
    &\qquad\qquad\qquad=
    \int d^3\!s\,
    \delta^3(\bs) 
    \left[
      \bW(\bp) \cdot \frac{\partial}{\partial\bs} 
      f(\bp-\bs)
      +
      f(\bp)
      \frac{\partial}{\partial\bs}
      \cdot \bW(\bp-\bs)
      \right]
    =
    -
    \frac{\partial}{\partial\bp}
    \cdot \bW(\bp) f(\bp)
    \,,
  \end{aligned}
\end{equation}
which appears similar to a force term, though already in flux-conservative form.  As such, it already is guaranteed to conserve phase-space density (i.e., satisfy Liouville's theorem).  The corresponding Boltzmann equation is then 
\begin{equation}
  \frac{\partial f}{\partial t}
  + \bv\cdot\frac{\partial f}{\partial\bx} 
  + \dot{\bp}\cdot\frac{\partial f}{\partial\bp} 
  + \frac{\partial}{\partial\bp} \cdot \bW f
  =
  0\,.
\end{equation}

From the perspective of the leptons, this scattering is indeed a dissipative (or accelerating) process.  This is evident from considering the evolution of the change in the energy induced by the scattering:
\begin{equation}
  \begin{aligned}
    \dot{E} 
    &=
    \frac{d}{dt} \int d^3\!x\,d^3\!p\, E f
    =
    \int d^3\!x\,d^3\!p\, pc \frac{\partial f}{\partial t}
    =
    - \int d^3\!x\,d^3\!p\,
    pc
    \frac{\partial}{\partial\bp} \cdot (\dot{\bp}+\bW) f\\
    &=
    \int d^3\!x\,d^3\!p\,
    f (\dot{\bp}+\bW) \cdot
    \frac{\partial}{\partial\bp} pc
    =
    \int d^3\!x\,d^3\!p\,
    f (\dot{\bp}+\bW) \cdot \bhp c\,.
  \end{aligned}
\end{equation}
Inserting $f=\delta^3(\bp-\bp_0)$, this gives for the average change in energy for a single particle
\begin{equation}
  \dot{E}(\bp_0)
  =
  \bhp\cdot\left[\dot{\bp}+\bW(\bp_0)\right]c\,.
\end{equation}
Note that this describes two contributions to the energy evolution of the particle: work done on the particle by the Hamiltonian forces acting on the particles, $\bhp\cdot\dot{\bp}c$, and the losses due to scattering $\bhp\cdot\bW c$.  In the case of primary case here, in which $\dot{\bp}$ describes the gyration about a background magnetic field, $\bhp\cdot\dot{\bp}=0$ identically, with the consequence that the energy evolves only as a result of scattering losses.  Thus,
\begin{equation}
  \dot{E}(\bp_0)
  =
  \bhp\cdot\bW(\bp_0)c\,,
  \label{eq:cont_diss}
\end{equation}
from which it will often be more convenient to determine $\bW(\bp)$.

\subsection{Restriction to Inverse Compton Scattering}
For inverse Compton scattering, choosing $\bW(\bp)=-(p^2/m_\e c t_{IC})\bhp$ provides the correct dissipation rate, though in this case we may derive the scattering terms explicitly.  We begin by noting that while we may follow the impact of scattering upon the photons explicitly, we will make the following simplifying assumptions:
\begin{enumerate}
\item Photons are injected, leave the region of relevance, or are so numerous that their distribution function, $g(\bq)$, is effectively unchanged by lepton scattering.
\item The photon distribution is mono-energetic, i.e.,
  \begin{equation}
    g(\bq) = \frac{u_s}{4\pi c q_s^3} \delta(q-q_s)\,.
  \end{equation}
\item The seed photon energy is sufficiently small that we can use the low-energy Compton formula, $\bq=2p^2q_s \bhp = 2 q_s p \bp$ where all momenta are measured in $m_\e c$.
\item The seed photon energy is sufficiently small that $q\ll p$, which is formally similar to the previous assumption.
\end{enumerate}
Adopting $\bp$ and $\bq$ for the incoming lepton and seed photon momenta, respectively, and $\bp'$ and $\bq'$ for the corresponding outgoing momenta, we obtain with the above assumptions
\begin{equation}
  \begin{aligned}
    \dot{f}_{\rm in}
    &=
    \int d^3\!q\,W(\bp,\bq) f(\bq)\\
    &=
    \int d^3\!p'\,d^3\!q\,d^3\!q'\,
    \sigma_T c \delta^3(\bp+\bq-\bp'-\bq') \delta^3(\bq-2q_s p'\bp') f(\bp') g(\bq')\\
    &=
    \sigma_T \frac{u_s}{4\pi q_s^3}
    \int d^3\!p'\,d^3\!q\,d^3\!q'\,
    f(\bp')
    \delta^3(\bp+\bq-\bp'-\bq') \delta^3(\bq-2q_s p'\bp')
    \delta(q'-q_s)\\
    &\approx
    \sigma_T \frac{u_s}{q_s}
    \int d^3\!p'\,d^3\!q\, f(\bp') \delta^3(\bp+\bq-\bp') \delta^3(\bq-2q_s p'\bp')\\
    &=
    \sigma_T \frac{u_s}{q_s}
    \int d^3\!p' f(\bp') \delta^3(\bp+2q_sp'\bp'-\bp')\\
    &\approx
    \sigma_T \frac{u_s}{q_s}
    \frac{f(\bp + 2q_sp\bp)}{\det\left[\bmath{1}(1-2q_sp)-2q_s p\bhp\bhp\right]}\\
    &\approx
    \sigma_T \frac{u_s}{q_s}
    \frac{f(\bp + 2q_sp\bp)}{1-8q_sp}
  \end{aligned}
\end{equation}
and
\begin{equation}
  \begin{aligned}
    \dot{f}_{\rm out}
    &=
    \int d^3\!q\,W(\bq,\bp) f(\bp)\\
    &=
    \int d^3\!p'\,d^3\!q\,d^3\!q'\,
    \sigma_T c\delta^3(\bp'+\bq-\bp-\bq') \delta^3(\bq-2q_s p\bp) f(\bp) g(\bq')\\
    &=
    \sigma_T \frac{u_s}{4\pi q_s^3}
    \int d^3\!p'\,d^3\!q\,d^3\!q'\,
    f(\bp)
    \delta^3(\bp'+\bq-\bp-\bq') \delta^3(\bq-2q_s p\bp)
    \delta(q'-q_s)\\
    &\approx
    \sigma_T \frac{u_s}{q_s}
    \int d^3\!p'\,d^3\!q\, f(\bp) \delta^3(\bp'+\bq-\bp) \delta^3(\bq-2q_s p\bp)\\
    &=
    \sigma_T \frac{u_s}{q_s}
    \int d^3\!p' f(\bp) \delta^3(\bp'+2q_sp\bp-\bp)\\
    &\approx
    \sigma_T \frac{u_s}{q_s} f(\bp)\,.
  \end{aligned}
\end{equation}
Therefore,
\begin{equation}
  \dot{f}_{\rm in} - \dot{f}_{\rm out}
  =
  \sigma_T \frac{u_s}{q_s}
  \left[
    \frac{f(\bp + 2q_sp\bp)}{1-8q_sp}
    -
    f(\bp)
    \right]
  \approx
  2\sigma_T u_s
  \left[
    4 p f
    +
    p \bp\cdot\frac{\partial f}{\partial\bp}
    \right]
  \approx
  \frac{3}{2}
  \frac{\partial}{\partial\bp}\cdot 
  \frac{p\bp}{m_\e c t_{IC}} f\,,
\end{equation}
where in the final step we used $4\sigma_T u_s/3 = m_\e c /t_{IC}$ and reinserted the $m_\e c$ in the momenta definitions.  The discrepancy of $3/2$ between this and the expression quoted following Equation (\ref{eq:cont_diss}) is due to the simplified Compton formula assumed, and in particular, the assumption of isotropic scattering.

\section{Solving the Boltzmann Equation by the Method of Characteristics} \label{app:chars}
Here we derive Equation (\ref{eq:BES}), explicitly solving the Boltzmann equation in the presence of a locally uniform magnetic field.  For compactness we assume near homogeneity and suppress the spatial dependence here. This is done transforming the PDE into a set of ODEs along characteristic curves, defined by
\begin{equation}
  \frac{d}{d\eta} = \left(\dot{\bp}+\bW\right)\cdot\grad_p
  \quad\Rightarrow\quad
  \frac{d\bp}{d\eta} = \dot{\bp}+\bW\,,
\end{equation}
along which we have the first order ODE,
\begin{equation}
  \frac{df}{d\eta}
  =
  \dot{f}_{\rm inj}[\bp(\eta,\bp_0)]
  -
  (\grad_p\cdot\bW) f\,,
\end{equation}
which may be solved directly to yield
\begin{equation}
  \begin{gathered}
    f(\bp) = e^{-\mu} \left[\int d\eta\, e^\mu \dot{f}_{\rm inj} + f_0\right]\\
    \text{where}\quad
    \mu = \int d\eta \grad_p\cdot\bW\,,
  \end{gathered}
\end{equation}
and $f_0$ is a constant of integration that may depend on $\bp_0$ and will usually set to zero.  The difficulty usually lies in the construction of the characteristic curves, $\bp(\eta,\bp_0)$, though in our case, these are particularly simple.

\begin{figure}
  \begin{center}
    \includegraphics[width=0.33\textwidth]{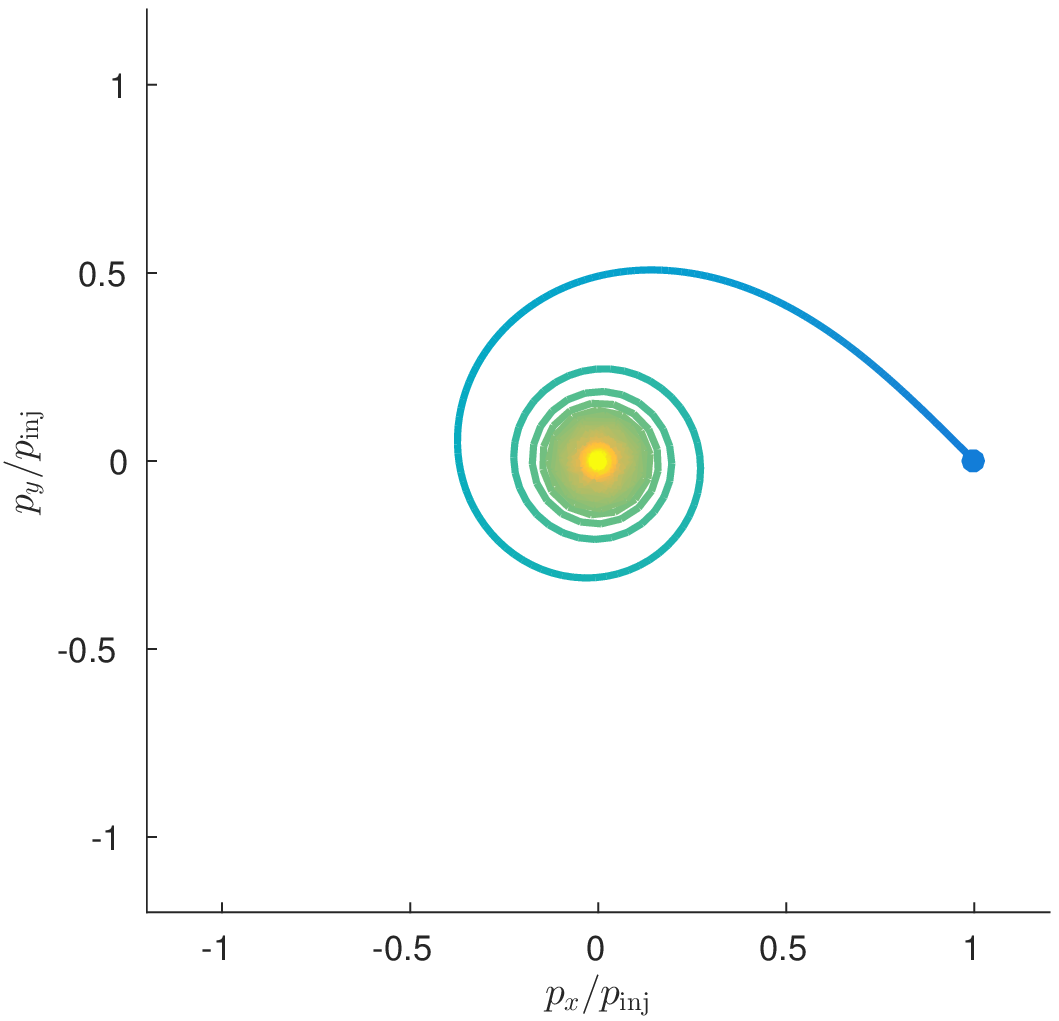}
    \includegraphics[width=0.33\textwidth]{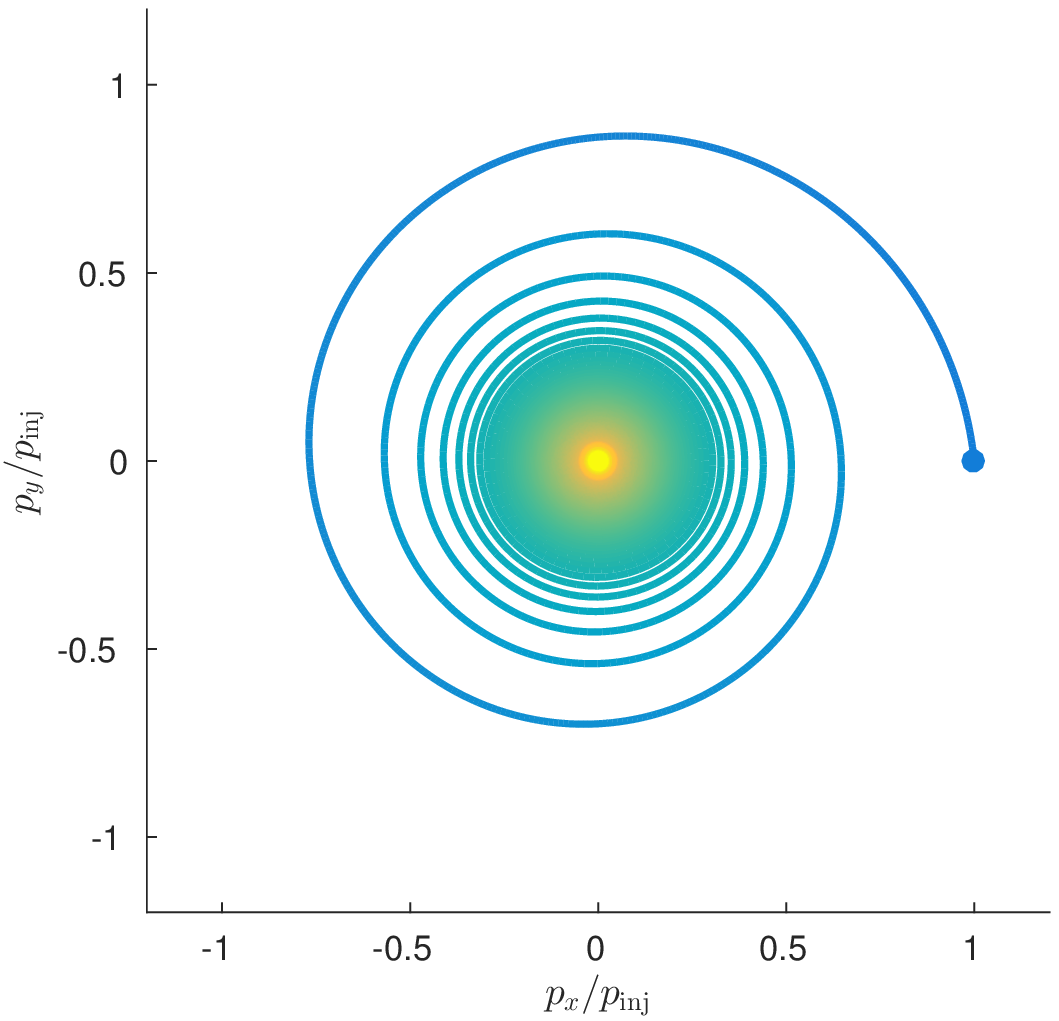}
    \includegraphics[width=0.33\textwidth]{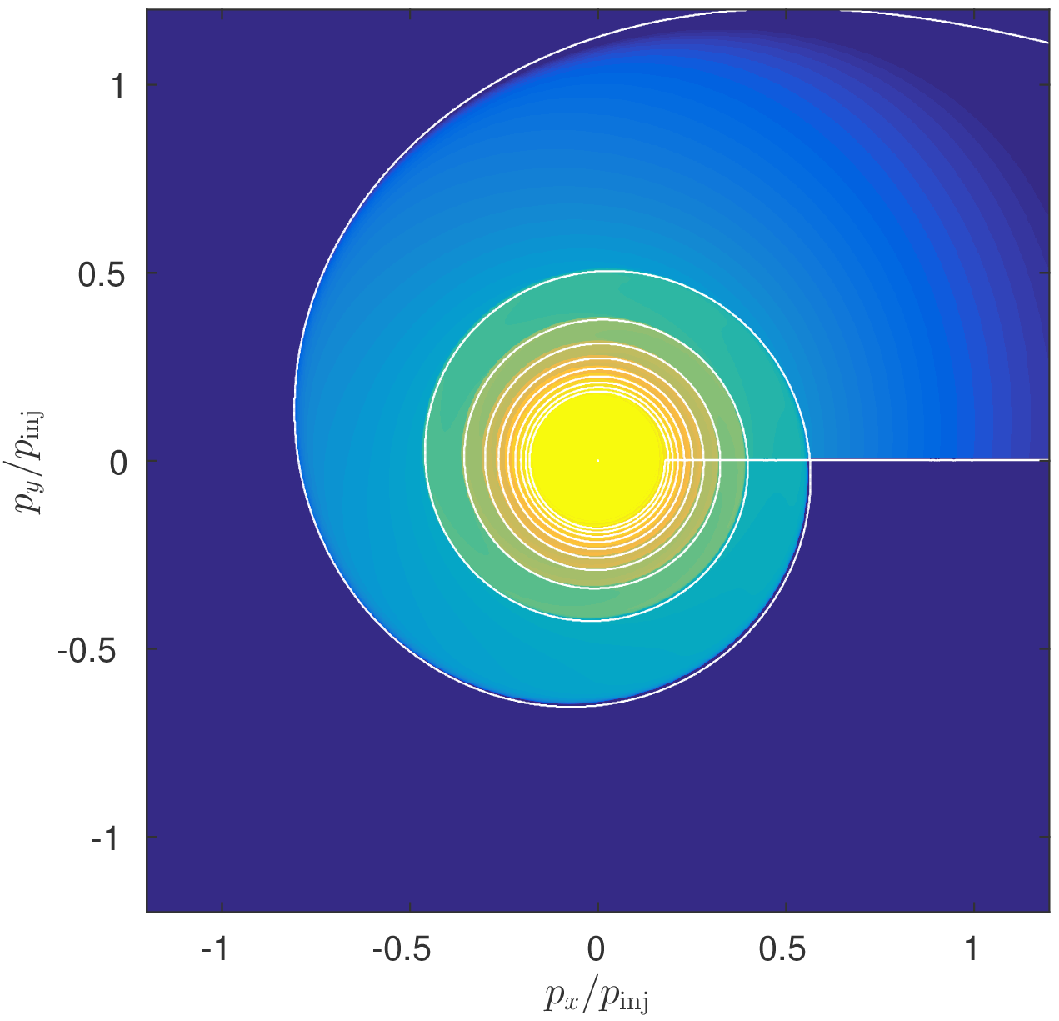}
  \end{center}
  \caption{Stationary lepton distribution function solutions.  Left and Center: Leptons are injected at the point indicated and spiral inwards along the characteristic.  These differ in the assumed value of $\omega_B t_{\rm IC} \sin\theta_p$, corresponding to 1 and 9, respectively.  Right: Leptons are injected along the positive $p_x$ axis with magnitude $p_x^{-2}$, for which $p_{\rm inj}$ is a characteristic momentum.  The white contours show the maximum gyration order contributing to each position in phase space up through 10.  In app plots the color indicates the magnitude of $f$ in arbitrary logarithmic units (9 orders of magnitude are shown). }\label{fig:exfsol}
\end{figure}

For deflection in a locally uniform magnetic field
\begin{equation}
\dot{\bp} + \bW = e \frac{\bp\times\bB}{p} - \frac{p \bp}{t_{\rm IC} m_\e c}\,.
\end{equation}
Since the Lorentz force is orthogonal to $\bp$, the pairs cool solely by inverse Compton.  Nevertheless, the magnetic field does redistribute momenta azimuthally about the field.  Defining polar momentum coordinates relative to $\bB$, the characteristics are given by
\begin{equation}
  \frac{\partial}{\partial\eta}
  =
  - \frac{p^2}{t_{\rm IC} m_\e c} \frac{\partial}{\partial p}
  +
  \omega_B \frac{m_\e c}{p} \sin\theta_p \frac{\partial}{\partial\phi_p}\,,
\end{equation}
where $\omega_B=eB/m_\e c$ is the cyclotron frequency, and thus
\begin{equation}
  \frac{\partial p}{\partial\eta} = - \frac{p^2}{t_{\rm IC} m_\e c}\,,~~
  \frac{\partial \theta_p}{\partial\eta} = 0\,,~~
  \frac{\partial \phi_p}{\partial\eta} = \omega_B \frac{m_\e c}{p} \sin\theta_p\,.
\end{equation}
These may be immediately integrated to give the desired characteristic
\begin{equation}
  \begin{gathered}
  p(\eta) = \frac{t_{\rm IC} m_\e c}{\eta + t_{\rm IC} m_\e c/p_0}\,,~~
  \theta_p(\eta) = \theta_{p,0}\,,\\
  \phi_p(\eta) = \phi_{p,0} + \eta \omega_B \frac{m_\e c}{p_0} \sin\theta_{p,0} + \frac{\eta^2}{2 t_{\rm IC}} \omega_B\sin\theta_{p,0}\,.
  \end{gathered}
\end{equation}
We have some freedom to choose $\bp_0$, and use this to set $p_0=\infty$, which produces the two simplifying results $p(\eta) = t_{\rm IC} m_\e c/\eta$ and $f(\bp_0)=0$.  In this case the integrating factor $\mu$ is
\begin{equation}
  \mu = - \int d\eta \frac{4 p}{t_{\rm IC} m_\e c} = - 4\log\eta
  \quad\Rightarrow\quad
  e^{\mu} = \eta^{-4} = \frac{p^4}{(t_{\rm IC} m_\e c)^4}\,.
\end{equation}
Therefore,
\begin{equation}
  f(\bp) = p^{-4} \int_0^{t_{\rm IC} m_\e c/p}  d\eta\,
  \left( \frac{t_{\rm IC} m_\e c}{\eta} \right)^4
  \dot{f}_{\rm inj}\left(\frac{t_{\rm IC} m_\e c}{\eta},\theta_p,\phi_p-\omega_B t_{\rm IC} \sin\theta_p\frac{m_\e^2c^2}{2p^2}+\frac{\eta^2}{2 t_{\rm IC}} \omega_B \sin\theta_p\right)\,,
\end{equation}
where we have taken care to set $\phi_{p,0}$ such that the leptons will have gyrated to $\phi_p$ when they have cooled to energy $pc$.  Upon affecting a change of variable from $\eta$ to $p$, this is identical to Equation (\ref{eq:BES}).

Figure \ref{fig:exfsol} shows example solutions for a $\delta$-function injection, i.e.,
\begin{equation}
  f_{\rm inj} = \frac{f_0}{p_{\rm inj}^2} \delta(p-p_{\rm inj}) \delta(\theta_p-\theta_{\rm inj}) \delta(\phi_p)\,.
\end{equation}
for different values of $\omega_B t_{\rm IC} \sin\theta_p$.  The leptons cool and gyrate, follwing the characteristic that passes through the injection site.  Larger values of $\omega_B t_{\rm IC} \sin\theta_p$ produce more rapid gyrations, and therefore approximate isotropization in $\phi_p$, at larger values of $p$.

A continuous injection model along the postitive $p_x$ axis, i.e.,
\begin{equation}
  f_{\rm inj} = \frac{f_0}{p_{\rm inj}^2} \left(\frac{p}{p_{\rm inj}}\right)^{-2} \delta(\theta_p-\theta_{\rm inj}) \delta(\phi_p)\,.
\end{equation}
is shown in the right panel of Figure \ref{fig:exfsol}.  The extended injection produces a correspondingly smooth $f$, though there are regions at which no particles are present as a result of the rapid cooling at large $p$.  In addition, as evident in Equation (\ref{eq:homogeneous_evolved_pdf}) there are potentially multiple gyration orders that contribute to $f$.  These increase as $p$ decreases at a rate that depends on $\omega_B t_{\rm IC}\sin\theta_p$, and jumps by one each time the characterstic that connects to $p=\infty$ crosses the injection axis.

\section{Derivation of the Inverse Compton Flux} \label{sec:DICF}
Here we derive the inverse Compton flux within the various limits described in Section \ref{sec:ICEG}.  Generally, the flux of the inverse Compton photons with momenta $\bmath{q}'$ is given in the low-density limit\footnote{We ignore the Fermi-blocking and Bose-enhancement terms that would be present at high occupancies.}
\begin{equation}
  \frac{dN_{\rm IC,\epm}}{dt' d^3\!x' d^3\!q'} (\bx',\bq')
  =
  \int d^3\!q'_s d^3\!p' d^3\!p'_f
  \frac{d\sigma}{d^3\!q' d^3\!q'_s d^3\!p' d^3\!p'_f}
  c
  f_s(\bx',\bq'_s) f_{\epm}(\bx',\bp')\,.
\end{equation}
Inserting the approximate differential cross section in Equation (\ref{eq:dsig_example}) and integrating over the second $\delta$-function gives
\begin{equation}
  \frac{dN_{\rm IC,\epm}}{dt' d^3\!x' d^3\!q'} (\bx',\bq')
  =
  \int d^3\!q'_s d^3\!p'
  \sigma_T c \,
  \delta^3\left(\bmath{q}'- \frac{2 q_s' p'}{m_\e^2 c^2} \bmath{p}'\right)
  f_s(\bx',\bq'_s) f_{\epm}(\bx',\bp')\,.
\end{equation}
The remaining $\delta$-function requires more care due to its more complicated argument.  In particular, note that generally
\begin{equation}
  \int f(\bmath{x}) \delta^n[\bmath{g}(\bmath{x})] d^n\!x
  =
  \sum_{{\bf r}\text{ s.t. }{\bf g}({\bf r})=0} 
  \frac{f(\bmath{r})}{ \left|\partial\bmath{g}/\partial\bmath{x} \right|(\bmath{r})}\,,
\end{equation}
where $\left|\partial\bmath{g}/\partial\bmath{x} \right|(\bmath{r})$ is the Jacobian of $\bmath{g}$, and therefore
\begin{equation}
  \left|\frac{\partial}{\partial\bp'} \frac{2 q'_s p'}{m_\e^2 c^2} \bp' \right|
  =
  \left(\frac{2 q'_s p'}{m_\e^2 c^2}\right)^3 \left|\bmath{1} +  \bhp' \bhp' \right|
  =
  2 \left(\frac{2 q'_s p'}{m_\e^2 c^2}\right)^3\,,
\end{equation}
in which we employed the freedom to evaluate the determinant in coordinates for which $\bhp'=(1,0,0)$, and note that
\begin{equation}
  \bp' = p' \bhp' = p' \bhq' = m_\e c \sqrt{\frac{q'}{2q'_s}} \bhq' =
  \frac{m_\e c \bq'}{\sqrt{2 q'_s q'}}
  \quad\Rightarrow\quad
  2 \left(\frac{2 q'_s p'}{m_\e^2 c^2}\right)^3 = 2 \left(\frac{2q_s'q'}{m_\e^2 c^2}\right)^{3/2}
  \,.
\end{equation}
Inserting these then gives generally
\begin{equation}
  \frac{dN_{\rm IC,\epm}}{dt' d^3\!x' d^3\!q'} (\bx',\bq')
  =
  \int d^3\!q'_s
  \frac{\sigma_T c}{2}
  \left(\frac{m_\e^2 c^2}{2 q_s' q'}\right)^{3/2}
  f_s(\bx',\bq'_s) f_{\epm}\left(\bx',\frac{m_\e c \bq'}{\sqrt{2q_s'q'}}\right)\,.
\end{equation}
Finally, we insert the mono-energetic, isotropic seed photon distribution in Equation (\ref{eq:fseed}), effect the final integral over the $\delta$-function and obtain
\begin{equation}
  \frac{dN_{{\rm IC},\epm}}{dt' d^3\!x' d^3\!q'}
  =
  \frac{\sigma_T c u_s}{2 q_s' c}
  \left(\frac{m_\e^2 c^2}{2 q_s' q'}\right)^{3/2}
  f_{\epm}\left.\left(\bx',\frac{m_\e c \bq'}{\sqrt{2q_s'q'}}\right)
  \right|_{q_s'=\frac{E_{\rm CMB}'}{c}}
  \,.
  \tag{\ref{eq:dNICC}}
\end{equation}

This is related to the surface brightness observed by \Fermi via Equation (\ref{eq:dNg}), which trivially gives the first line of Equation (\ref{eq:image_master}).  Inserting the expression above gives
\begin{equation}
  \begin{aligned}
    \frac{dN_\gamma}{dt dE d^2\!\alpha}
    &=
    \frac{dt'}{dt}
    \frac{dE'}{dE}
    \sum_{S=e^+,e^-}
    \int_{\Omega'_{\rm{Fermi}}} \frac{q'^2 d\Omega'_q}{c}
    \frac{d^2\!x'}{d^2\!\alpha}
    \int d\ell' \frac{dN_{{\rm IC},S}}{dt' d^3\!x' d^3\!q'} \left(\bx',\frac{E'\bhl'}{c}\right)\\
    &=
    \sum_{S=e^+,e^-}
    \frac{q'^2}{c}
    \frac{A_{\rm eff}}{D_A^2(1+z)^2} D_A^2
    \int d\ell' \frac{dN_{{\rm IC},S}}{dt' d^3\!x' d^3\!q'} \left(\bx',\frac{E'}{c}\bhl'\right)\\
    &=
    \sum_{S=e^+,e^-}
    \frac{A_{\rm eff}}{(1+z)^2}
    \frac{\sigma_T u_s}{2 E_{\rm CMB}' c^2}
    \left(\frac{m_\e^2 c^4}{2 E_{\rm CMB}' E'}\right)^{3/2}
    E'^2
    \int d\ell' 
    f_{S}\left(\bx', m_\e c \sqrt{\frac{E'}{2E_{\rm CMB}'}}\bhl'\right)\,.\\
  \end{aligned}
\end{equation}
where $E'=(1+z)E$.  Noting that $E'/E_{\rm CMB}'=E/E_{\rm CMB}$ produces the second line in Equation (\ref{eq:image_master}).

\section{Applicability on the Forward Scattering Limit} \label{app:fsl}
We have made a number of explicit and implicit assumptions regarding the direction of the inverse Compton gamma rays and their impact on the responsible leptons.  These fundamentally relate to the forward scattering limit, i.e., the assumption that the up-scattered photon propagates only in the direction of the of scattering lepton.  This has two consequences: first it produces a one-to-one map between observed halo photons and leptons propagating towards us, and second it simplifies the evolution of the pairs.  For a single scattering event the forward-scattering limit is an exceedingly good approximation -- the typical range of angles about the forward direction is $\gamma^{-1}$, and thus small for high-energy leptons.  However, many scatterings provide a means to evolve not just the magnitude of the lepton momentum but also its direction.  Here we show that this is generally small for the cases of interest and thus may be neglected in the construction of ICC halo images.

The typical kick a lepton receives orthogonal to its motion at each inverse Compton scattering is 
\begin{equation}
  \delta p_\perp \approx \frac{2\gamma^2 E_{\rm CMB}}{\gamma c} \approx \frac{2 \gamma E_{\rm CMB}}{c}\,,
\end{equation}
where we have assumed that the inverse Compton up-scattered gamma rays are propagating within $\gamma^{-1}$ of the original direction of motion of the lepton.  Due to the small value of $E_{\rm CMB}$ in comparison to the energy of pairs there will be many such scatterings over the cooling time of the pairs, i.e., 
\begin{equation}
  N_{\rm scat} \approx \frac{\gamma m_\e c^2}{2\gamma^2 E_{\rm CMB}} \gg 1\,.
\end{equation}
These are stochastic, scattering in random directions, and thus the net deflection grows as
\begin{equation}
  \Delta p_\perp
  \approx
  N_{\rm scat}^{1/2} \delta p_{\perp}
  \approx
  \sqrt{2\gamma m_\e E_{\rm CMB}}\,.
\end{equation}
The angular deflection, and thus the width over which halo features will be smeared by violations of the forward-scattering limit is then
\begin{equation}
  \delta\alpha \approx \frac{\Delta p_\perp}{p} =
  \sqrt{\frac{2 E_{\rm CMB}}{\gamma m_\e c^2}}
  \approx
  3^\circ\times10^{-6} \left(\frac{\gamma}{10^6}\right)^{-1/2}\,.
\end{equation}
This is much smaller than the other angular scales of relevance, including the range of angles over which the up-scattered gamma rays are distributed, $\gamma^{-1}$.  More importantly, this is small in comparison with the smearing induced by convolution with typical \Fermi PSFs, justifying its neglect.

\end{appendix}

\bibliography{bigmh,bigmh_orig}
\bibliographystyle{apj}

\end{document}